# Classical Klein-Gordon solutions, symplectic structures and isometry actions on AdS spacetimes

Max Dohse[*]

*Centro de Ciencias Matemáticas,*
*Universidad Nacional Autónoma de México,*
*Campus Morelia, C.P. 58190,*
*Morelia, Michoacán, Mexico*
(Dated: October 19, 2018)

We study classical, real Klein-Gordon theory on Lorentzian Anti de Sitter ($\text{AdS}_{1,d}$) spacetimes with spatial dimension $d$. We give a complete list of well defined and bounded Klein-Gordon solutions for three types of regions on AdS: slice (time interval times all of space), rod hypercylinder (all of time times solid ball in space), and tube hypercylinder (all of time times solid shell in space). Hypercylinder regions are of natural interest for AdS since the neighborhood of the AdS-boundary is a tube. For the solution spaces of our regions we find the actions induced by the AdS isometry group $\text{SO}(2, d)$. For all three regions we find one-to-one correspondences between initial data and solutions on the regions. For rod and tube regions this initial data can also be given on the AdS boundary. We calculate symplectic structures associated to the solution spaces, and show their invariance under the isometry actions. We compare our results to the corresponding expressions for $(3+1)$-dimensional Minkowski spacetime, arising from $\text{AdS}_{1,3}$ in the limit of large curvature radius.

## I. INTRODUCTION

There are three classic spacetimes of constant curvature in Mathematical Physics: Minkowski $\mathbb{R}^{1,d}$, de Sitter $\text{dS}_{1,d}$ and Anti de Sitter $\text{AdS}_{1,d}$ (wherein $d$ is the spatial dimension). They all have constant (zero, positive and negative) curvature. The particular interest for field theory on these spacetimes is due to their high degree of symmetry: each of them possesses the maximum number of $(d+1)(d+2)/2$ (linear independent) Killing vector fields, that is, spacetime isometries, and therefore they are called maximally symmetric spacetimes.

Apart from QFT on curved spacetime, current research related to AdS concerns mainly two topics: Black Holes and String Theory. In classical General Relativity, scalar fields on AdS spacetime containing a Schwarzschild Black Hole are studied for example by Holzegel and Smulevici in [1]. Yagdjian and Galstian in [2] investigate the limit of vanishing black hole mass in Schwarzschild-AdS spacetime (that is: pure AdS), and find the solution of the Cauchy problem for the Klein-Gordon (KG) equation $(\square - m^2)\phi(x) = f(x)$ with source term $f(x)$. AdS has been one of the most studied spacetimes in String Theory since the late 90's. This was caused by the famous conjecture of Maldacena [3], about a correspondence between type IIB string theory on $\text{AdS}_5 \times \mathbb{S}^5$ background and four-dimensional $\mathcal{N} = 4$ Super Yang-Mills theory on this spacetime's boundary $\partial(\text{AdS}_5 \times \mathbb{S}^5) = \partial \text{AdS}_5 = \mathbb{R} \times \mathbb{S}^3$. Further, in [4] Witten argues that a version of this correspondence is related to the thermodynamics of AdS black holes.

Despite AdS being such an object of interest, we found in the literature only the standard symplectic structure for standard Klein-Gordon solutions (well defined and bounded on all of AdS). Its time-independence is well known [5]. Some studies have been done of solutions that are not regular on all of space, e.g. [6]. These nonstandard solutions are well defined and bounded on (rod respectively tube) hypercylinder regions, see Section III. However, we have found no mention of a symplectic structure for these nonstandard solutions in the literature. Neither have we found addressed the issues of isometry actions on the solutions, nor the isometry-invariance of the symplectic structure(s). This work fills this gap by introducing a natural symplectic structure for the nonstandard KG solutions, and showing the isometry invariance of both standard and new symplectic structure. To this end we calculate the actions of isometries on the solutions, and as a byproduct find some contiguous relations for hyperspherical harmonics and Jacobi polynomials. Moreover, we compare our results to the corresponding cases for KG theory on Minkowski spacetime. We find correspondences between the flat limit of AdS Killing vectors, field expansions and symplectic structures and the respective Minkowski counterparts. In particular, we give the symplectic structure for hypercylinder surfaces $\Sigma_\rho$, which turns out to be independent of the radius $\rho$. The hypercylinder regions and surfaces are of interest for the conjectured AdS/CFT correspondence, because the AdS boundary is a hypercylinder surface, and its neighborhood is a tube region. Another byproduct is the Wronskian for the involved hypergeometric functions.

---

[*] max@matmor.unam.mx

In the following we give an overview of past results concerning KG theory on AdS, and briefly outline the structure of this article. The earliest publication on Quantum Theory on spacetimes with constant curvature that we found in the literature dates back to 1935 and is by Dirac [7]. He studies scalar and electron wave equations and Maxwell equations for deSitter $dS_{1,3}$ and Anti-deSitter $AdS_{1,3}$ spacetime but without giving solutions. Next we mention an article by Fronsdal from 1965 [8]. Therein, he conjectures that *"a physical theory in flat space is obtainable as the limit of a physical theory in a curved space"*, with limit being understood as that of zero curvature. In the spirit of this we compare (the limit of large curvature radius, that is: zero curvature, of) our results for AdS KG theory with the corresponding Minkowski results.

The earliest solution of the KG equation for AdS that we could spot is in the article [9] by Limic, Niederle and Raczka from 1966. Although written with hypergeometric functions, their solutions are what we call AdS-Jacobi modes. These modes have a discrete set of frequencies, dubbed "magic frequencies" in [10]. In their next article [11] they also present one of the nonstandard hypergeometric solutions, which we call hypergeometric $S^a$-modes. In [12] Fronsdal considers KG theory on $AdS_{1,3}$ and constructs wave functions using the Jacobi solutions found in [9]. Moreover, he includes a beautiful section about the geometry of AdS and provides historical references. Avis, Isham and Storey in [5] study KG theory on $AdS_{1,3}$ as well, with another clarifying discussion of AdS geometry. They introduce an "inner product" that is actually the standard symplectic structure for KG solutions on AdS. Although this symplectic structure is defined using an equal-time surface $\Sigma_t$, it turns out to be time-independent. In order to set up a covariant canonical quantization, the authors use both Jacobi and hypergeometric $S^a$-modes. More about AdS geometry and its Penrose diagram can be found in Sections V-VII of [13] by Podolsky and Griffiths, in the Chapter "AdS" of Bengtsson's [14], in Section 2.2 of [15] by Aharony et al., and in [16].

In [17, 18] Breitenlohner and Freedman for $AdS_{1,3}$ study the energy-momentum tensor and the energy functional for a KG solution at fixed time $t$. They find that the energy is positive only for the Jacobi solutions (which we denote by $J^{(+)}_{nl}(\rho)$ and $J^{(-)}_{nl}(\rho)$, and for the $J^{(-)}_{nl}(\rho)$ an "improved" version of the usual energy momentum tensor must be employed). In [19] their result is generalized in a detailed presentation by Mezincescu and Townsend for $AdS_{1,d}$ of arbitrary spatial dimension $d$. We also mention the works [20] of Burgess and Lutken and [21] of Dullemond and van Beveren about the Feynman propagator for scalar fields on AdS.

In Section 3.2 of [6] Balasubramanian, Kraus and Lawrence show how to find more types of KG solutions on AdS. In particular, they distinguish solutions which depend on $\sin^2 \rho$ from those which depend on $\cos^2 \rho$ (therein $\rho \in [0, \pi/2)$ is a compact version of a radial coordinate on AdS). We make use of their idea, because the former characterizes solutions according to their behaviour on the time axis $\rho \equiv 0$ and the latter according to their behaviour near the timelike boundary of AdS at $\rho \equiv \pi/2$. Further, they give a list of KG solutions that is nearly complete (beware: small typo in their equation (30)). In [22], Giddings proposes an S-matrix for KG fields on AdS spacetime using canonical quantization. It is well known that in AdS no temporally asymptotical free states exist, due to the periodic convergence of timelike geodesics. Giddings avoids this problem by placing states on the timelike boundary of AdS. This boundary is a hypercylinder, and its neighborhood is what we call a tube region. Gary and Giddings in [23] investigate the relation between flat space S-matrix and the AdS/CFT correspondence. In Section 3.1 they discuss the flat limit of AdS, where its curvature radius $R_{\text{AdS}}$ tends towards infinity.

Last but not least we mention the work of Dorn et al., who in Section 3 of [24] develop a quantization for particle dynamics for $AdS_{1,d}$. They construct a Schrödinger wave function representation, and obtain as energy eigenvalues what we call magic frequencies. Moreover, they construct an equivalent covariant quantization.

This article is structured as follows: in Section II we start with a compact summary of the relevant results for the cases in Minkowski spacetime corresponding to the ones we later study in AdS. We give a concise review of AdS geometry in Section III. Next, in Section IV we give the KG solutions bounded respectively on three types of regions on AdS: slice regions, and tube and rod hypercylinder regions. We then calculate the actions of the AdS isometry group on the solution spaces of these regions. One-to-one correspondences between initial data on hypersurfaces and bounded solutions are established for all three regions in Section V. In Section VI we then give the symplectic structures for the regions, and show their invariance under isometry actions. While trying to keep this article self-contained, we keep our elaborations rather compact to avoid overlength. We allocate the technical parts in the appendices to make the main part more readable.

We will often refer by e.g. AS[4.2.42] to formulas from the Handbook [25] of Abramowitz and Stegun, and by e.g. DLMF[4.2.42] to its online reincarnation, the Digital Library of Mathematical Functions [26].

## II. MINKOWSKI SPACETIME

For later comparison let us recall some quantities on Minkowski spacetime $\mathbb{R}^{1,3}$. In standard spherical coordinates $(t, r, \Omega)$ with $\Omega = (\theta, \varphi)$ and $\Box_{\mathbb{S}^2}$ the Laplacian on $\mathbb{S}^2$, the metric and Laplace-Beltrami operator write:

$$\mathrm{d}s^2_{\mathrm{Mink}} = -\mathrm{d}t^2 + \mathrm{d}r^2 + r^2\,\mathrm{d}\Omega^2 \qquad \Box_{\mathrm{Mink}} = -\partial_t^2 + \partial_r^2 + \tfrac{2}{r}\,\partial_r + r^{-2}\,\Box_{\mathbb{S}^2}\ . \tag{II.1}$$

The Killing vector fields in cartesian coordinates $(x^0 = t, x^1, x^2, x^3)$ result to be the translations $T_\mu^{\mathrm{Mink}} := \partial_\mu$ and Lorentz rotations $K_{\mu\nu}^{\mathrm{Mink}} := x_\mu \partial_\nu - x_\nu \partial_\mu$ with $\mu = 0, 1, 2, 3$. In spherical coordinates with $\xi_k = x_k/r$ for $i, k = 1, 2, 3$ and using Einstein's summation they write

$$T_0^{\mathrm{Mink}} = \partial_t \qquad\qquad T_k^{\mathrm{Mink}} = \xi_k\,\partial_r + \tfrac{1}{r}\left(\partial_{\xi_k} - \xi_k \xi_i\,\partial_{\xi_i}\right) \tag{II.2}$$
$$K_{jk}^{\mathrm{Mink}} = \xi_j\,\partial_{\xi_k} - \xi_k\,\partial_{\xi_j} \qquad\qquad K_{0k}^{\mathrm{Mink}} = -r\xi_k\,\partial_t - t\xi_k\,\partial_r - \tfrac{t}{r}\left(\partial_{\xi_k} - \xi_k\xi_i\,\partial_{\xi_i}\right)\ .$$

### A. Minkowski slice regions

On a slice region $\mathbb{M}^{\mathrm{Mink}}_{[t_1,t_2]} = [t_1, t_2] \times \mathbb{R}^3$ we can expand any bounded KG solution in the following way:

$$\phi(t, r, \Omega) = \int_0^\infty \mathrm{d}p \sum_{l, m_l} 2p\,(2\pi)^{-1/2} j_l(pr) \left\{ \phi^+_{plm_l}\,\mathrm{e}^{-\mathrm{i}E_p t}\,Y_l^{m_l}(\Omega) + \overline{\phi^-_{plm_l}}\,\mathrm{e}^{\mathrm{i}E_p t}\,\overline{Y_l^{m_l}(\Omega)} \right\}\ , \tag{II.3}$$

wherein $j_l(pr)$ denotes the spherical Bessel functions. The modes $p\mathrm{e}^{-\mathrm{i}E_p t}Y_l^{m_l}(\Omega)j_l(pr)$ are called positive frequency modes and $p\mathrm{e}^{\mathrm{i}E_p t}Y_l^{m_l}(\Omega)j_l(pr)$ negative frequency modes. The momentum space is $l \in \mathbb{N}_0$ with $m_l \in \{-l, \ldots, +l\}$, and $p \geq 0$ such that $E_p \geq m$. The symplectic structure for such solutions is given by

$$\omega_{\Sigma_t}(\eta, \zeta) = -\frac{1}{2} \int_{\Sigma_t} \mathrm{d}r\,\mathrm{d}^2\Omega\,r^2\,(\eta\,\partial_t\zeta - \zeta\,\partial_t\eta) = +\mathrm{i} \int_0^\infty \mathrm{d}p \sum_{l, m_l} E \left\{ \overline{\eta^-_{plm_l}}\,\zeta^+_{plm_l} - \eta^+_{plm_l}\,\overline{\zeta^-_{plm_l}} \right\}\ . \tag{II.4}$$

This shows that it is independent of $t$: it depends only on the mode content of the solutions. The positive and negative frequency modes form Lagrangian subspaces of the space of KG solutions in a neighborhood of the equal-time plane $\Sigma_t$, see (1.2.3) in [27]. The full space of KG solutions on this neighborhood is the direct sum of these subspaces.

### B. Minkowski tube and rod regions

For the results of this subsection consult Section 5.3 in [28] by Oeckl. For the tube region $\mathbb{M}^{\mathrm{Mink}}_{[r_1,r_2]} = \mathbb{R} \times [r_1, r_2] \times \mathbb{S}^2$, with $n_l(pr)$ denoting the spherical Neumann functions, $D := E^2 - m^2$, and $p_E^{\mathbb{R}} := \sqrt{|D|}$ we define the following functions (which are real for real $p_E^{\mathbb{R}} r$):

$$\check{\jmath}_{El}(r) := \begin{cases} j_l(p_E^{\mathbb{R}} r) & D \geq 0 \\ \mathrm{i}^{-l}\,j_l(\mathrm{i}p_E^{\mathbb{R}} r) & D < 0 \end{cases} \qquad \check{n}_{El}(r) := \begin{cases} n_l(p_E^{\mathbb{R}} r) & D \geq 0 \\ \mathrm{i}^{l+1}\,n_l(\mathrm{i}p_E^{\mathbb{R}} r) & D < 0 \end{cases}\ . \tag{II.5}$$

On tube regions we expand any bounded KG solution as:

$$\phi(t, r, \Omega) = \int \mathrm{d}E \sum_{l, m_l} \frac{p_E^{\mathbb{R}}}{4\pi} \left\{ \phi^a_{Elm_l}\mathrm{e}^{-\mathrm{i}Et} Y_l^{m_l}(\Omega)\,\check{\jmath}_{El}(r) + \phi^b_{Elm_l}\mathrm{e}^{-\mathrm{i}Et} Y_l^{m_l}(\Omega)\,\check{n}_{El}(r) \right\}\ . \tag{II.6}$$

We call $p_E^{\mathbb{R}}\mathrm{e}^{-\mathrm{i}Et}Y_l^{m_l}(\Omega)\check{\jmath}_{El}(r)$ the Bessel modes and $p_E^{\mathbb{R}}\mathrm{e}^{-\mathrm{i}Et}Y_l^{m_l}(\Omega)\check{n}_{El}(r)$ the Neumann modes. The momentum space is $E \in \mathbb{R}$ and $l \in \mathbb{N}_0$ with $m_l \in \{-l, \ldots, +l\}$. The Bessel and Neumann modes with $|E| > m$ are propagating modes: they decay inversely to the metric distance from the time axis ($r$ equals this metric distance). The Bessel and Neumann modes with $|E| < m$ are evanescent modes: they grow exponentially for metric distance. The singularity of the Neumann modes on the time axis is a power of the metric distance from the time axis: $n_l(r) \approx r^{-(l+1)}$, and thus does not behave exponentially. Energies with $|E| < m$ need to be included here because an orthogonal system on an equal-radius hypercylinder $\Sigma_r$ is given by the functions $\mathrm{e}^{-\mathrm{i}Et}\,Y_l^{m_l}(\Omega)$ with the energies ranging over all $\mathbb{R}$. The symplectic structure for such solutions is

$$\omega_{\Sigma_r}(\eta,\zeta) \;=\; \frac{r^2}{2}\int_{\Sigma_r}\mathrm{d}t\,\mathrm{d}^2\Omega\,(\eta\,\partial_r\zeta - \zeta\,\partial_r\eta) \;=\; \int\mathrm{d}E\sum_{l,m_l}\frac{p}{16\pi}\left\{\eta^a_{Elm_l}\,\zeta^b_{-E,l,-m_l} - \eta^b_{Elm_l}\,\zeta^a_{-E,l,-m_l}\right\}. \qquad \text{(II.7)}$$

This shows that it is independent of $r$, and depends only on the mode content of the solutions. The $a$-modes and $b$-modes form Lagrangian subspaces of the space of KG solutions on the tube region. The full space of KG solutions is the direct sum of these subspaces.

The solutions for the rod region $\mathbb{M}^{\mathrm{Mink}}_{r_0} = \mathbb{R}\times[0,r_0]\times\mathbb{S}^2$ are those with $\phi^b_{Elm_l}\equiv 0$, since only the spherical Bessel functions are regular on the time axis $r\equiv 0$. The symplectic structure $\omega_{\Sigma_r}$ vanishes for these solutions.

## III.   BASIC ADS GEOMETRY

This section briefly summarizes the aspects of AdS geometry relevant in this work. We denote by AdS what is more precisely denoted as $\mathrm{CAdS}_{1,d}$, that is, $(1+d)$-dimensional Anti-deSitter spacetime with Lorentzian signature in the universal covering version. AdS then has the topology of $\mathbb{R}^{1+d}$ and no closed timelike curves. In contrast to Minkowski spacetime, AdS at $\rho = \pi/2$ has a timelike boundary, which we denote by $\partial\mathrm{AdS}$. Its topology is that of a hypercylinder: $\partial\mathrm{AdS} = \mathbb{R}\times\mathbb{S}^{d-1}$. Where not explicitly stated otherwise, we shall only consider AdS with odd spatial dimension $d \geq 3$.

We use global coordinates with the time coordinate $t \in (-\infty,+\infty)$, a radial coordinate $\rho \in [0,\frac{\pi}{2})$, and denote the $(d-1)$ angular coordinates on $\mathbb{S}^{d-1}$ collectively by $\Omega := (\theta_1,\ldots,\theta_{d-1})$. With $R_{\mathrm{AdS}}$ denoting the curvature radius of AdS and $\mathrm{d}s^2_{\mathbb{S}^{d-1}}$ the area element on the unit sphere $\mathbb{S}^{d-1}$, the metric writes

$$\mathrm{d}s^2_{\mathrm{AdS}} \;=\; \frac{R^2_{\mathrm{AdS}}}{\cos^2\rho}\left(-\mathrm{d}t^2 + \mathrm{d}\rho^2 + \sin^2\rho\,\mathrm{d}s^2_{\mathbb{S}^{d-1}}\right) \;=\; R^2_{\mathrm{AdS}}\left(-\cosh^2\widetilde{\rho}\,\mathrm{d}t^2 + \mathrm{d}\widetilde{\rho}^2 + \sinh^2\widetilde{\rho}\,\mathrm{d}s^2_{\mathbb{S}^{d-1}}\right). \qquad \text{(III.1)}$$

The metric distance of a point from the time axis $\rho \equiv 0$ coincides with a modified version $\widetilde{\rho}$ of our radial coordinate, given by $\cosh\widetilde{\rho} = 1/\cos\rho$. The Laplace-Beltrami operator on AdS is given by

$$\Box_{\mathrm{AdS}} \;:=\; |g|^{-1/2}\,\partial_\mu\,|g|^{1/2}\,g^{\mu\nu}\partial_\nu \;=\; R^{-2}_{\mathrm{AdS}}\left\{-\cos^2\rho\,\partial_t^2 + \cos^2\rho\,\partial_\rho^2 + \frac{(d-1)}{\tan\rho}\,\partial_\rho + \tan^{-2}\rho\,\Box_{\mathbb{S}^{d-1}}\right\}, \qquad \text{(III.2)}$$

We distinguish between three types of regions on AdS, on which different types of Klein-Gordon solutions are well defined and bounded. The first type of region denoted by $\mathbb{M}^{\mathrm{AdS}}_{[t_1,t_2]}$ is called slice region and consists of a time interval $[t_1,t_2]$ times all of space. The second, denoted by $\mathbb{M}^{\mathrm{AdS}}_{\rho_0}$, is called solid hypercylinder or rod region, and consists of all of time times a solid ball $\mathbb{B}_{\rho_0}$ of radius $\rho_0$ in space. The third is denoted by $\mathbb{M}^{\mathrm{AdS}}_{[\rho_1,\rho_2]}$ and called a pierced hypercylinder or tube region: it consists of all of time times a spherical shell $\mathbb{B}_{[\rho_1,\rho_2]}$ with inner radius $\rho_1$ and outer radius $\rho_2$ in space.

Two of these regions arise naturally as (infinitesimal or finite) neighborhoods of hypersurfaces (submanifolds) of AdS. The slice region is a neighborhood of an equal-time hyperplane $\Sigma_{t_0}$, and the tube region is a neighborhood of an equal-radius hypercylinder $\Sigma_{\rho_0}$ (if the neighborhood is chosen big enough, such that it covers the whole region enclosed by $\Sigma_{\rho_0}$, then the tube becomes a rod region). In particular, neighborhoods of the boundary hypercylinder $\partial\mathrm{AdS}= \Sigma_{\rho=\pi/2}$ are tube regions. We remark that the "region" of all of AdS can be obtained in two ways: the first is the limit $t_0 \to \infty$ of the slice region $\mathbb{M}^{\mathrm{AdS}}_{[-t_0,+t_0]}$, and the second is the limit $\rho_0 \to \pi/2$ of the rod region $\mathbb{M}^{\mathrm{AdS}}_{\rho_0}$. Our three regions are generically not type-invariant under isometries. That is, for example, after applying an isometry, what previously was a rod region will not be a rod in the new coordinates, but some deformed version of it. In particular, our regions are only type-invariant under time translations and spatial rotations, but not under boosts.

### A.   The flat limit $R_{\mathrm{AdS}} \to \infty$

Defining for later use the rescaled global coordinates and parameters (see Section IV):

$$r := R_{\mathrm{AdS}}\,\rho \qquad\qquad \tau := R_{\mathrm{AdS}}\,t \qquad\qquad \widetilde{\omega} := \omega/R_{\mathrm{AdS}} \qquad \text{(III.3)}$$

$$p^{\mathbb{R}}_\omega := \sqrt{|\omega^2 - m^2 R^2_{\mathrm{AdS}}|} \qquad\qquad \widetilde{p}^{\mathbb{R}}_{\widetilde{\omega}} := p^{\mathbb{R}}_\omega/R_{\mathrm{AdS}} \;=\; \sqrt{|\widetilde{\omega}^2 - m^2|}\;,$$

the AdS metric (III.1) for $\rho = r/R_{\mathrm{AdS}} \ll 1$ approximates the flat metric: $\mathrm{d}s^2_{\mathrm{AdS}} \approx -\mathrm{d}\tau^2 + \mathrm{d}r^2 + r^2\mathrm{d}s^2_{\mathbb{S}^{d-1}} = \mathrm{d}s^2_{\mathrm{Mink}}$ of Section II, see Section 3.1 in [23]. Therefore the large-$R_{\mathrm{AdS}}$ limit is also called flat limit. With increasing curvature radius the AdS region where $\rho \ll 1$ approximates an increasing part of Minkowski spacetime, covering all of it in the flat limit $R_{\mathrm{AdS}} \to \infty$. Wishing to consider some fixed $r$, for sufficiently large $R_{\mathrm{AdS}}$ we have $r \ll R_{\mathrm{AdS}}$ and the flat approximation holds. Hence in the flat limit $R_{\mathrm{AdS}} \to \infty$ it holds for all $r$.

In the flat limit the AdS Laplace-Beltrami operator approximates the Minkowski one for all $r \ll R_{\text{AdS}}$, that is: $\Box_{\text{AdS}} \approx +\partial_\tau^2 + \partial_r^2 + (d-1)/r\, \partial_r + r^{-2}\, \Box_{\mathbb{S}^{d-1}} = \Box_{\text{Mink}}$. Thus in the flat limit the AdS Klein-Gordon equation with mass parameter $m^2$ approximates the Minkowski KG equation with the same $m^2$. Hence we define the flat limit by letting $R_{\text{AdS}}$ grow larger and larger while keeping fixed the coordinates $\tau = t\, R_{\text{AdS}}$ and $r = \rho\, R_{\text{AdS}}$ and the parameters $m^2$, $\tilde{\omega}$ and $\tilde{p}^{\mathbb{R}}_{\tilde{\omega}}$.

### B. Killing vector fields on $\text{AdS}_{1,d}$

It is sometimes useful to regard $\text{AdS}_{1,d}$ as embedded in $\mathbb{R}^{2,d}$, with (covariant) cartesian coordinates usually denoted by $X = (X_0, \underline{X}, X_{d+1})$ with $\underline{X} = (X_1, ..., X_d)$ and embedding space metric $\mathrm{d}s^2_{(2,d)} = -\mathrm{d}X_0^2 + \mathrm{d}\underline{X}^2 - \mathrm{d}X_{d+1}^2$. We can now introduce (see p.17 in [29]) so-called orispherical coordinates $(R, t, \rho, \Omega)$ on the part of embedding space on which $R^2 = -X^2 = X_0^2 + X_{d+1}^2 - \underline{X}^2 > 0$:

$$X_0 = -R\,\sin t\,\cos^{-1}\rho \qquad X_{d+1} = +R\,\cos t\,\cos^{-1}\rho \qquad X_k = +R\,\xi_k(\Omega)\,\tan\rho\,. \tag{III.4}$$

$\underline{\xi}(\Omega) = (\xi_1, ..., \xi_d)$ are standard constrained cartesian coordinates on the unit sphere with $\underline{\xi}^2 = 1$. The hyperboloid obtained by fixing some $R = R_{\text{AdS}}$ then is AdS with curvature radius $R_{\text{AdS}}$ and time coordinate $t \in [0, 2\pi)$. This version of AdS is called hyperboloidal AdS and contains closed timelike curves. The version of AdS which we use is the universal covering space of this hyperboloid. It is obtained by extending the range of time to $t \in (-\infty, +\infty)$ and unidentifying the points in embedding space obtained by $t \to t + 2\pi$, thereby avoiding closed timelike curves. The AdS metric (III.1) is then induced by the embedding space metric. With Latin uppercase indices having the range $A = 0, \ldots, (d+1)$, and Latin lowercase indices ranging as $k = 1, \ldots, d$, we can write the embedding space Killing vector fields $K(X) := X_A\, \partial_B - X_B\, \partial_A$ in orispherical coordinates $(R, t, \rho, \underline{\xi}(\Omega))$. All of them leave the $R$-coordinate invariant. Since $\text{AdS}_{1,d}$ is a submanifold of $\mathbb{R}^{(2,d)}$ with fixed $R = R_{\text{AdS}}$ and induced metric, the embedding space Killing vectors $K_{AB}$ are Killing vector fields on $\text{AdS}_{1,d}$ as well (therefore the label AdS), and using Einstein's summation they write:

$$K^{\text{AdS}}_{d+1,0} = X_{d+1}\,\partial_0 - X_0\,\partial_{d+1} = \partial_t \tag{III.5}$$

$$K^{\text{AdS}}_{jk} = X_j\,\partial_k - X_k\,\partial_j = \xi_j\partial_{\xi_k} - \xi_k\partial_{\xi_j} \tag{III.6}$$

$$K^{\text{AdS}}_{0j} = X_0\,\partial_j - X_j\,\partial_0 = -\xi_j\cos t\,\sin\rho\,\partial_t - \xi_j\sin t\,\cos\rho\,\partial_\rho - (\sin t)/(\sin\rho)\left(\partial_{\xi_j} - \xi_j\xi_i\,\partial_{\xi_i}\right) \tag{III.7}$$

$$K^{\text{AdS}}_{d+1,j} = X_{d+1}\,\partial_j + X_j\,\partial_{d+1} = -\xi_j\sin t\,\sin\rho\,\partial_t + \xi_j\cos t\,\cos\rho\,\partial_\rho + (\cos t)/(\sin\rho)\left(\partial_{\xi_j} - \xi_j\xi_i\,\partial_{\xi_i}\right)\,. \tag{III.8}$$

The above Killing vectors are the generators of the Lie algebra $\mathbf{so}\,(2,d)$ of the isometry group $\text{SO}\,(2,d)$ of $\text{AdS}_{1,d}$. The above formulas generalize the corresponding ones for Bengtsson's favourite "sausage" coordinates given for $\text{AdS}_{1,2}$ in equations (134-138) of [14]. The AdS Killing vectors are of three different types: only one translation $K^{\text{AdS}}_{d+1,0}$, plus $d(d-1)/2$ spatial rotations $K^{\text{AdS}}_{jk}$, and $(2d)$ boosts $K^{\text{AdS}}_{0j}$ and $K^{\text{AdS}}_{d+1,j}$. In total we thus have $(d+1)(d+2)/2$ Killing vectors on $\text{AdS}_{1,d}$, making it a maximally symmetric space(time). Since for $\rho = \frac{\pi}{2}$ the $\partial_\rho$-component of the boosts vanishes, all AdS Killing vectors map the boundary to itself. Between the Killing vectors of AdS and Minkowski spacetime there exists a correspondence: we can first switch to the coordinates $\tau = R_{\text{AdS}} t$ and $r = R_{\text{AdS}}\rho$, and then take the flat limit $R_{\text{AdS}} \to \infty$, giving us:

$$K^{\text{AdS}}_{d+1,0} \xrightarrow{\text{flat lim.}} R_{\text{AdS}}\,\partial_\tau \qquad\qquad K^{\text{AdS}}_{0j} \xrightarrow{\text{flat lim.}} -\xi_j\,r\,\partial_\tau - \xi_j\,\tau\,\partial_r - \frac{\tau}{r}\left(\partial_{\xi_j} - \xi_j\xi_i\,\partial_{\xi_i}\right)$$

$$K^{\text{AdS}}_{jk} \xrightarrow{\text{flat lim.}} \xi_j\partial_{\xi_k} - \xi_k\partial_{\xi_j} \qquad\qquad K^{\text{AdS}}_{d+1,j} \xrightarrow{\text{flat lim.}} R_{\text{AdS}}\left(\xi_j\,\partial_r + \frac{1}{r}\left(\partial_{\xi_j} - \xi_j\xi_i\,\partial_{\xi_i}\right)\right)\,.$$

Comparing these to (II.2) gives the AdS-Minkowski Killing vector correspondence:

| **AdS (flat limit)** | $R_{\text{AdS}}^{-1} K^{\text{AdS}}_{d+1,0}$ | $R_{\text{AdS}}^{-1} K^{\text{AdS}}_{d+1,j}$ | $K^{\text{AdS}}_{jk}$ | $K^{\text{AdS}}_{0j}$ |
|---|---|---|---|---|
| **Minkowski** | $T^{\text{Mink}}_0$ | $T^{\text{Mink}}_j$ | $K^{\text{Mink}}_{jk}$ | $K^{\text{Mink}}_{0j}$ |

## IV. KG SOLUTIONS IN ADS SPACETIME

With $m$ denoting the field's mass, the action for a free, real scalar field $\phi(x)$ living in AdS is $S[\phi] = \int \mathrm{d}^{d+1}x\,\sqrt{|g|}\,\frac{1}{2}\left[-g^{\mu\nu}(\partial_\mu\phi)(\partial_\nu\phi) - m^2\phi^2\right]$, and its Euler-Lagrange equation is the free Klein-Gordon equation $0 = (-\Box_{\text{AdS}} + m^2)\phi$. Below we list the solutions which are well defined and bounded on the respective

AdS regions. On tube regions $\mathbb{M}^{\text{AdS}}_{[\rho_1,\rho_2]}$ the form of some solutions depends on whether the spatial dimension $d$ is odd or even, and the form of other solutions on whether the quantity $\nu$ is integer or not, with $\nu = \sqrt{d^2/4 + m^2 R^2_{\text{AdS}}}$. However, the form of the solutions on slice regions $\mathbb{M}^{\text{AdS}}_{[t_1,t_2]}$ is always the same, ditto for the rod regions $\mathbb{M}^{\text{AdS}}_{\rho_0}$. Unless stated otherwise we shall assume that $d$ is odd and $\nu$ noninteger (see Appendix B 2 for the other cases). The most general solutions of the KG equation on AdS are four types of modes which we call hypergeometric $a$ and $b$-modes of type $S$ respectively $C$ (quantities relating to the $S^a$ and $S^b$-modes carry superscripts $a$ and $b$, e.g.: $\mu^{(a)}$, while quantities relating to the $C$-modes carry an additional $C$, e.g.: $\mu^{(C,a)}$):

$$\mu^{(a)}_{\omega \underline{l} m_l}(t,\rho,\Omega) = e^{-i\omega t}\, Y^{m_l}_{\underline{l}}(\Omega)\, S^a_{\omega l}(\rho) \qquad\qquad \mu^{(C,a)}_{\omega \underline{l} m_l}(t,\rho,\Omega) = e^{-i\omega t}\, Y^{m_l}_{\underline{l}}(\Omega)\, C^a_{\omega l}(\rho) \qquad\qquad \text{(IV.1)}$$
$$\mu^{(b)}_{\omega \underline{l} m_l}(t,\rho,\Omega) = e^{-i\omega t}\, Y^{m_l}_{\underline{l}}(\Omega)\, S^b_{\omega l}(\rho) \qquad\qquad \mu^{(C,b)}_{\omega \underline{l} m_l}(t,\rho,\Omega) = e^{-i\omega t}\, Y^{m_l}_{\underline{l}}(\Omega)\, C^b_{\omega l}(\rho)\,.$$

Therein, $Y^{m_l}_{\underline{l}}(\Omega)$ denote the hyperspherical harmonics, see Appendix A. With $F(a, b; c; x)$ denoting Gauss's hypergeometric function, we use the radial functions

$$S^a_{\omega l}(\rho) = \sin^l \rho \;\; \cos^{\widetilde{m}_+}\rho \; F(\alpha^a,\beta^a;\gamma^a;\sin^2\rho) \qquad C^a_{\omega l}(\rho) = \sin^l \rho\, \cos^{\widetilde{m}_+}\rho\, F(\alpha^{C,a},\beta^{C,a};\gamma^{C,a};\cos^2\rho) \qquad \text{(IV.2)}$$
$$S^b_{\omega l}(\rho) = -(\sin\rho)^{2-l-d}\cos^{\widetilde{m}_+}\rho\, F(\alpha^b,\beta^b;\gamma^b;\sin^2\rho) \qquad C^b_{\omega l}(\rho)=\sin^l\rho\, \cos^{\widetilde{m}_-}\rho\, F(\alpha^{C,b},\beta^{C,b};\gamma^{C,b};\cos^2\rho).$$

The hypergeometric parameters therein are given by

$$\begin{aligned}
\alpha^a &= \alpha^{C,a} = \tfrac{1}{2}(l+\widetilde{m}_+-\omega) & \alpha^b &= \alpha^a-\gamma^a+1 & \alpha^{C,b} &= \alpha^{C,a}-\gamma^{C,a}+1 \\
\beta^a &= \beta^{C,a} = \tfrac{1}{2}(l+\widetilde{m}_++\omega) & \beta^b &= \beta^a-\gamma^a+1 & \beta^{C,b} &= \beta^{C,a}-\gamma^{C,a}+1 \\
\gamma^a &= l+\tfrac{d}{2} & \gamma^{C,a} &= 1+\nu & \gamma^b &= 2-\gamma^a & \gamma^{C,b} &= 2-\gamma^{C,a}\,.
\end{aligned} \qquad \text{(IV.3)}$$

With $m$ denoting the field mass, we further use

$$\widetilde{m}_\pm = \tfrac{d}{2}\pm\nu \qquad\qquad \nu = \sqrt{d^2/4+m^2 R^2_{\text{AdS}}} \qquad\qquad \begin{matrix}\widetilde{m}_+>0 & \forall\,\nu\geq 0 \\ \widetilde{m}_->0 & \forall\,\nu\in(0,1)\end{matrix}\,.$$

The value of $m^2$ for which $\nu$ vanishes is called Breitenlohner-Freedman mass $m^2_{\text{BF}} := -d^2/(4R^2_{\text{AdS}})$. Whenever the frequency $\omega$ is one of the discrete values dubbed magic frequencies in [10]: $\omega^\pm_{nl} = 2n+l+\widetilde{m}_\pm$ (nonnegative for $\widetilde{m}_\pm \geq 0$), then the $S^a$-modes take on a special form. The hypergeometric function then can be written as a Jacobi polynomial, and therefore we call the two discrete sets of modes below Jacobi modes:

$$\mu^{(\pm)}_{n\underline{l}m_l}(t,\rho,\Omega) = \mu^{(a)}_{\omega^+_{nl}\underline{l}m_l}(t,\rho,\Omega) = e^{-i\omega^\pm_{nl}t}\, Y^{m_l}_{\underline{l}}(\Omega)\, J^{(\pm)}_{nl}(\rho)\,. \qquad \text{(IV.4)}$$

We call $\mu^{(+)}_{n\underline{l}m_l}(t,\rho,\Omega)$ ordinary and $\mu^{(-)}_{n\underline{l}m_l}(t,\rho,\Omega)$ exceptional Jacobi modes. Moreover, we call $\mu^{(\pm)}_{n\underline{l}m_l}(t,\rho,\Omega)$ the positive frequency modes and $\overline{\mu^{(\pm)}_{n\underline{l}m_l}(t,\rho,\Omega)}$ the negative frequency modes. The exceptional AdS-Jacobi modes are only well defined for $\nu\in(0,1)$, and for this case $\widetilde{m}_->0$. By $P^{(a,b)}_n(x)$ we denote the Jacobi polynomials, by $(a)_n$ the Pochhammer symbols, and $J^{(\pm)}_{nl}(\rho) := \frac{n!}{(l+d/2)_n}\sin^l\rho\,\cos^{\widetilde{m}_+}\rho\, P^{(l+d/2-1,\pm\nu)}_n(\cos 2\rho)$.

We sketch how to find all these modes in Appendix B 1, and in B 2 give a complete list of solutions including the complementary cases of even $d$ and integer $\nu$. The Jacobi modes usually are the only modes used in the literature, e.g. [18], [19], [5] and [20]. The earliest mention of the Jacobi modes that we could spot is equation (4.8) of the first article [9] by Limic, Niederle and Raczka. The correspondence to our notation is given by $H^q_p = \text{AdS}_{1,d}$ for $(q,p) = (2,d)$, with AdS coordinates $\tilde{\varphi}^1 = t$ and $\theta = \tilde{\rho}$ and thus $\tanh\theta = \sin\rho$ while $\cosh\theta = 1/\cos\rho$. The parameters are $\tilde{m}_1 = -\omega$, $L = -\widetilde{m}_\pm$ and $\lambda = -m^2 R^2_{\text{AdS}}$. Then, their relation (4.5) corresponds to our magic frequencies. The first mention of the $S^a_{\omega l}(\rho)$-modes we found in equation (2.5) of [11]. The correspondence to our notation is the same as above plus $\Lambda^2 = \lambda - (d^2/4)$, that is, $-\Lambda^2 = \nu^2$. The authors of [18], [19] and [5] mention both $S^a$ and $S^b$-modes but then discard the latter, and use only the special case of Jacobi modes of the $S^a$-modes. Giddings in [22] then actually makes use of *all* $S^a$-modes. The first appearance of the $C^{a,b}$-modes we found in [6], wherein the $S^b$-modes are discarded once more.

Dorn et al. in Section 3 of [24] find another physical meaning of the magic frequencies. Constructing a Schrödinger wave function representation for particle dynamics for $\text{AdS}_{1,d}$, they obtain as energy eigenvalues precisely the magic frequencies: $\omega^\pm_{nl} = E^\pm_0 + 2n + l$ with ground state energy $E^\pm_0 = \widetilde{m}_\pm = d/2 \pm \sqrt{d^2/4 + m^2 R^2_{\text{AdS}}}$. To see this, note that their $N$ is our $d$, their $\mathcal{M}^2$ our $m^2$, and combine their expressions (3.13), (4.3) and $\tilde{a} = (N-1)/4N$ below (4.13).

## A. Properties of the AdS Klein-Gordon solutions

The $S^a$ and $S^b$-modes form one pair of linear independent solutions, and the $C^a$ and $C^b$-modes form another. They are related through ("on" stands for odd-noninteger):

$$\begin{pmatrix} S^a_{\omega l} \\ S^b_{\omega l} \end{pmatrix} = M^{\mathrm{on}}_{\omega l} \begin{pmatrix} C^a_{\omega l} \\ C^b_{\omega l} \end{pmatrix}, \qquad (\mathrm{IV.5})$$

see Appendix B 3 for the matrix elements of $M^{\mathrm{on}}_{\omega l}$. The hypergeometric $S^a$ and $S^b$-modes and the $C^b$-modes are evanescent modes (except for the magic frequencies): when approaching the boundary $\rho = \frac{\pi}{2}$ they grow exponentially with metric distance $\tilde{\rho}$ from the time axis. (This holds for $\tilde{m}_- < 0$, that is, positive mass square $m^2$. For $\tilde{m}_- = 0$ their value for large $\tilde{\rho}$ becomes some finite constant, and for $\tilde{m}_- > 0$ they decay exponentially with $\tilde{\rho}$). The $C^a$-modes are also evanescent: they decay exponentially with $\tilde{\rho}$ when approaching the boundary. On the time axis $\rho \equiv 0 \equiv \tilde{\rho}$ the $S^a$-modes are regular, like the Bessel modes on Minkowski spacetime. However, the $C^a$ and $C^b$-modes and $S^b$-modes are singular there: they behave like $\tilde{\rho}^{-l-(d-2)}$, that is, inverse power of metric distance, like the Neumann modes on Minkowski spacetime.

It is necessary to study all these solutions for the following reasons. First, consider the case of KG theory with some source field. We let the source field have compact support, say within some hypercylinder surface $\Sigma_{\rho_0}$. Any solution of this inhomogeneous KG equation coincides with some free solution outside of the source region. If we take this free solution and continue it, then generically it is *not* well defined and bounded on the whole region enclosed by $\Sigma_{\rho_0}$, and therefore we need to study divergent solutions, too. Second, the $S^a$ and $S^b$-modes and the $C^a$ and $C^b$-modes provide us two different parametrizations of KG solutions. The $S^a$ and $S^b$-modes parametrize solutions according to their behaviour near the time axis $\rho \equiv 0$, on which the former are regular and the latter diverge. The $C^a$ and $C^b$-modes parametrize solutions according to their behaviour near the timelike boundary of AdS at $\rho \equiv \pi/2$, on which the former are regular and the latter diverge.

The Jacobi modes are well defined and bounded both on time axis and boundary. Thus they are the only modes that are $L^2$-normalizable on an equal-time surface. For later use we define the following normalization constant for all $\nu \geq 0$ (that is: all masses $m^2 \geq m^2_{\mathrm{BF}}$ above the Breitenlohner-Freedman mass):

$$\mathcal{N}^{\pm}_{nl} := \int_0^{\pi/2} \mathrm{d}\rho \; \tan^{d-1}\rho \left(J^{(\pm)}_{nl}(\rho)\right)^2 = \frac{n! \, \Gamma(\gamma^{\mathrm{S}})^2 \, \Gamma(n \pm \nu + 1)}{2\omega^{\pm}_{nl} \, \Gamma(n + \gamma^{\mathrm{S}}) \, \Gamma(n \pm \nu + \gamma^{\mathrm{S}})} \; . \qquad (\mathrm{IV.6})$$

## B. KG solutions on tube regions

For the tube region $\mathbb{M}^{\mathrm{AdS}}_{[\rho_1, \rho_2]} = \mathbb{R} \times [\rho_1, \rho_2] \times \mathbb{S}^{d-1}$, that is: all of time times a spherical shell, we need KG solutions that are bounded for all of time while in space we only need them bounded on $[\rho_1, \rho_2]$. Thus we can use all four hypergeometric modes here, with the frequency $\omega$ being real. We expand an arbitrary complex(ified) KG solution (see e.g. Section 2.3 in [30]) on the tube region as an integral over these modes, where we call the upper line $S$-expansion and the lower line $C$-expansion:

$$\phi(t, r, \Omega) = \int \mathrm{d}\omega \sum_{\underline{l}, m_l} \left\{ \phi^a_{\omega \underline{l} m_l} \, \mu^{(a)}_{\omega \underline{l} m_l}(t, \rho, \Omega) + \phi^b_{\omega \underline{l} m_l} \, \mu^{(b)}_{\omega \underline{l} m_l}(t, \rho, \Omega) \right\} \qquad (\mathrm{IV.7})$$

$$= \int \mathrm{d}\omega \sum_{\underline{l}, m_l} \left\{ \phi^{C,a}_{\omega \underline{l} m_l} \, \mu^{(C,a)}_{\omega \underline{l} m_l}(t, \rho, \Omega) + \phi^{C,b}_{\omega \underline{l} m_l} \, \mu^{(C,b)}_{\omega \underline{l} m_l}(t, \rho, \Omega) \right\} . \qquad (\mathrm{IV.8})$$

If and only if $\overline{\phi^a_{\omega \underline{l} m_l}} = \phi^a_{-\omega, \underline{l}, -m_l}$ and the same for $\phi^b_{\omega \underline{l} m_l}$ (respectively for $\phi^{C,a}_{\omega \underline{l} m_l}$ and $\phi^{C,b}_{\omega \underline{l} m_l}$), then the solution $\phi(t, \rho, \Omega)$ is real. We can also use the modified momentum representation $(\tilde{\phi}^{\mathrm{M},a}_{\omega \underline{l} m_l}, \tilde{\phi}^{\mathrm{M},b}_{\omega \underline{l} m_l})$, (where the label M stands for Minkowski) in the $S$-expansion: $\phi^a_{\omega \underline{l} m_l} = \tilde{\phi}^{\mathrm{M},a}_{\tilde{\omega} \underline{l} m_l} \tilde{p}^{\mathbb{R}}_{\tilde{\omega}} (p^{\mathbb{R}}_\omega)^l / [4\pi R_{\mathrm{AdS}}(2l+d-2)!!]$ and $\phi^b_{\omega \underline{l} m_l} = \tilde{\phi}^{\mathrm{M},b}_{\tilde{\omega} \underline{l} m_l} \tilde{p}^{\mathbb{R}}_{\tilde{\omega}} (p^{\mathbb{R}}_\omega)^{-(l+1)} (2l+d-4)!! / (4\pi R_{\mathrm{AdS}})$. Then, the flat limit of the $S$-expansion for $d=3$ yields the Minkowski tube expansion (II.6), see Appendix B 5:

$$\phi(t, r, \Omega) \xrightarrow[\mathrm{lim.}]{\mathrm{flat}} \int \mathrm{d}\tilde{\omega} \sum_{l, m_l} \frac{\tilde{p}^{\mathbb{R}}_{\tilde{\omega}}}{4\pi} \left\{ \tilde{\phi}^{\mathrm{M},a}_{\tilde{\omega} \underline{l} m_l} e^{-\mathrm{i}\tilde{\omega}\tau} Y^{m_l}_l(\Omega) \check{j}_{\tilde{\omega}l}(r) + \tilde{\phi}^{\mathrm{M},b}_{\tilde{\omega} \underline{l} m_l} e^{-\mathrm{i}\tilde{\omega}\tau} Y^{m_l}_l(\Omega) \check{n}_{\tilde{\omega}l}(r) \right\} . \qquad (\mathrm{IV.9})$$

### C. KG solutions on rod regions

For the rod region $\mathbb{M}^{\mathrm{AdS}}_{\rho_0} = \mathbb{R} \times [0, \rho_0] \times \mathbb{S}^{d-1}$, that is: all of time times a solid ball, we need KG solutions that are bounded for all of time while in space we only need them bounded on $[0, \rho_0]$. Therefore, the hypergeometric $S^a$-modes (IV.1) alone span the space of KG solutions for the rod region. We expand any KG solution on the rod region as an integral over these modes, which we call rod expansion (again, $\phi(t, \rho, \Omega)$ is real iff $\phi^a_{\omega \underline{l} m_l} = \overline{\phi^a_{-\omega, \underline{l}, -m_l}}$)

$$\phi(t, r, \Omega) = \int\!\mathrm{d}\omega \sum_{\underline{l}, m_l} \phi^a_{\omega \underline{l} m_l}\, \mu^{(a)}_{\omega \underline{l} m_l}(t, \rho, \Omega)\,. \tag{IV.10}$$

### D. KG solutions on slice regions

For the slice region $\mathbb{M}^{\mathrm{AdS}}_{[t_1, t_2]} = [t_1, t_2] \times [0, \pi/2) \times \mathbb{S}^{d-1}$, that is: time interval times all of space, we need KG solutions that are bounded on all of space. Thus we can only use Jacobi modes here, and expand any complex KG solution on the slice region as a sum of ordinary Jacobi modes, calling it (ordinary) Jacobi expansion:

$$\phi(t, \rho, \Omega) = \sum_{n \underline{l} m_l} \left\{ \phi^+_{n \underline{l} m_l}\, \mu^{(+)}_{n \underline{l} m_l}(t, \rho, \Omega) + \overline{\phi^-_{n \underline{l} m_l}}\, \overline{\mu^{(+)}_{n \underline{l} m_l}(t, \rho, \Omega)} \right\}. \tag{IV.11}$$

Only for $\nu \in (0, 1)$ we can equivalently expand any solution using the exceptional AdS-Jacobi modes given above, see e.g. (3.22) in [19]. Since these behave like the ordinary ones, we do not study them in this work. $\phi^+_{n \underline{l} m_l}$ determines the positive frequency part of the KG solution, and $\phi^-_{n \underline{l} m_l}$ the negative frequency part. If and only if $\phi^+_{n \underline{l} m_l} = \phi^-_{n \underline{l} m_l}$ then the solution $\phi(t, \rho, \Omega)$ is real. The ordinary Jacobi modes are propagating modes, well defined on the whole spacetime. Since the Jacobi modes are special cases of the $S^a$-modes, the space of KG solutions on slice regions is contained in the spaces of KG solutions on tube and rod regions as a subspace. Again, we can use a modified momentum representation $(\tilde{\phi}^{\mathrm{M},+}_{n \underline{l} m_l}, \tilde{\phi}^{\mathrm{M},-}_{n \underline{l} m_l})$ with now $\phi^\pm_{n \underline{l} m_l} \equiv \phi^\pm_{\omega^+_{nl} \underline{l} m_l} = \tilde{\phi}^{\mathrm{M},\pm}_{n \underline{l} m_l} 2\omega^+_{nl} (p^\mathbb{R}_\omega)^l / \bigl[R_{\mathrm{AdS}} \sqrt{2\pi}\, (2l+d-2)!!\bigr]$. Then, in the flat limit for $d = 3$ the ordinary Jacobi expansion becomes the Minkowski slice expansion (II.3), see Appendix B 5:

$$\phi(t, r, \Omega) \xrightarrow[\mathrm{lim.}]{\mathrm{flat}} \int_0^\infty\!\mathrm{d}\tilde{p} \sum_{l, m_l} 2\tilde{p}\,(2\pi)^{-1/2} j_l(\tilde{p} r) \left\{ \tilde{\phi}^{\mathrm{M},+}_{\tilde{p} l m_l}\, \mathrm{e}^{-\mathrm{i}\tilde{\omega}_{\tilde{p}}\tau}\, Y^{m_l}_l(\Omega) + \overline{\tilde{\phi}^{\mathrm{M},-}_{\tilde{p} l m_l}}\, \mathrm{e}^{\mathrm{i}\tilde{\omega}_{\tilde{p}}\tau}\, \overline{Y^{m_l}_l(\Omega)} \right\}. \tag{IV.12}$$

### E. Action of isometries on solution space

In this section we consider the action of the isometry group $\mathrm{SO}(2, d)$ of $\mathrm{AdS}_{1,d}$. Uppercase Latin indices range as $A = 0, 1, ..., d, (d+1)$, and lowercase Latin indices as $k = 1, ..., d$. The generators of the Lie algebra $\mathbf{so}(2, d)$ are the Killing vectors $K_{AB}$ of Section III (think of $AB$ as *one* index, not two): $K_{AB} = (X_A \partial_B - X_B \partial_A)$. This choice is the same as in (4.18) in [31] up to an overall sign. The (representations of) finite group elements are denoted by $g$. The Lie algebra of $\mathrm{SO}(2, d)$ is determined by the Lie bracket

$$[K_{AB}, K_{CD}] = -\eta_{AC}\, K_{BD} + \eta_{BC}\, K_{AD} - \eta_{BD}\, K_{AC} + \eta_{AD}\, K_{BC}\,, \tag{IV.13}$$

which corresponds to (4.21) in [31]. All combinations of time translation, rotations and boosts are:

$$\begin{aligned}
&[K_{d+1,0}, K_{jk}] = 0 & &[K_{0j}, K_{0k}] = \eta_{00} K_{kj} & &[K_{0q}, K_{jk}] = \eta_{jq} K_{0k} - \eta_{kq} K_{0j} \\
&[K_{d+1,j}, K_{d+1,k}] = \eta_{d+1,d+1} K_{kj} & &[K_{d+1,0}, K_{0k}] = \eta_{00} K_{d+1,k} & &[K_{d+1,q}, K_{jk}] = \eta_{jq} K_{d+1,k} - \eta_{kq} K_{d+1,j} \\
&[K_{d+1,k}, K_{d+1,0}] = \eta_{d+1,d+1} K_{0k} & &[K_{0k}, K_{d+1,j}] = \eta_{jk} K_{d+1,0} & &[K_{jk}, K_{pq}] = \eta_{kp} K_{jq} - \eta_{jp} K_{kq} + \eta_{jq} K_{kp} - \eta_{kq} K_{jp}.
\end{aligned} \tag{IV.14}$$

On solution space, we denote the (infinitesimal) action of a generator $K_{AB}$ respectively group element $k$ on a field $\phi(x)$ by $K_{AB} \triangleright \phi$ and $k \triangleright \phi$. Requiring the transformed field at transformed coordinates to agree with the original field at original coordinates, we get for the transformed field at the original coordinates:

$$\bigl(K_{AB} \triangleright \phi\bigr)(x) = (-K_{AB}\phi)(x) \qquad\qquad (k \triangleright \phi)(x) = \phi(k^{-1} x)\,. \tag{IV.15}$$

Therein, $K_{AB}\phi$ means letting the generator(Killing vector) act as differential operator on the field(function) $\phi(x)$. In the following sections we calculate these actions for the time translation, rotations and boosts for KG solutions on the slice and tube regions. Our goal will always be to translate the action from the coordinate representation to the momentum representation. We thus start from a field expansion over modes $\mu^{(a,b)}_{\omega \underline{l} m_l}(x)$ of momentum $(\omega, \underline{l}, m_l)$ wherein $\phi^{a,b}_{\omega \underline{l} m_l}$ are the momentum representation of the KG solution. What we want to find is an explicit expression for the momentum representations $(k \triangleright \phi)^{a,b}_{\omega \underline{l} m_l}$ of the transformed field, such that we can directly write the transformed field in the original coordinates as in:

$$(k \triangleright \phi)(x) = \int d\omega \sum_{\underline{l}, m_l} \left\{ (k \triangleright \phi)^a_{\omega \underline{l} m_l} \mu^{(a)}_{\omega \underline{l} m_l}(x) + (k \triangleright \phi)^b_{\omega \underline{l} m_l} \mu^{(b)}_{\omega \underline{l} m_l}(x) \right\} . \tag{IV.16}$$

### F. Action of time translations on KG solutions

Infinitesimal time translations arise nicely from the finite ones, thus we only deal with the latter here. Denoting finite time translations by $k_{\Delta t}: t \to t + \Delta t$, its action on the hypergeometric modes is

$$\left(k_{\Delta t} \triangleright \mu^{(a)}_{\omega \underline{l} m_l}\right)(t,\rho,\Omega) = e^{i\omega \Delta t} \mu^{(a)}_{\omega \underline{l} m_l}(t,\rho,\Omega) \qquad \left(k_{\Delta t} \triangleright \mu^{(b)}_{\omega \underline{l} m_l}\right)(t,\rho,\Omega) = e^{i\omega \Delta t} \mu^{(b)}_{\omega \underline{l} m_l}(t,\rho,\Omega) , \tag{IV.17}$$

and the action on the Jacobi modes as a special case of $S^a$-modes is the same:

$$\left(k_{\Delta t} \triangleright \mu^{(+)}_{n \underline{l} m_l}\right)(t,\rho,\Omega) = e^{i\omega^+_{nl} \Delta t} \mu^{(+)}_{n \underline{l} m_l}(t,\rho,\Omega) \qquad \left(k_{\Delta t} \triangleright \overline{\mu^{(+)}_{n \underline{l} m_l}}\right)(t,\rho,\Omega) = e^{-i\omega^+_{nl} \Delta t} \overline{\mu^{(+)}_{n \underline{l} m_l}(t,\rho,\Omega)} . \tag{IV.18}$$

Applying these to the tube expansion (IV.7) respectively slice expansion (IV.11), we can read off

$$\left(k_{\Delta t} \triangleright \phi\right)^a_{\omega \underline{l} m_l} = e^{i\omega \Delta t} \phi^a_{\omega \underline{l} m_l} \qquad \qquad \left(k_{\Delta t} \triangleright \phi\right)^b_{\omega \underline{l} m_l} = e^{i\omega \Delta t} \phi^b_{\omega \underline{l} m_l} \tag{IV.19}$$

$$\left(k_{\Delta t} \triangleright \phi\right)^+_{n \underline{l} m_l} = e^{i\omega^+_{nl} \Delta t} \phi^+_{n \underline{l} m_l} \qquad \qquad \overline{\left(k_{\Delta t} \triangleright \phi\right)^-_{n \underline{l} m_l}} = e^{-i\omega^+_{nl} \Delta t} \overline{\phi^-_{n \underline{l} m_l}} . \tag{IV.20}$$

### G. Action of rotations on KG solutions

Let $\hat{R}(\underline{\alpha})$ denote a finite rotation, with $\underline{\alpha}$ denoting the rotation angles. We recall that rotated spherical harmonics are a linear combination of unrotated ones, with elements of Wigner's $D$-matrix as coefficients:

$$Y^{m_l}_{(l,\underline{l})}(\hat{R}(\underline{\alpha})\Omega) = \sum_{\underline{l}', m'_l} Y^{m'_l}_{(l,\underline{l}')}(\Omega) \overline{\left(D^l_{\underline{l},\underline{l}'}(\underline{\alpha})\right)}_{m_l m'_l} ,$$

see Appendix A. For the hypergeometric modes this induces the action

$$\mu^{(a,b)}_{\omega l \underline{l} m_l}(t,\rho,\hat{R}(\underline{\alpha})\Omega) = \sum_{\underline{l}', m'_l} \mu^{(a,b)}_{\omega l \underline{l}' m'_l}(t,\rho,\Omega) \overline{\left(D^l_{\underline{l},\underline{l}'}(\underline{\alpha})\right)}_{m_l m'_l} . \tag{IV.21}$$

Since the Jacobi modes are special cases of the $S^a$-modes, the action is the same for them. For tube solution we can apply (IV.15) to expansion (IV.7), giving:

$$\left(\hat{R}(\underline{\alpha})^{-1} \triangleright \phi\right)^{a,b}_{\omega l \underline{l} m_l} = \sum_{\underline{l}', m'_l} \phi^{a,b}_{\omega l \underline{l}' m'_l} \overline{\left(D^l_{\underline{l}',\underline{l}}(\underline{\alpha})\right)}_{m'_l m_l} . \tag{IV.22}$$

For slice solutions, we apply (IV.15) to expansion (IV.11), yielding:

$$\left(\hat{R}(\underline{\alpha})^{-1} \triangleright \phi\right)^{\pm}_{nl \underline{l} m_l} = \sum_{\underline{l}', m'_l} \phi^{\pm}_{nl \underline{l}' m'_l} \overline{\left(D^l_{\underline{l}',\underline{l}}(\underline{\alpha})\right)}_{m'_l m_l} . \tag{IV.23}$$

### H. Action of boosts on KG solutions

Here the goal is to calculate the action of the AdS boost generators $K_{d+1,j}$ and $K_{0j}$ on the bounded KG modes on slice and tube regions. We only consider infinitesimal boosts. The effect of boosts on the AdS time axis is qualitatively different from its Minkowski counterpart. For seeing this, consider e.g. a boost with finite rapidity $\lambda$ in the $(0,j)$-plane in the embedding space of AdS, see (III.4). The boosted coordinates are $X'_0 = X_0 \cosh \lambda + X_j \sinh \lambda$ and $X'_j = X_0 \sinh \lambda + X_j \cosh \lambda$, with the other coordinates unchanged. Thus all points with $X_0 = X_j = 0$ are preserved under these boosts, while other points are moved a finite distance on the AdS-hyperboloid. This means that, in contrast to Minkowski spacetime, on AdS the boosts do not rotate the time axis but deform it periodically (into some timelike geodesic). Therefore, small boosts on AdS move the time axis only a small distance away from the unboosted one. In contrast, on Minkowski spacetime even an arbitrarily small boost for sufficiently large times separates the boosted time axis an arbitrary distance from the unboosted one.

Recall now that the $S^a$-modes are regular on the time axis $\rho \equiv 0$ while the $S^b$-modes diverge there. A finite boost moves the boosted time axis off the unboosted one. Therefore the $S^b$-modes of the boosted coordinates now have singularities off the unboosted time axis, and thus cannot be well defined linear combinations of the original $S^a$ and $S^b$-modes (because these are regular off the unboosted time axis). Hence only for infinitesimal boosts there is a chance that we might find a well defined action on KG solutions.

For the two $d$-boosts $K_{d+1,d}$ and $K_{0d}$ we recall the expressions

$$K_{d+1,d} = -\sin t \sin\rho \cos\theta_{d-1}\, \partial_t + \cos t \cos\rho \cos\theta_{d-1}\, \partial_\rho + \sin^2\theta_{d-1} \cos t/(\sin\rho)\, \partial_{\cos\theta_{d-1}} \quad \text{(IV.24)}$$

$$K_{0d} = -\cos t \sin\rho \cos\theta_{d-1}\, \partial_t - \sin t \cos\rho \cos\theta_{d-1}\, \partial_\rho - \sin^2\theta_{d-1} \sin t/(\sin\rho)\, \partial_{\cos\theta_{d-1}}\,. \quad \text{(IV.25)}$$

Letting these act directly on Jacobi or hypergeometric modes results in rather complicated expressions. However, in [24] Dorn et al. note below equation (3.18) that a complex linear combination of these boosts increases the frequency $\omega$ exactly by one. Inspired by their equation (2.7) we thus define the $Z$-generators of the complexified Lie algebra:

$$\begin{aligned}
Z_d &:= K_{0d} + \mathrm{i} K_{d+1,d} = -\mathrm{e}^{\mathrm{i}t} \sin\rho \cos\theta_{d-1}\, \partial_t + \mathrm{i}\,\mathrm{e}^{\mathrm{i}t} \cos\rho \cos\theta_{d-1}\, \partial_\rho + \mathrm{i}\,\mathrm{e}^{\mathrm{i}t} \sin^{-1}\rho \sin^2\theta_{d-1}\, \partial_{\cos\theta_{d-1}} \\
\overline{Z_d} &:= K_{0d} - \mathrm{i} K_{d+1,d} = -\mathrm{e}^{-\mathrm{i}t} \sin\rho \cos\theta_{d-1}\, \partial_t - \mathrm{i}\,\mathrm{e}^{-\mathrm{i}t} \cos\rho \cos\theta_{d-1}\, \partial_\rho - \mathrm{i}\,\mathrm{e}^{-\mathrm{i}t} \sin^{-1}\rho \sin^2\theta_{d-1}\, \partial_{\cos\theta_{d-1}}.
\end{aligned} \quad \text{(IV.26)}$$

We first calculate the action of $\overline{Z_d}$ and $Z_d$ on the modes, and then the action of the $d$-boosts $K_{d+1,d}$ and $K_{0d}$. It is enough to know the actions of these two boosts, because the actions of the other boosts can be obtained from the Lie brackets of these two boosts with some rotators: from (IV.14) we have $[K_{0d}, K_{dk}] = \eta_{dd} K_{0k}$ and $[K_{d+1,d}, K_{dk}] = \eta_{dd} K_{d+1,k}$. We do not treat the Jacobi modes as special case of the $S^a$-modes in this section, because it is more direct to give the results in terms of $n$ for them, as opposed to $\omega$ for the hypergeometric modes.

Since for tube regions we have the ranges $\omega \in \mathbb{R}$ and $l \in \mathbb{N}_0$, for notational convenience for negative values of $l$ we set to zero all $z_{\omega l}^{(a,b)--}$, $z_{\omega l}^{(a,b)-+}$, $\tilde{z}_{\omega l}^{(a,b)+-}$, $\tilde{z}_{\omega l}^{(a,b)++}$ and $\phi_{\omega,l,\underline{\tilde{l}},m_l}^{a,b}$. The actions of the $d$-boosts write

$$\left(K_{0d} \triangleright \phi\right)^a_{\omega \underline{l} m_l} = \tfrac{\mathrm{i}}{2} \tilde{z}_{\omega-1,l+1}^{(a)+-}\, \phi^a_{\omega-1,l+1,\underline{\tilde{l}},m_l} + \tfrac{\mathrm{i}}{2} \tilde{z}_{\omega-1,l-1}^{(a)++}\, \phi^a_{\omega-1,l-1,\underline{\tilde{l}},m_l} + \tfrac{\mathrm{i}}{2} z_{\omega+1,l+1}^{(a)--}\, \phi^a_{\omega+1,l+1,\underline{\tilde{l}},m_l} + \tfrac{\mathrm{i}}{2} z_{\omega+1,l-1}^{(a)-+}\, \phi^a_{\omega+1,l-1,\underline{\tilde{l}},m_l} \quad \text{(IV.27)}$$

$$\left(K_{0d} \triangleright \phi\right)^b_{\omega \underline{l} m_l} = \tfrac{\mathrm{i}}{2} \tilde{z}_{\omega-1,l+1}^{(b)+-}\, \phi^b_{\omega-1,l+1,\underline{\tilde{l}},m_l} + \tfrac{\mathrm{i}}{2} \tilde{z}_{\omega-1,l-1}^{(b)++}\, \phi^b_{\omega-1,l-1,\underline{\tilde{l}},m_l} + \tfrac{\mathrm{i}}{2} z_{\omega+1,l+1}^{(b)--}\, \phi^b_{\omega+1,l+1,\underline{\tilde{l}},m_l} + \tfrac{\mathrm{i}}{2} z_{\omega+1,l-1}^{(b)-+}\, \phi^b_{\omega+1,l-1,\underline{\tilde{l}},m_l}$$

$$\left(K_{d+1,d} \triangleright \phi\right)^a_{\omega \underline{l} m_l} = -\tfrac{1}{2} \tilde{z}_{\omega-1,l+1}^{(a)+-}\, \phi^a_{\omega-1,l+1,\underline{\tilde{l}},m_l} - \tfrac{1}{2} \tilde{z}_{\omega-1,l-1}^{(a)++}\, \phi^a_{\omega-1,l-1,\underline{\tilde{l}},m_l} + \tfrac{1}{2} z_{\omega+1,l+1}^{(a)--}\, \phi^a_{\omega+1,l+1,\underline{\tilde{l}},m_l} + \tfrac{1}{2} z_{\omega+1,l-1}^{(a)-+}\, \phi^a_{\omega+1,l-1,\underline{\tilde{l}},m_l} \quad \text{(IV.28)}$$

$$\left(K_{d+1,d} \triangleright \phi\right)^b_{\omega \underline{l} m_l} = -\tfrac{1}{2} \tilde{z}_{\omega-1,l+1}^{(b)+-}\, \phi^b_{\omega-1,l+1,\underline{\tilde{l}},m_l} - \tfrac{1}{2} \tilde{z}_{\omega-1,l-1}^{(b)++}\, \phi^b_{\omega-1,l-1,\underline{\tilde{l}},m_l} + \tfrac{1}{2} z_{\omega+1,l+1}^{(b)--}\, \phi^b_{\omega+1,l+1,\underline{\tilde{l}},m_l} + \tfrac{1}{2} z_{\omega+1,l-1}^{(b)-+}\, \phi^b_{\omega+1,l-1,\underline{\tilde{l}},m_l}.$$

The above infinitesimal actions can be derived from those of $K_{d+1,d}$ and $K_{0d}$ on the hypergeometric modes. Applying actions (B.59)-(B.62) to expansion (IV.7) and shifting $\omega$ and $l$ by $\pm 1$ depending on the respective term yields the above actions.

Since for slice regions we have the ranges $n, l \in \mathbb{N}_0$, now for notational convenience we set to zero all quantities where $n$ or $l$ take values outside this range: all $\omega_{nl}^+$, $z_{nl}^{(+)-+}$, $z_{nl}^{(+)0-}$, $\tilde{z}_{nl}^{(+)+-}$, $\tilde{z}_{nl}^{(+)0+}$ and $\phi^\pm_{n,l,\underline{\tilde{l}},m_l}$ are set to zero for negative $n$ or $l$. Then the actions of the infinitesimal $d$-boosts write

$$\left(K_{0d} \triangleright \phi\right)^+_{n \underline{l} m_l} = \tfrac{\mathrm{i}}{2} z_{n,l+1}^{(+)0-}\, \phi^+_{n,l+1,\underline{\tilde{l}},m_l} + \tfrac{\mathrm{i}}{2} z_{n+1,l-1}^{(+)-+}\, \phi^+_{n+1,l-1,\underline{\tilde{l}},m_l} + \tfrac{\mathrm{i}}{2} \tilde{z}_{n-1,l+1}^{(+)+-}\, \phi^+_{n-1,l+1,\underline{\tilde{l}},m_l} + \tfrac{\mathrm{i}}{2} \tilde{z}_{n,l-1}^{(+)0+}\, \phi^+_{n,l-1,\underline{\tilde{l}},m_l} \quad \text{(IV.29)}$$

$$\overline{\left(K_{0d} \triangleright \phi\right)^-_{n \underline{l} m_l}} = -\tfrac{\mathrm{i}}{2} z_{n,l+1}^{(+)0-}\, \overline{\phi^-_{n,l+1,\underline{\tilde{l}},m_l}} - \tfrac{\mathrm{i}}{2} z_{n+1,l-1}^{(+)-+}\, \overline{\phi^-_{n+1,l-1,\underline{\tilde{l}},m_l}} - \tfrac{\mathrm{i}}{2} \tilde{z}_{n-1,l+1}^{(+)+-}\, \overline{\phi^-_{n-1,l+1,\underline{\tilde{l}},m_l}} - \tfrac{\mathrm{i}}{2} \tilde{z}_{n,l-1}^{(+)0+}\, \overline{\phi^-_{n,l-1,\underline{\tilde{l}},m_l}}$$

$$\left(K_{d+1,d} \triangleright \phi\right)^+_{n\underline{l}m_l} = \tfrac{1}{2} z^{(+)0-}_{n,l+1} \phi^+_{n,l+1,\underline{l},m_l} + \tfrac{1}{2} z^{(+)-+}_{n+1,l-1} \phi^+_{n+1,l-1,\underline{l},m_l} - \tfrac{1}{2} \tilde{z}^{(+)+-}_{n-1,l+1} \phi^+_{n-1,l+1,\underline{l},m_l} - \tfrac{1}{2} \tilde{z}^{(+)0+}_{n,l-1} \phi^+_{n,l-1,\underline{l},m_l}$$
(IV.30)

$$\overline{\left(K_{d+1,d} \triangleright \phi\right)^-_{n\underline{l}m_l}} = \tfrac{1}{2} z^{(+)0-}_{n,l+1} \overline{\phi^-_{n,l+1,\underline{l},m_l}} + \tfrac{1}{2} z^{(+)-+}_{n+1,l-1} \overline{\phi^-_{n+1,l-1,\underline{l},m_l}} - \tfrac{1}{2} \tilde{z}^{(+)+-}_{n-1,l+1} \overline{\phi^-_{n-1,l+1,\underline{l},m_l}} - \tfrac{1}{2} \tilde{z}^{(+)0+}_{n,l-1} \overline{\phi^-_{n,l-1,\underline{l},m_l}}$$

These can be derived from the action of $K_{d+1,d}$ and $K_{0d}$ on the Jacobi modes: applying actions (B.51)-(B.54) to expansion (IV.11) and shifting $n,l$ by $\pm 1$ depending on the respective term yields the above actions.

## V.  CORRESPONDENCE BETWEEN KG SOLUTIONS AND BOUNDARY DATA

In this section we develop evidence for one-to-one correspondences between initial/boundary data and solutions on the interior of our three types of AdS regions. For a complete analysis of this situation it would be necessary to specify the (equivalence) classes of solutions forming the solution spaces. Since we do not do this at this point, our calculations remain of formal nature, although we can derive explicit formulas.

### A.  AdS slice region

On slice regions any bounded free KG solution $\phi(t,\rho,\Omega)$ is a linear combination of Jacobi modes. Independently of what kind of boundary conditions one chooses at spatial infinity, once that $\phi(t,\rho,\Omega)$ has been fixed fulfilling them, it is completely determined by its momentum representation $(\phi^+_{n\underline{l}m_l}, \phi^-_{n\underline{l}m_l})$. Solutions on a slice region are determined by "initial" data on an equal-time hypersurface $\Sigma_{t_0}$ located at the (early or late) boundary of the slice or inside the region. The necessary data are the values of the field and its derivative on this hypersurface. The momentum representation of a solution $\phi(t,\rho,\Omega)$ can be calculated from these initial data on $\Sigma_{t_0}$ by formally inverting the Jacobi expansion (IV.11)

$$\phi^+_{n\underline{l}m_l} = \int dt \int d^{d-1}\Omega \ \tan^{d-1}\rho \ \overline{Y^{m_l}_{\underline{l}}(\Omega)} \ J^{(+)}_{nl}(\rho) \left(\hat{f}(t_0)\phi + \hat{d}(t_0)\partial_t \phi\right)(t_0,\rho,\Omega)$$

$$\overline{\phi^-_{n\underline{l}m_l}} = \int dt \int d^{d-1}\Omega \ \tan^{d-1}\rho \ \overline{Y^{m_l}_{\underline{l}}(\Omega)} \ J^{(+)}_{nl}(\rho) \left(\overline{\hat{f}(t_0)}\phi + \overline{\hat{d}(t_0)}\partial_t \phi\right)(t_0,\rho,\Omega) \ ,$$
(V.1)

with the operators $\hat{f}(t_0)$ and $\hat{d}(t_0)$ having $Y^{m_l}_{\underline{l}}(\Omega) \ J^{(+)}_{nl}(\rho)$ as eigenfunctions with eigenvalues $f_{n\underline{l}m_l}(t_0) = e^{i\omega^\pm_{nl}t_0} \tfrac{1}{2}\left(R^{d-1}_{\text{AdS}}\mathcal{N}^\pm_{nl}\right)^{-1/2}$ and $d_{n\underline{l}m_l}(t_0) = e^{i\omega^\pm_{nl}t} \tfrac{i}{2\omega^\pm_{nl}}\left(R^{d-1}_{\text{AdS}}\mathcal{N}^\pm_{nl}\right)^{-1/2}$. Because $Y^{m_l}_{\underline{l}}(\Omega) \ J^{(+)}_{nl}(\rho)$ form a complete orthogonal basis on $\Sigma_{t_0}$, the above formula provides a one-to-one correspondence between initial data $(\varphi(\rho,\Omega), \dot{\varphi}(\rho,\Omega))$ on $\Sigma_{t_0}$ and bounded solutions $\phi(t,\rho,\Omega)$ on the interior of the slice. The dot in $\dot{\varphi}$ is just a label, meaning that $\varphi$ represents values of the field $\phi$ and $\dot{\varphi}$ the values of its derivative $\partial_t \phi$ on $\Sigma_{t_0}$.

### B.  AdS tube region: well-behaved cases

Solutions on a tube region are also determined by "initial" data on a hypersurface: the hypercylinder $\Sigma_{\rho_0}$. Again, this "initial" hypersurface can be located at the (inner or outer) boundary of the tube or inside the region. Only when we consider initial/boundary data on the boundary $\rho_0 = \tfrac{\pi}{2}$ of AdS, we need to proceed more carefully. In any case, since the bounded tube KG solutions (like the slice solutions) are determined by *two* functions $\phi^a_{\omega l m_l}$ and $\phi^b_{\omega l m_l}$, the necessary data again consists of two pieces: the field configuration and the field derivative. We recall that bounded solutions for tube regions are the hypergeometric $S^a$ and $S^b$-modes which are regular everywhere, except for the time axis $\rho = 0$ where the $S^b$-modes diverge, and for spatial infinity $\rho = \tfrac{\pi}{2}$ where both $S^a$ and $S^b$-modes diverge. The latter divergence occurs only for $\widetilde{m}_- < 0$, that is, for masses $(m^2 > 0) \Leftrightarrow (\nu > d/2)$. For $\widetilde{m}_- \geq 0$, that is, $(m^2 \leq 0) \Leftrightarrow (\nu \leq d/2)$ the field (and its derivative) remain regular at $\rho = \tfrac{\pi}{2}$. Therefore, except for the case of both the "initial" hypersurface being located at spatial infinity $\rho = \tfrac{\pi}{2}$ with positive mass square $m^2 > 0$, we have a similar formula as for the slice region. Formally inverting the $S$-expansion (IV.7) determines the momentum representation of a free KG solution $\phi(t,\rho,\Omega)$ via its initial values and derivatives on $\Sigma_{\rho_0}$:

$$\phi^a_{\omega\underline{l}m_l} = \int_{\Sigma_{\rho_0}} dt\, d^{d-1}\Omega\, e^{i\omega t}\, \overline{Y^{m_l}_{\underline{l}}(\Omega)}\, \frac{\tan^{d-1}\rho}{2\pi(2\hat{l}+d-2)} \left( +(\partial_\rho S^b_{\hat{\omega}\hat{l}})(\rho_0)\,\phi(t,\rho_0,\Omega) - S^b_{\hat{\omega}\hat{l}}(\rho_0)\,(\partial_\rho\phi)(t,\rho_0,\Omega) \right)$$

$$\phi^b_{\omega\underline{l}m_l} = \int_{\Sigma_{\rho_0}} dt\, d^{d-1}\Omega\, e^{i\omega t}\, \overline{Y^{m_l}_{\underline{l}}(\Omega)}\, \frac{\tan^{d-1}\rho}{2\pi(2\hat{l}+d-2)} \left( -(\partial_\rho S^a_{\hat{\omega}\hat{l}})(\rho_0)\,\phi(t,\rho_0,\Omega) + S^a_{\hat{\omega}\hat{l}}(\rho_0)\,(\partial_\rho\phi)(t,\rho_0,\Omega) \right).$$

(V.2)

An equivalent formula for the expansion in $C$-modes is

$$\phi^{C,a}_{\omega\underline{l}m_l} = \int_{\Sigma_{\rho_0}} dt\, d^{d-1}\Omega\, e^{i\omega t}\, \overline{Y^{m_l}_{\underline{l}}(\Omega)}\, \frac{\tan^{d-1}\rho}{2\pi(2\nu)} \left( +(\partial_\rho C^b_{\hat{\omega}\hat{l}})(\rho_0)\,\phi(t,\rho_0,\Omega) - C^b_{\hat{\omega}\hat{l}}(\rho_0)\,(\partial_\rho\phi)(t,\rho_0,\Omega) \right)$$

$$\phi^{C,b}_{\omega\underline{l}m_l} = \int_{\Sigma_{\rho_0}} dt\, d^{d-1}\Omega\, e^{i\omega t}\, \overline{Y^{m_l}_{\underline{l}}(\Omega)}\, \frac{\tan^{d-1}\rho}{2\pi(2\nu)} \left( -(\partial_\rho C^a_{\hat{\omega}\hat{l}})(\rho_0)\,\phi(t,\rho_0,\Omega) + C^a_{\hat{\omega}\hat{l}}(\rho_0)\,(\partial_\rho\phi)(t,\rho_0,\Omega) \right).$$

(V.3)

The operators $\hat{\omega}$ and $\hat{l}$ have the eigenfunctions $e^{-i\omega t}\, Y^{m_l}_{\underline{l}}(\Omega)$ with eigenvalues $\omega$ and $l$. Again, since $e^{-i\omega t}\, Y^{m_l}_{\underline{l}}(\Omega)$ form a complete system on $\Sigma_{\rho_0}$, each pair of formulas provides a one-to-one correspondence between initial data $(\varphi(t,\Omega),\dot{\varphi}(t,\Omega))$ on $\Sigma_{\rho_0}$ and solutions $\phi(t,\rho,\Omega)$ on the interior of the tube. The dot in $\dot{\varphi}$ is again a label, now indicating that $\varphi$ represents the field values of $\phi$ and $\dot{\varphi}$ the values of the derivative $\partial_\rho\phi$ on $\Sigma_{\rho_0}$.

### C. Tube region: data on the boundary of AdS

For the "initial" hypersurface at spatial infinity $\rho = \frac{\pi}{2}$ with the mass square being positive $m^2 > 0$, we can try a version of the formula above, but since both $S^a_{\omega l}(\rho)$ and $S^b_{\omega l}(\rho)$ diverge for $\rho = \frac{\pi}{2}$ the "raw" boundary data will now be divergent, too. Since both radial functions diverge like $\cos^{\tilde{m}_-}\rho$ (and their derivatives like $\cos^{\tilde{m}_--1}\rho$), we could try to use the *rescaled* boundary data $\varphi^\partial(t,\Omega) = \cos^{-\tilde{m}_-}\rho\,\phi(t,\rho,\Omega)|_{\rho=\pi/2}$ and $\dot{\varphi}^\partial(t,\Omega) = \cos^{1-\tilde{m}_-}\rho\,(\partial_\rho\phi(t,\rho,\Omega))|_{\rho=\pi/2}$. We recall that the $S$ and $C$-modes are not linear independent, see IV.5. Using the $S$-expansion IV.7, let a first solution $\phi(t,\rho,\Omega)$ be given by its momentum representation $(\phi^a_{\omega\underline{l}m_l},\phi^b_{\omega\underline{l}m_l})$, and a second solution $\phi'(t,\rho,\Omega)$ by $(\phi'^a_{\omega\underline{l}m_l},\phi'^b_{\omega\underline{l}m_l})$, wherein $\phi'^a_{\omega\underline{l}m_l} = \phi^a_{\omega\underline{l}m_l} + \phi^0_{\omega\underline{l}m_l}\,(M^{no}_{11})_{\omega l}$ and $\phi'^b_{\omega\underline{l}m_l} = \phi^b_{\omega\underline{l}m_l} + \phi^0_{\omega\underline{l}m_l}\,(M^{no}_{12})_{\omega l}$, with $\phi^0_{\omega\underline{l}m_l}$ arbitrary. Then, both $\phi$ and $\phi'$ are different on the interior of AdS, but have the same boundary field values and derivatives (with or without rescaling as above). That is, this boundary data cannot distinguish between $\phi$ and $\phi'$. Using the $C$-expansion IV.8 makes the problem even more obvious: both rescaled boundary data $\varphi^\partial(t,\Omega) = \cos^{-\tilde{m}_-}\rho\,\phi(t,\rho,\Omega)|_{\rho=\pi/2}$ and $\dot{\varphi}^\partial(t,\Omega) = \cos^{1-\tilde{m}_-}\rho\,(\partial_\rho\phi(t,\rho,\Omega))|_{\rho=\pi/2}$ now only depend on one half of the momentum representation, namely on $\phi^{C,b}_{\omega\underline{l}m_l}$. This boundary data is blind to the $C^a$-mode content of any KG solution, since these modes vanish on the boundary whereas the $C^b$-modes diverge. The rescaling cures the divergence, but makes the vanishing even faster. We solve this problem in the following.

In [32], Claude Warnick investigates boundary conditions for (asymptotically) AdS spacetimes. However, as he noted below equations (3.4) and (4.6) therein, his method only works for a rather narrow range of mass: $\nu \in (0, 1)$, that is: $m^2 R^2_{\text{AdS}} \in (m^2_{\text{BF}}, m^2_{\text{BF}}+1)$. He introduces a "twisted derivative" $\partial^\alpha_{\cos\rho}$ that in our coordinates acts as $\partial^\alpha_{\cos\rho}\, f(\cos\rho) := (\cos\rho)^{-\alpha}\, \partial_{\cos\rho}(\cos^\alpha\rho\, f(\cos\rho))$. Motivated by his work we now construct a method that works for all mass values. To this end we define a higher order twisted derivative $\partial^{(\nu)}_\rho$ by

$$\partial^{(\nu)}_\rho f(\cos\rho) := (\cos\rho)^{1+2\lfloor\nu\rfloor-2\nu}\, \partial_{\cos\rho}\left(\frac{1}{\cos\rho}\partial_{\cos\rho}\right)^{\lfloor\nu\rfloor}\left\{(\cos\rho)^{-\tilde{m}_-}\, f(\cos\rho)\right\}.$$

(V.4)

Therein $\lfloor\nu\rfloor$ (read: floor) denotes the largest natural number that is smaller than $\nu$. Thus our twisted derivative is of order $\lfloor\nu\rfloor+1$, that is, its order depends on the value of the mass parameter $m^2$. In order to write its action on the radial functions $C^a_{\omega l}(\rho)$ and $C^b_{\omega l}(\rho)$, we perform a Taylor expansion of them near the boundary where $\rho \to \pi/2$ and thus $\cos\rho \to 0$. For this we need the definition of the hypergeometric function and the Taylor expansion of $\sin^l\rho$ around $\rho = \pi/2$, which for $\rho \in [0, \pi/2]$ is given by

$$\sin^l\rho = (1-\cos^2\rho)^{l/2} = \sum_{j=0}^{\infty} \frac{(-1)^j}{j!}\, (\cos\rho)^{2j}\, (\tfrac{l}{2}+1-j)_j.$$

(V.5)

The Pochhammer symbols therein are defined as $(a)_k := \Gamma(a+k)/\Gamma(a)$ for all complex $a, k$ that leave the Gamma functions well defined. For $k \in \mathbb{N}_0$ this definition becomes $(a)_k = a \cdot (a+1) \cdot \ldots \cdot (a+k-1)$. We also define the double Pochhammer symbol as $((a))_k := 2^k \Gamma(\frac{a}{2}+k)/\Gamma(\frac{a}{2})$ for all complex $a, k$ that leave the Gamma functions well defined. For $k \in \mathbb{N}_0$ this definition becomes $((a))_k = a\,(a+2)\,(a+4)\cdot\ldots\cdot(a+2k-2)$. Although we did not find it in the literature, this definition is quite natural. We thus have the relation $((2a))_n = 2^n (a)_n$ (a similar relation between factorial and double factorial is $(2n)!! = 2^n\,n!$).

Now back to (V.5): for even $l$ this sum only has $(\frac{l}{2}+1)$ terms, while for odd $l$ it has infinitely many. We find for the Taylor expansion of $C^a_{\omega l}$ and $C^b_{\omega l}$ near the boundary:

$$C^a_{\omega l}(\rho) = \sum_{a=0}^{\infty}(\cos\rho)^{\tilde{m}_+ + 2a}\,d_a^{(+)} \qquad\qquad C^b_{\omega l}(\rho) = \sum_{a=0}^{\infty}(\cos\rho)^{\tilde{m}_- + 2a}\,d_a^{(-)}\;,$$

wherein the coefficients are given by the finite sums

$$d_a^{(+)} = \sum_{b=0}^{a}\frac{(-1)^b}{b!}\,(\tfrac{l}{2}+1-b)_b\,\frac{(\alpha_+)_{a-b}\,(\beta_+)_{a-b}}{(\gamma_+^C)_{a-b}\,(a-b)!} \qquad\qquad d_a^{(-)} = \sum_{b=0}^{a}\frac{(-1)^b}{b!}\,(\tfrac{l}{2}+1-b)_b\,\frac{(\alpha_-)_{a-b}\,(\beta_-)_{a-b}}{(\gamma_-^C)_{a-b}\,(a-b)!}\;.$$

Letting the twisted derivative act on the Taylor expansions of $C^a_{\omega l}$ and $C^b_{\omega l}$ results in

$$\partial_\rho^{(\nu)} C^a_{\omega l}(\rho) = \sum_{a=0}^{\infty}(\cos\rho)^{2a}\,d_a^{(+)}\,((2\nu+2a-2\lfloor\nu\rfloor))_{\lfloor\nu\rfloor+1} \tag{V.6}$$

$$\partial_\rho^{(\nu)} C^b_{\omega l}(\rho) = \sum_{a=0}^{\infty}(\cos\rho)^{-2\nu+2a}\,d_a^{(-)}\,((2a-2\lfloor\nu\rfloor))_{\lfloor\nu\rfloor+1} \tag{V.7}$$

$$= \sum_{a=\lfloor\nu\rfloor+1}^{\infty}(\cos\rho)^{-2\nu+2a}\,d_a^{(-)}\,((2a-2\lfloor\nu\rfloor))_{\lfloor\nu\rfloor+1}\;. \tag{V.8}$$

The first terms in (V.7) vanish, because for low values of $a$ the double Pochhammer symbol contains a zero factor, caused by the factor $(1/\cos\rho)$ in the centre of the twisted derivative. Studying the limit $\rho \to \frac{\pi}{2}$, in (V.6) only the $(a=0)$-term survives (with $d_0^{(+)} = 1$), and thus the limit is finite :

$$\left[\partial_\rho^{(\nu)} C^a_{\omega l}\right]_{\rho\to\pi/2} = ((2\nu-2\lfloor\nu\rfloor))_{\lfloor\nu\rfloor+1} \qquad\qquad \forall\,\nu \notin \mathbb{Z}\;. \tag{V.9}$$

(If $\nu \in \mathbb{Z}$, then this limit vanishes, because the double Pochhammer symbol then contains a zero factor.) By contrast in (V.8) we always have $a \geq \lfloor\nu\rfloor+1$, and thus $a > \nu$. Hence the factor of $(\cos\rho)$ always appears with positive power. Therefore each summand vanishes in the limit $\rho \to \pi/2$ and we get

$$\left[\partial_\rho^{(\nu)} C^b_{\omega l}\right]_{\rho\to\pi/2} = 0\;. \tag{V.10}$$

For constructing a relation between a KG solution and its boundary behaviour, first we define $\varphi^{\partial-}$ as rescaled boundary field value, and $\varphi^{\partial+}_{(\nu)}$ as the boundary value of the twisted derivative of the field:

$$\varphi^{\partial-}(t,\rho,\Omega) := \left[\cos^{-\tilde{m}_-}\rho\,\phi(t,\rho,\Omega)\right]_{\rho\to\pi/2} \qquad\qquad \varphi^{\partial+}_{(\nu)}(t,\rho,\Omega) := \left[\partial_\rho^{(\nu)}\phi(t,\rho,\Omega)\right]_{\rho\to\pi/2}. \tag{V.11}$$

Plugging into this the $C$-expansion (IV.8) of the solution and using (V.9), we can formally invert this relation, thus recovering the full momentum representation as a function of the (rescaled and hence finite) boundary data:

$$\phi^{C,a}_{\omega\underline{l}m_l} = \int\!dt\,d^{d-1}\Omega\,\frac{e^{i\omega t}\,\overline{Y^{m_l}_{\underline{l}}(\Omega)}}{((2\nu-2\lfloor\nu\rfloor))_{\lfloor\nu\rfloor+1}}\,\varphi^{\partial+}_{(\nu)}(t,\Omega)/(2\pi) \qquad \phi^{C,b}_{\omega\underline{l}m_l} = \int\!dt\,d^{d-1}\Omega\,e^{i\omega t}\,\overline{Y^{m_l}_{\underline{l}}(\Omega)}\,\varphi^{\partial-}(t,\Omega)/(2\pi)\;. \tag{V.12}$$

## D. AdS rod region

Free KG solutions on a rod region are also determined by "initial" data on a hypercylinder. Again, this "initial" hypersurface can be located at the (one and only outer) boundary of the rod region, or in its interior. The bounded solutions for the rod regions are only the hypergeometric $a$-modes which are regular everywhere, except for spatial infinity $\rho = \frac{\pi}{2}$ where they diverge. Thus the rod region's free KG solutions are determined by *only one* function $\phi^a_{\omega \underline{l} m_l}$ on momentum space, and therefore the necessary "initial" data is now only the field configuration on a hypercylinder. The rod expansion (IV.10) can be written as

$$\phi(t,r,\Omega) = \int d\omega \sum_{\underline{l},m_l} e^{-i\omega t} Y_{\underline{l}}^{m_l}(\Omega) \phi^a_{\omega \underline{l} m_l} \left[ (M_{11}^{\text{on}})_{\omega l} C^a_{\omega l}(\rho) + (M_{12}^{\text{on}})_{\omega l} C^b_{\omega l}(\rho) \right] . \tag{V.13}$$

see (B.23). If the initial hypersurface $\Sigma_{\rho_0}$ is not at the boundary: $\rho_0 < \frac{\pi}{2}$, we can formally invert the definition

$$\varphi^{\rho_0}(t,\Omega) := \phi(t,\rho_0,\Omega) = \int d\omega \sum_{\underline{l},m_l} \phi^a_{\omega \underline{l} m_l} e^{-i\omega t} Y_{\underline{l}}^{m_l}(\Omega) S^a_{\omega l}(\rho_0) , \tag{V.14}$$

$$\Longrightarrow \quad \phi^a_{\omega \underline{l} m_l} = \int dt\, d^{d-1}\Omega\, \frac{1}{(2\pi) S^a_{\omega l}(\rho_0)} e^{i\omega t} \overline{Y_{\underline{l}}^{m_l}(\Omega)}\, \varphi^{\rho_0}(t,\Omega) . \tag{V.15}$$

The zeros of $S^a_{\omega l}(\rho_0)$ do not pose a problem, because the set of $(\omega,l)$ causing them is of measure zero in momentum space. Singularities caused by zeros of $S^a_{\omega l}(\rho_0)$ are canceled by their second appearance when we expand the initial data in terms of $S^a$-modes as in (V.14). If the initial data is placed on the boundary $\rho_0 = \pi/2$, then only the $C^b_{\omega l}$-part of $S^a_{\omega l}$ survives, as in (V.13), and we can rescale as follows:

$$\varphi^\partial_{\rho_0}(t,\Omega) := \cos^{-\tilde{m}_-}\rho\, \phi(t,\rho \to \pi/2,\Omega) = \int d\omega \sum_{\underline{l},m_l} e^{-i\omega t} Y_{\underline{l}}^{m_l}(\Omega)\, \phi^a_{\omega \underline{l} m_l} (M_{12}^{\text{on}})_{\omega l} . \tag{V.16}$$

Thus finite boundary data $\varphi^\partial(t,\Omega)$ is the rescaled boundary field value. Formal inversion recovers the momentum representation of a bounded free KG solution on a rod region:

$$\phi^a_{\omega \underline{l} m_l} = \int dt\, d^{d-1}\Omega\, \frac{1}{(2\pi)(M_{12}^{\text{on}})_{\omega l}} e^{i\omega t} \overline{Y_{\underline{l}}^{m_l}(\Omega)}\, \varphi^\partial_{\rho_0}(t,\Omega) . \tag{V.17}$$

## VI. SYMPLECTIC STRUCTURES ON SPACES OF ADS KG SOLUTIONS

We now establish symplectic structures for the spaces of KG solutions for our three types of regions. Recalling that the slice is a neighborhood of an equal-time hyperplane $\Sigma_t$, and the tube of an equal-$\rho$ hypercylinder $\Sigma_\rho$, we associate a symplectic structure to each of these hypersurfaces. These symplectic structures are antisymmetric, bilinear forms on the spaces of those KG solutions that are well defined and bounded in a neighborhood of their associated hypersurface. As discussed in Section 3 of [28] and Chapter 7 of [27], the symplectic structure arises naturally from the Lagrange density form $\Lambda(x)$ in the following way. Choose a (compact or noncompact) region $\mathbb{M}$ of $(d+1)$-dimensional spacetime. Then, for solutions of the equations of motion (EOM) the one-form $dS^{\mathbb{M}}$ associated to $\mathbb{M}$ by $S^{\mathbb{M}} = \int_{\mathbb{M}} \Lambda(x)$ is an integral over the boundary $\partial\mathbb{M}$, and we define the symplectic potential $\theta$ associated to a hypersurface $\Sigma$ to be just this integral: let $\phi$ and $\eta$ two solutions, then

$$dS^{\mathbb{M}}_\phi(\eta) = \int_{\partial\mathbb{M}} \eta\, \partial_\mu \lrcorner \left. \frac{\delta\Lambda}{\delta(\partial_\mu \varphi)} \right|_{\varphi = \phi} \qquad \theta^\Sigma_\phi(\eta) = \int_\Sigma \eta\, \partial_\mu \lrcorner \left. \frac{\delta\Lambda}{\delta(\partial_\mu \varphi)} \right|_{\varphi = \phi} . \tag{VI.1}$$

Thus, for solutions on a compact region $\mathbb{M}$ with only one connected boundary surface $\partial\mathbb{M}$ the symplectic potential is precisely the differential of the action: $\theta^{\partial\mathbb{M}}_\phi(\eta) = dS^{\mathbb{M}}_\phi(\eta)$. The symplectic structure associated to a hypersurface $\Sigma$ is the differential of its symplectic potential: for solutions $\phi, \eta, \zeta$ well defined in a neighborhood of $\Sigma$ this means that $\omega^\Sigma_\phi(\eta,\zeta) = d\theta^\Sigma_\phi(\eta,\zeta)$. Thus, for a compact region with only one connected boundary surface the symplectic structure is the second differential of the action and therefore zero for solutions. For compact regions bounded by two hypersurfaces $\Sigma_1$ and $\Sigma_2$ the symplectic structure is not identically zero for each hypersurface. What is more, since the difference $\omega_{\Sigma_2} - \omega_{\Sigma_1}$ now is the second differential of the action, it vanishes. This implies that the symplectic structure is actually independent of the hypersurface: $\omega_{\Sigma_2} = \omega_{\Sigma_1}$. Our explicit expressions below show that for AdS the symplectic structures associated to $\Sigma_t$ and $\Sigma_\rho$ are independent of their associated hypersurface, too: $\omega^{\Sigma_t}_\phi(\eta,\zeta)$ and $\omega^{\Sigma_\rho}_\phi(\eta,\zeta)$ depend only on the two solutions $\eta$ and $\zeta$. Moreover, the symplectic structure vanishes on the space of solutions on rod regions (which possesses only one connected boundary surface). These results are obtained despite neither of our regions being compact nor our solutions being compactly supported.

### A. Equal-time hypersurfaces

Let $\eta(t,\rho,\Omega)$ and $\zeta(t,\rho,\Omega)$ two KG solutions well defined and bounded on a neighborhood of an equal-time hyperplane $\Sigma_{t_0}$, that is, of Jacobi type (IV.11). For this type of solutions a symplectic structure can be associated to $\Sigma_{t_0}$ by:

$$\omega_{\Sigma_{t_0}}(\eta,\zeta) = -\frac{1}{2}\int_0^{\pi/2}\!\!\mathrm{d}\rho\int_{\mathbb{S}^{d-1}}\!\!\mathrm{d}^{d-1}\Omega\; R_{\mathrm{AdS}}^{d-1}\tan^{d-1}\rho\,\big(\eta\,\partial_t\zeta - \zeta\,\partial_t\eta\big)(t_0,\rho,\Omega) \tag{VI.2}$$

$$= +\mathrm{i}\sum_{n l m_l}\omega_{nl}^+ R_{\mathrm{AdS}}^{d-1}\mathcal{N}_{nl}^+\left\{\overline{\eta_{n\underline{l}m_l}^-}\,\zeta_{n\underline{l}m_l}^+ - \eta_{n\underline{l}m_l}^+\,\overline{\zeta_{n\underline{l}m_l}^-}\right\}. \tag{VI.3}$$

It was introduced for $\mathrm{AdS}_{1,3}$ in equation (2.5) of [5] as "inner product" $B(\alpha,\beta)$. The first line is the coordinate representation of the symplectic structure, the second the momentum representation. The latter can be calculated from the former by plugging in expansion (IV.11) for both $\eta$ and $\zeta$, and then using orthogonality properties of the solutions. From this result we can read off that the positive and negative frequency modes form Lagrangian subspaces of the space of KG solutions on the AdS slice region, see (1.2.3) in [27]. The full space of KG solutions on the slice region is the direct sum of the positive frequency subspace and the negative frequency subspace. The flat limit of this symplectic structure for $d=3$ is the symplectic structure (II.4) associated to an equal-time plane in Minkowski spacetime (for $\tilde\eta^{\mathrm{M},\pm}_{\tilde p l m_l}, \tilde\zeta^{\mathrm{M},\pm}_{\tilde p l m_l}$ see IV.12):

$$\omega_{\Sigma_t}(\eta,\zeta) \xrightarrow[\text{lim.}]{\text{flat}} \mathrm{i}\int_0^\infty\!\!\mathrm{d}\tilde p\sum_{l,m_l}\tilde\omega_{\tilde p}\left\{\overline{\tilde\eta^{\mathrm{M},-}_{\tilde p l m_l}}\,\tilde\zeta^{\mathrm{M},+}_{\tilde p l m_l} - \tilde\eta^{\mathrm{M},+}_{\tilde p l m_l}\,\overline{\tilde\zeta^{\mathrm{M},-}_{\tilde p l m_l}}\right\}.$$

It is calculated by plugging the flat limit of the ordinary Jacobi expansion into the coordinate representation of the symplectic structure, and again using orthogonality property of the modes.

### B. Hypercylinders

Let now $\eta(t,\rho,\Omega)$ and $\zeta(t,\rho,\Omega)$ two KG solutions well defined and bounded on a neighborhood of an equal-radius hypercylinder $\Sigma_{\rho_0}$, that is, of hypergeometric type (IV.7) respectively (IV.8). For this type of solutions a symplectic structure associated to $\Sigma_{\rho_0}$ is given by

$$\omega_{\Sigma_{\rho_0}}(\eta,\zeta) = \frac{1}{2}\int\mathrm{d}t\,\mathrm{d}^{d-1}\Omega\; R_{\mathrm{AdS}}^{d-1}\tan^{d-1}\rho_0\,\big(\eta\,\partial_\rho\zeta - \zeta\,\partial_\rho\eta\big)(t,\rho_0,\Omega) \tag{VI.4}$$

$$= \pi R_{\mathrm{AdS}}^{d-1}\int\mathrm{d}\omega\sum_{\underline{l},m_l}\left\{\eta^a_{\omega\underline{l}m_l}\zeta^b_{-\omega,\underline{l},-m_l} - \eta^b_{\omega\underline{l}m_l}\zeta^a_{-\omega,\underline{l},-m_l}\right\}(2l+d-2) \tag{VI.5}$$

$$= \pi R_{\mathrm{AdS}}^{d-1}\int\mathrm{d}\omega\sum_{\underline{l},m_l}\left\{\eta^{C,a}_{\omega\underline{l}m_l}\zeta^{C,b}_{-\omega,\underline{l},-m_l} - \eta^{C,b}_{\omega\underline{l}m_l}\zeta^{C,a}_{-\omega,\underline{l},-m_l}\right\}(2\nu). \tag{VI.6}$$

The first line is again the coordinate representation of the symplectic structure, the second and third are momentum space representations. The latter two can be calculated from the first by plugging in expansions (IV.7) respectively (IV.8), plus using orthogonality properties of the solutions and the Wronskians computed in Appendix B 4. We can read off that the $S^a$ and $S^b$-modes form Lagrangian subspaces of the space of KG solutions on the AdS tube region. The $C^a$ and $C^b$-modes form a different pair of Lagrangian subspaces. In both cases, the full space of KG solutions on the tube region is the direct sum of the Lagrangian ($a$)-subspace and the Lagrangian ($b$)-subspace. Recalling the results of Section V on initial/boundary data, one subspace is related to the field values and another one to the field derivatives/momenta. Lagrangian subspaces corresponding to field values and momenta play an important role in the Schrödinger representation of Quantum Field Theory, see e.g. [33]. The flat limit of this symplectic structure is the symplectic structure (II.7) associated to a hypercylinder $\Sigma_r$ in Minkowski spacetime, see Section 5.3 in [28] by Oeckl. It can be calculated by plugging the flat limit of the $S$-expansion into the coordinate representation of the symplectic structure, and using orthogonality properties of the modes (for $\tilde\eta^{\mathrm{M},a,b}_{\tilde\omega l m_l}, \tilde\zeta^{\mathrm{M},a,b}_{\tilde\omega l m_l}$ see (IV.9)):

$$\omega_{\Sigma_\rho}(\eta,\zeta) \xrightarrow[\text{lim.}]{\text{flat}} \int\mathrm{d}\tilde\omega\sum_{l,m_l}\frac{\tilde p^{\mathbb{R}}_{\tilde\omega}}{16\pi}\left\{\tilde\eta^{\mathrm{M},a}_{\tilde\omega l m_l}\,\tilde\zeta^{\mathrm{M},b}_{-\tilde\omega,l,-m_l} - \tilde\eta^{\mathrm{M},b}_{\tilde\omega l m_l}\,\tilde\zeta^{\mathrm{M},a}_{-\tilde\omega,l,-m_l}\right\}.$$

Recalling that the KG solutions on the rod region are purely $S^a$-modes, we find confirmed the initial consideration that the symplectic structure associated to the boundary of any rod region vanishes. And further, since the AdS-Jacobi modes (the only allowed modes for slice regions) are merely special cases of the $S^a$-modes, the symplectic structure $\omega_{\rho_0}$ returns zero for any two such modes.

### C. Invariance of symplectic structures under isometries

For showing the invariance of the symplectic structures associated to equal-time hyperplanes and equal-radius hypercylinders under time translation, rotations and boosts, we use the action of these isometries on the KG solutions. For boosts, it is sufficient to check that the $d$-boosts leave the symplectic structures invariant: from the Lie algebra (IV.14) we know that the commutators of rotations $K_{dk}$ and $d$-boosts are the remaining boosts: $[K_{0d}, K_{dk}] = \eta_{dd} K_{0k}$ and $[K_{d+1,d}, K_{dk}] = \eta_{dd} K_{d+1,k}$, and if two generators leave the symplectic structures invariant, then their commutator does so as well. The action of an (infinitesimal or finite) isometry $k$ on the symplectic structures $\omega$ fulfills

$$(k \triangleright \omega)(\eta, \zeta) = \omega(k^{-1} \triangleright \eta, k^{-1} \triangleright \zeta). \qquad (K_{AB} \triangleright \omega)(\eta, \zeta) = \omega(-K_{AB} \triangleright \eta, \zeta) + \omega(\eta, -K_{AB} \triangleright \zeta). \tag{VI.7}$$

#### 1. Invariance of symplectic structures under time translations

For finite time translations we denote the corresponding element of $\mathrm{SO}(2,d)$ by $k_{\Delta t}$. For both equal-time surface and hypercylinder we then have the action on the symplectic structure given by (VI.7). For the hypercylinder $\Sigma_\rho$ plugging action (IV.19) of $k_{-\Delta t}$ into the symplectic structure (VI.5) shows its invariance under time translations (for $\Sigma_t$ we plug action (IV.20) of $k_{-\Delta t}$ into the symplectic structure (VI.3)):

$$(k_{\Delta t} \triangleright \omega_{\Sigma_\rho})(\eta, \zeta) = \pi R_{\mathrm{AdS}}^{d-1} \int d\omega \sum_{\underline{l}, m_l} (2l+d-2) \left\{ \mathrm{e}^{-\mathrm{i}\omega\Delta t} \eta^a_{\omega \underline{l} m_l} \, \mathrm{e}^{\mathrm{i}\omega\Delta t} \, \zeta^b_{-\omega, \underline{l}, -m_l} - \mathrm{e}^{-\mathrm{i}\omega\Delta t} \eta^b_{\omega \underline{l} m_l} \, \mathrm{e}^{\mathrm{i}\omega\Delta t} \, \zeta^a_{-\omega, \underline{l}, -m_l} \right\}$$

$$= \pi R_{\mathrm{AdS}}^{d-1} \int d\omega \sum_{\underline{l}, m_l} (2l+d-2) \left\{ \eta^a_{\omega \underline{l} m_l} \zeta^b_{-\omega, \underline{l}, -m_l} - \eta^b_{\omega \underline{l} m_l} \zeta^a_{-\omega, \underline{l}, -m_l} \right\} = \omega_{\Sigma_\rho}(\eta, \zeta) \qquad \forall \, \eta, \zeta.$$

#### 2. Invariance of symplectic structures under rotations

For finite rotations we denote the corresponding element of $\mathrm{SO}(2,d)$ by $\hat{R}(\underline{\alpha})$, see Appendix A. For $\Sigma_t$ and $\Sigma_\rho$ the action on the symplectic structure is given by (VI.7). For the hypercylinder plugging the action (IV.22) of $\hat{R}(\underline{\alpha})^{-1}$ into the symplectic structure (VI.5) (and applying completeness relation (A.11) for Wigner's $D$-matrix) shows its invariance under rotations (for $\Sigma_t$ we plug action (IV.23) into (VI.3)):

$$(\hat{R}(\underline{\alpha}) \triangleright \omega_{\Sigma_\rho})(\eta, \zeta) = \pi R_{\mathrm{AdS}}^{d-1} \int d\omega \sum_{\underline{l}, m_l} (2l+d-2) \sum_{\underline{l}' m_l'} \sum_{\underline{l}'' m_l''} \left\{ \eta^a_{\omega l \, \underline{l}' m_l'} \, \overline{(D^l_{\underline{l}', \underline{l}}(\underline{\alpha}))_{m_l' m_l}} \, \zeta^b_{-\omega, l, \underline{l}'', -m_l''} \, (D^l_{\underline{l}'', \underline{l}}(\underline{\alpha}))_{m_l'' m_l} \right.$$

$$\left. - \eta^b_{\omega l \, \underline{l}' m_l'} \, \overline{(D^l_{\underline{l}', \underline{l}}(\underline{\alpha}))_{m_l' m_l}} \, \zeta^a_{-\omega, l, \underline{l}'', -m_l''} \, (D^l_{\underline{l}'', \underline{l}}(\underline{\alpha}))_{m_l'' m_l} \right\}$$

$$= \pi R_{\mathrm{AdS}}^{d-1} \int d\omega \sum_{\underline{l}, m_l} (2l+d-2) \left\{ \eta^a_{\omega \underline{l} m_l} \zeta^b_{\omega, \underline{l}, -m_l} - \eta^b_{\omega \underline{l} m_l} \zeta^a_{\omega, \underline{l}, -m_l} \right\} = \omega_{\Sigma_\rho}(\eta, \zeta) \qquad \forall \, \eta, \zeta.$$

#### 3. Invariance of symplectic structures under $d$-boosts

For showing the invariance of the symplectic structures under the actions of $K_{0d}$ and $K_{d+1,d}$ the calculations are essentially the same (up to some factors of $\pm \mathrm{i}$). For $\Sigma_t$ and $\Sigma_\rho$ we have the action on the symplectic structure given by (VI.7) For the hypercylinder $\Sigma_\rho$ we plug action (IV.27) of $K_{0d}$ respectively (IV.28) of $K_{d+1,d}$ into the symplectic structure (VI.5). Then, shifting indexes appropriately and using the equalities

$$\tilde{z}^{(a)+-}_{\omega-1,l+1} = \tfrac{2l+d}{2l+d-2} z^{(b)-+}_{\omega l} \qquad\qquad \tilde{z}^{(a)++}_{\omega-1,l-1} = \tfrac{2l+d-4}{2l+d-2} z^{(b)--}_{\omega l} \tag{VI.8}$$

$$z^{(a)--}_{\omega+1,l+1} = \tfrac{2l+d}{2l+d-2} \tilde{z}^{(b)++}_{\omega l} \qquad\qquad z^{(a)-+}_{\omega+1,l-1} = \tfrac{2l+d-4}{2l+d-2} \tilde{z}^{(b)+-}_{\omega l},$$

it is somewhat lengthy but straightforward to check that $0 = \omega_{\Sigma_\rho}(-K_{0d} \triangleright \eta, \zeta) + \omega_{\Sigma_\rho}(\eta, -K_{0d} \triangleright \zeta)$, making the hypercylinder's symplectic structure invariant under infinitesimal $d$-boosts. Checking (VI.8) is again somewhat lengthy but straightforward, plugging in the definitions (B.57) and (B.58) plus using $\chi^{(d-1)}_-(l+1, l_{d-2}) = \chi^{(d-1)}_+(l, l_{d-2})$ is sufficient.

For the equal-time surface $\Sigma_t$ we plug the action (IV.29) of $K_{0d}$ respectively (IV.30) of $K_{d+1,d}$ into the symplectic structure (VI.3). The calculation is similar as for $\Sigma_\rho$, now using the equalities

$$\omega_{nl}^+ \mathcal{N}_{nl}^+ z_{n,l+1}^{(+)0-} = \omega_{n,l+1}^+ \mathcal{N}_{n,l+1}^+ \tilde{z}_{nl}^{(+)0+} \qquad \omega_{nl}^+ \mathcal{N}_{nl}^+ z_{n+1,l-1}^{(+)-+} = \omega_{n+1,l-1}^+ \mathcal{N}_{n+1,l-1}^+ \tilde{z}_{nl}^{(+)+-} . \tag{VI.9}$$

Thus $0 = \omega_{\Sigma_t}(-K_{0d}\triangleright\eta,\,\zeta) + \omega_{\Sigma_t}(\eta,\,-K_{0d}\triangleright\zeta)$, and hence the equal-time plane's symplectic structure is also invariant under infinitesimal $d$-boosts. Checking the above equalities is done by plugging in the definitions (B.9), (IV.6), (B.47) and (B.50).

## VII. SUMMARY AND OUTLOOK

We study Klein-Gordon (KG) theory for three types of regions on AdS: the slice region, and rod and tube hypercylinder regions. Slice regions $[t_1,t_2]\times[0,\infty)_\rho\times\mathbb{S}^{d-1}$ are neighborhoods of equal-time surfaces $\Sigma_t$, tubes $\mathbb{R}_t\times[\rho_1,\rho_2]\times\mathbb{S}^{d-1}$ are neighborhoods of equal-radius hypercylinders $\Sigma_\rho$, and rods $\mathbb{R}_t\times[0,\rho_0]\times\mathbb{S}^{d-1}$ are $\Sigma_\rho$-neighborhoods containing the whole region (containing the time axis) enclosed by $\Sigma_\rho$. We present a complete list of KG solutions on AdS spacetimes. These solutions are well known already, except for two exceptional cases that we present here to make the list complete. The KG solutions well defined and bounded on slice regions are the ordinary and exceptional AdS-Jacobi modes. For tube regions, there are the hypergeometric modes of types $S^a$, $S^b$ and $C^a$, $C^b$, while on rod regions the hypergeometric $S^a$-modes give the only well defined and bounded solutions. Between (the flat limit of the) Killing vectors of AdS and Minkowski spacetime we find a correspondence which among others relates Minkowski translations to some AdS boosts. We also find the actions of the Killing vector fields of AdS (that is, the generators of its isometry group's Lie algebra $\mathbf{so}\,(2,d)$) on the modes. For time translations this action is a phase factor. For rotations it is a sum over modes of same top angular momentum number $l$ with elements of Wigner's $D$-matrix as coefficients. For infinitesimal $d$-boosts the action is a sum over contiguous modes with certain coefficients that we calculate explicitly. As a byproduct we find some contiguous relations for hyperspherical harmonics and Jacobi polynomials.

We also find one-to-one correspondences between initial data and classical solutions on the regions. For slice regions, this initial data consists of the values of the field and its derivative on an equal-time surface $\Sigma_t$, despite AdS not being globally hyperbolic. For tube regions, the initial data can be given on an equal-radius hypercylinder $\Sigma_\rho$ that is inside of AdS. Then it consists again of field and derivative. If the hypercylinder is the boundary $\Sigma_{\rho=\pi/2}$ of AdS, then we need to rescale: the initial data consists now of rescaled field values together with a rescaled "twisted" derivative of the field. For rod regions, a one-to-one correspondence can be established with the field values as sufficient initial data on a hypercylinder. For the boundary hypercylinder we need to rescale again.

Next, for equal-time surfaces $\Sigma_t$ and equal-radius hypercylinder $\Sigma_\rho$ we give the symplectic structures determined naturally by the Lagrange density of the theory. The well known symplectic structure for slice regions is defined using an equal-time surface $\Sigma_t$, but turns out to be actually $t$-independent. Our new symplectic structure for tube regions is defined using an equal-radius hypercylinder $\Sigma_\rho$, and turns out to be $\rho$-independent. These independencies are not trivial, since usually they only hold for boundaries of compact regions (which our slice, rod and tube regions are not), respectively compactly supported solutions, (which our slice, rod and tube solutions are not). The symplectic structure for rod regions vanishes identically. Again, for regions with a connected boundary this usually holds only if the region is compact respectivvely for compactly supported solutions. As a byproduct we find the Wronskian for the hypergeometric functions involved. Then we proceed checking the invariance of the symplectic structures under the actions of the $\mathbf{so}\,(2,d)$-generators. It turns out, that both symplectic structures are invariant under all AdS isometries, that is, under time translation, spatial rotations and boosts.

One motivation for this study is that our goal in future work will be to construct an S-matrix for AdS using Oeckl's General Boundary Formulation (GBF) of Quantum Theory [34, 35]. To date two quantization methods are used in this framework: Schrödinger-Feynman (SFQ) and Holomorphic Quantization (HQ), see [28]. The equivalence of SFQ and HQ has been proven in [33] by Oeckl. Previous results related to our goal are the following. Using Schrödinger-Feynman Quantization for real Klein-Gordon theory in Minkowski spacetime, Colosi and Oeckl in [36, 37] rederived the standard S-matrix and moreover an equivalent radial S-matrix. In [38] Colosi then calculated both types of S-matrices also for deSitter spacetimes, again using SFQ. Lastly, the author and Colosi in [39] derived the Schrödinger-Feynman expressions for both types of S-matrices for a wide class of spacetimes, with their results being applied to AdS in [40]. Now wishing to calculate this S-matrix also with Holomorphic Quantization, as an essential ingredient on the spaces of Klein-Gordon solutions we need the symplectic structures that we calculate in this article. The second necessary ingredient will be complex structures, to be presented in future work. The $SO(2,d)$-invariance of the symplectic structures then will be one ingredient for assuring the isometry invariance of the AdS S-matrix that we aim to construct.

### ACKNOWLEDGMENTS

The author is very grateful to Robert Oeckl and Daniele Colosi of CCM-UNAM at Morelia for many stimulating and critical discussions and thorough proofreading. This work was supported by CONACyT scholarship 213531 and UNAM-DGAPA-PAPIIT Project Nr. IN100212.

### Appendix A: Hyperspherical harmonics

On the unit sphere $\mathbb{S}^{d-1}$ the metric tensor is denoted by $g_{\mathbb{S}^{d-1}}$. The $(d-1)$ angular coordinates $\theta_i$ on $\mathbb{S}^{d-1}$ with $i \in \{1,...,(d-1)\}$ are denoted collectively by $\Omega = (\theta_{d-1},...,\theta_1)$. On the two-sphere, traditionally $\theta_2$ is just denoted by $\theta$, and $\theta_1$ by $\varphi$. Defining $|g_{\mathbb{S}}| := |\det(g_{\mathbb{S}^{d-1}})_{\mu\nu}|$, we denote the volume element on $\mathbb{S}^{d-1}$ by $\mathrm{d}^{d-1}\Omega = \mathrm{d}\Omega \sqrt{|g_{\mathbb{S}}|}$, where $\mathrm{d}\Omega$ abbreviates $\mathrm{d}\theta_1 \ldots \mathrm{d}\theta_{d-1}$.

Since there are different ways of defining them, see e.g. [41], we here give the hyperspherical harmonics used in this work. The hyperspherical harmonics $Y_{\underline{l}}^{m_l}(\Omega)$ are the eigenfunctions of the Laplacian on $\mathbb{S}^{d-1}$ with $d \geq 3$. The eigenvalues are: $\square_{\mathbb{S}^{d-1}} Y_{\underline{l}}^{m_l}(\Omega) = -l(l+d-2) Y_{\underline{l}}^{m_l}(\Omega)$. The $\underline{l}$ represents the multiindex $\underline{l} = (l_{d-1}, \tilde{\underline{l}})$ with $\tilde{\underline{l}} = (l_{d-2},...,l_2)$. We often write simply $l$ for $l_{d-1}$. The $m_l$ carries its subindex in order to distinguish it from the mass $m$, and $\overline{Y_{\underline{l}}^{m_l}(\Omega)} = Y_{\underline{l}}^{-m_l}(\Omega)$. The indices take the following integer values: $l \equiv l_{d-1} \in \{0, 1, 2, ...\}$, while $l_i \in \{0, 1, ..., l_{i+1}\}$ for all $i \in \{2,...,(d-2)\}$, and $m_l \in \{-l_2, -l_2+1, ..., l_2\}$.

The spherical harmonics on the two-sphere are $Y_l^{m_l}(\Omega) = \mathcal{N}_l^{m_l} \mathrm{e}^{\mathrm{i}m_l \varphi} P_l^{m_l}(\cos\theta)$, with $P_l^{m_l}$ denoting the associated Legendre polynomials. An orthonormal system is obtained by choosing the standard normalisation (using AS[8.14.11+13]) $\mathcal{N}_l^{m_l} = \sqrt{(2l+1)(l-m_l)! / (4\pi (l+m_l)!)}$.

We need to decompose spherical harmonics $Y_l^{m_l}$ and their derivative into contiguous spherical harmonics $Y_{l\pm 1}^{m_l}$. Using AS [8.5.3+4] we find the following relations for contiguous spherical harmonics:

$$\cos\theta \, Y_l^{m_l}(\Omega) = \chi_-^{(2)}(l,m_l) \, Y_{l-1}^{m_l}(\Omega) + \chi_+^{(2)}(l,m_l) \, Y_{l+1}^{m_l}(\Omega) \tag{A.1}$$

$$(1-\cos^2\theta) \frac{\mathrm{d}}{\mathrm{d}\cos\theta} Y_l^{m_l}(\Omega) = \delta_-^{(2)}(l,m_l) \, Y_{l-1}^{m_l}(\Omega) + \delta_+^{(2)}(l,m_l) \, Y_{l+1}^{m_l}(\Omega) \tag{A.2}$$

with the raising and lowering coefficients

$$\chi_-^{(2)}(l,m_l) = \sqrt{\frac{l^2-m_l^2}{(2l-1)(2l+1)}} \qquad \chi_+^{(2)}(l,m_l) = \sqrt{\frac{(l+1)^2-m_l^2}{(2l+1)(2l+3)}} \tag{A.3}$$

$$\delta_-^{(2)}(l,m_l) = (l+1) \, \chi_-^{(2)}(l,m_l) \qquad \delta_+^{(2)}(l,m_l) = -l \, \chi_+^{(2)}(l,m_l) \, .$$

We know from formula (3.101) in Avery's book [41] and formula (21) in [42] by Aquilanti et al. that $Y_{\underline{l}}^{m_l}(\Omega) = \mathcal{N}_{\underline{l}}^{(d-1)} P_{l_2}^{m_l}(\cos\theta_2) \mathrm{e}^{\mathrm{i}m_l \theta_1} \prod_{d-1}^{k=3} \left( (\sin\theta_k)^{l_{k-1}} C_{l_k-l_{k-1}}^{(l_{k-1}+(k-1)/2)}(\cos\theta_k) \right)$ for the (hyper)spherical harmonics on the $(d-1)$-sphere, wherein the $C_{l-l_{d-2}}^{(l_{d-2}+d/2-1)}$ denote the Gegenbauer (ultraspherical) polynomials. For obtaining an orthonormal system on the $(d-1)$-sphere we have to fix the normalisation to $\mathcal{N}_{\underline{l}}^{(d-1)} = \prod_{d-1}^{k=2} \mathcal{N}_{l_k,l_{k-1}}^{(k)}$ wherein $\mathcal{N}_{l_2,l_1}^{(2)} = \mathcal{N}_{l_2}^{m_l}$ and $\mathcal{N}_{l,l_{d-2}}^{(d-1)} = 2^{l_{d-2}+d/2-2} \Gamma(l_{d-2}+d/2-1) \sqrt{(l-l_{d-2})! \, (2l+d-2) / (\pi (l+l_{d-2}+d-3))}$ for $(d-1) \geq 3$. Our normalisation can be derived using AS [22.2.3] and agrees with Aquilanti's. We find analogs of the contiguous relations (A.1) and (A.2) for higher dimensions using AS[22.7.3] and AS [22.8.2]:

$$\cos\theta_{d-1} Y_{\underline{l}}^{m_l}(\Omega) = \chi_-^{(d-1)}(l,l_{d-2}) \, Y_{(l-1,\tilde{\underline{l}})}^{m_l}(\Omega) + \chi_+^{(d-1)}(l,l_{d-2}) \, Y_{(l+1,\tilde{\underline{l}})}^{m_l}(\Omega) \tag{A.4}$$

$$(1-\cos^2\theta_{d-1}) \frac{\mathrm{d}}{\mathrm{d}\cos\theta_{d-1}} Y_{\underline{l}}^{m_l}(\Omega) = \delta_-^{(d-1)}(l,l_{d-2}) \, Y_{(l-1,\tilde{\underline{l}})}^{m_l}(\Omega) + \delta_+^{(d-1)}(l,l_{d-2}) \, Y_{(l+1,\tilde{\underline{l}})}^{m_l}(\Omega) \tag{A.5}$$

with the raising and lowering coefficients

$$\chi_-^{(d-1)}(l,l_{d-2}) = \sqrt{\frac{(l-l_{d-2})(l+l_{d-2}+d-3)}{(2l+d-4)(2l+d-2)}} \qquad \chi_+^{(d-1)}(l,l_{d-2}) = \sqrt{\frac{(l-l_{d-2}+1)(l+l_{d-2}+d-2)}{(2l+d-2)(2l+d)}} \tag{A.6}$$

$$\delta_-^{(d-1)}(l,l_{d-2}) = (l+d-2) \, \chi_-^{(d-1)}(l,l_{d-2}) \qquad \delta_+^{(d-1)}(l,l_{d-2}) = -l \, \chi_+^{(d-1)}(l,l_{d-2}) \, .$$

For $d = 3$ these coefficients reproduce exactly those of the two-sphere (with $l_1 \equiv m_l$). For all $d \geq 3$, for $l = 0$ automatically $l_{d-2} = 0$ and thus the lowering coefficients vanish in this case: $\chi_-^{(d-1)}(0,0) = \delta_-^{(d-1)}(0,0) = 0$, and therefore hyperspherical harmonics with negative $l$ do not appear. Further, for $d \geq 4$ the lowering coefficients also vanish whenever $l_{d-2}$ has its top value $l_{d-2} = l$, that is, $\chi_-^{(d-1)}(l,l) = \delta_-^{(d-1)}(l,l) = 0$. For $d = 3$ this happens whenever $m_l$ has its extremal values $m_l = \pm l$: then $\chi_-^{(2)}(l, \pm l) = \delta_-^{(2)}(l, \pm l) = 0$. Thus, for all $d \geq 3$ there appear no hyperspherical harmonics $Y_{(l-1,l,l_{d-3}...,l_2)}^{m_l}$ where $|l_{d-2}|$ is bigger than $l_{d-1}$.

A relation connecting raising and lowering coefficients is $\chi_-^{(d-1)}(l+1, l_{d-2}) = \chi_+^{(d-1)}(l, l_{d-2})$. Further, we have a generalization of DLMF [1.17.25]: $\sum_{\underline{l}, m_l} \overline{Y_{\underline{l}}^{m_l}(\Omega)} Y_{\underline{l}}^{m_l}(\Omega') = \delta^{(d-1)}(\Omega - \Omega')/\sqrt{|g_{\mathbb{S}^{d-1}}|}$, wherein $\delta^{(d-1)}(\Omega - \Omega') := \delta(\theta_1 - \theta_1') \cdot \ldots \cdot \delta(\theta_{d-1} - \theta_{d-1}')$.

### 1. Hyperspherical harmonics: transformation under rotations

Any SO$(d)$-rotation on $\mathbb{R}^d$ is determined by $n = d(d-1)/2$ Euler angles $\underline{\alpha} = (\alpha_1, ..., \alpha_n)$ and a choice of $n$ axes $\{\underline{e}_i\}_{i=1,...,n}$ around which to rotate. The action of the rotation operator $(\hat{R}(\underline{\alpha}))$ on hyperspherical harmonics is given by Wigner's $D$-matrix. In Chapter 15 of [43] it is defined for $\mathbb{S}^2$ in $\mathbb{R}^3$ using one-particle states $|l \, m_l\rangle$ with angular momentum numbers $l$ and $m_l$. Its generalization to $\mathbb{S}^{d-1}$ in $\mathbb{R}^d$ is straightforward: $\left(D_{\underline{\tilde{l}}', \underline{\tilde{l}}}^l(\underline{\alpha})\right)_{m_l' m_l} := \langle l \, \underline{\tilde{l}}' m_l' | \hat{R}(\underline{\alpha}) | l \, \underline{\tilde{l}} \, m_l \rangle$, with $|l \, \underline{\tilde{l}} \, m_l\rangle$ denoting one-particle states with angular momentum numbers $l, \underline{\tilde{l}}$ and $m_l$. They fulfill the usual orthonormality and completeness relations: $\langle l \, \underline{\tilde{l}} \, m_l | l' \, \underline{\tilde{l}}' \, m_l' \rangle = \delta_{ll'} \delta_{\underline{\tilde{l}} \, \underline{\tilde{l}}'} \delta_{m_l m_l'}$ and $\mathbb{1} = \sum_{l, \underline{\tilde{l}}, m_l} |l \, \underline{\tilde{l}} \, m_l\rangle \langle l \, \underline{\tilde{l}} \, m_l|$, while $\mathbb{1} = \int_{\mathbb{S}^{d-1}} d^{d-1}\Omega \, |\Omega\rangle \langle \Omega|$ and $\langle \Omega | \Omega' \rangle = \delta^{(d-1)}(\Omega, \Omega')/\sqrt{g_{\mathbb{S}^{d-1}}}$. The coordinate representation of the wave function of $|l \, \underline{\tilde{l}} \, m_l\rangle$ is $\langle \Omega | l \, \underline{\tilde{l}} \, m_l \rangle = Y_{(l, \underline{\tilde{l}})}^{m_l}(\Omega)$. Thus

$$\left(D_{\underline{\tilde{l}}', \underline{\tilde{l}}}^l(\underline{\alpha})\right)_{m_l' m_l} = \int d^2 \Omega' \, \overline{Y_{(l, \underline{\tilde{l}}')}^{m_l'}(\hat{R}(\underline{\alpha})\Omega')} \, Y_{(l, \underline{\tilde{l}})}^{m_l}(\Omega') = \langle Y_{(l, \underline{\tilde{l}}')}^{m_l'} \circ \hat{R}(\underline{\alpha}), \, Y_{(l, \underline{\tilde{l}})}^{m_l} \rangle_{\mathbb{S}^{d-1}}. \qquad (A.7)$$

$\langle \cdot, \cdot \rangle_{\mathbb{S}^{d-1}}$ is the inner product of two functions on $\mathbb{S}^{d-1}$. The total angular momentum $l$ on $\mathbb{S}^{d-1}$ is conserved under SO$(d)$ rotations around the center of the sphere. Thus rotated spherical harmonics are linear combinations of unrotated ones, with elements of Wigner's $D$-matrix as coefficients:

$$Y_{(l, \underline{\tilde{l}}')}^{m_l'}(\hat{R}(\underline{\alpha})\Omega) = \sum_{\underline{\tilde{l}}, m_l} Y_{(l, \underline{\tilde{l}})}^{m_l}(\Omega) \, \langle Y_{(l, \underline{\tilde{l}})}^{m_l}, \, Y_{(l, \underline{\tilde{l}}')}^{m_l'} \circ \hat{R}(\underline{\alpha}) \rangle = \sum_{\underline{\tilde{l}}, m_l} Y_{(l, \underline{\tilde{l}})}^{m_l}(\Omega) \, \overline{\left(D_{\underline{\tilde{l}}', \underline{\tilde{l}}}^l(\underline{\alpha})\right)_{m_l' m_l}}. \qquad (A.8)$$

Since for SO$(d)$-rotations we have $(\hat{R}(\underline{\alpha}))^{-1} = (\hat{R}(\underline{\alpha}))^\top = (\hat{R}(\underline{\alpha}))^\dagger$, we can also write

$$\left(D_{\underline{\tilde{l}}', \underline{\tilde{l}}}^l(\underline{\alpha})\right)_{m_l' m_l} = \int d^{d-1}\Omega \, \overline{Y_{(l, \underline{\tilde{l}}')}^{m_l'}(\Omega)} \, Y_{(l, \underline{\tilde{l}})}^{m_l}((\hat{R}(\underline{\alpha}))^{-1}\Omega) = \langle Y_{(l, \underline{\tilde{l}}')}^{m_l'}, \, Y_{(l, \underline{\tilde{l}})}^{m_l} \circ (\hat{R}(\underline{\alpha}))^{-1} \rangle_{\mathbb{S}^{d-1}}, \qquad (A.9)$$

and (mind the position of the prime!)

$$Y_{(l, \underline{\tilde{l}}')}^{m_l'}((\hat{R}(\underline{\alpha}))^{-1}\Omega) = \sum_{\underline{\tilde{l}}, m_l} Y_{(l, \underline{\tilde{l}})}^{m_l}(\Omega) \, \langle Y_{(l, \underline{\tilde{l}})}^{m_l}, \, Y_{(l, \underline{\tilde{l}}')}^{m_l'} \circ (\hat{R}(\underline{\alpha}))^{-1} \rangle = \sum_{\underline{\tilde{l}}, m_l} Y_{(l, \underline{\tilde{l}})}^{m_l}(\Omega) \, \left(D_{\underline{\tilde{l}}, \underline{\tilde{l}}'}^l(\underline{\alpha})\right)_{m_l m_l'}. \qquad (A.10)$$

Most important for us is the following completeness relation fulfilled by Wigner's $D$-matrix, which can be obtained via their definition and $\mathbb{1} = \sum_{l, \underline{\tilde{l}}, m_l} |l \, \underline{\tilde{l}} \, m_l\rangle \langle l \, \underline{\tilde{l}} \, m_l|$:

$$\sum_{\underline{\tilde{l}}, m_l} (D_{\underline{\tilde{l}}', \underline{\tilde{l}}}^l(\underline{\alpha}))_{m_l' m_l} \, \overline{(D_{\underline{\tilde{l}}'', \underline{\tilde{l}}}^l(\underline{\alpha}))_{m_l'' m_l}} = \delta_{\underline{\tilde{l}}', \underline{\tilde{l}}''} \, \delta_{m_l' m_l''}. \qquad (A.11)$$

### Appendix B: Anti-deSitter spacetime

#### 1. From Klein-Gordon to hypergeometric equation

Following Mezincescu and Townsend in [19], a separation ansatz $\phi(t, \rho, \Omega) = T(t) \, f(\rho) \, Y(\Omega)$ for solutions of the free KG equation $0 = (-\Box_{\text{AdS}} + m^2) \phi$ leads to $\phi(t, \rho, \Omega) = e^{-i\omega t} Y_{\underline{l}}^{m_l}(\Omega) \, f(\rho)$. By $Y_{\underline{l}}^{m_l}$ we denote the hyperspherical harmonics, see Appendix A. The reality of the field is assured by expanding it over the complex modes and their complex conjugates. $\omega$ is the frequency of the solution and a priori allowed to be complex, but it turns out that only real frequencies are needed. The remaining radial differential equation to be obeyed by the function $f(\rho)$ then is

$$0 = \left(\omega^2 \cos^2\rho - l(l+d-2)\tan^{-2}\rho - m^2 R_{\text{AdS}}^2\right) f(\rho) + (d-1)/(\tan\rho)\, \partial_\rho f(\rho) + \cos^2\rho\, \partial_\rho^2 f(\rho)\,.$$

We shall call the solutions $f(\rho)$ of this equation radial solutions, not to be mixed up with Klein-Gordon solutions $\phi(t,\rho,\Omega)$. As shown by Balasubramanian et al. in [6], we can make two ansatzes for $f(\rho)$, which we shall call the sine respectively cosine ansatz:

$$f(\rho) \to S(\rho) = \sin^{\tilde{l}}\rho \, \cos^{\widetilde{m}}\rho \, \widetilde{S}(\sin^2\rho) \qquad\qquad f(\rho) \to C(\rho) = \sin^{\tilde{l}}\rho \, \cos^{\widetilde{m}}\rho \, \widetilde{C}(\cos^2\rho)\,. \qquad (\text{B.1})$$

We denote $\tilde{l}$ and $\widetilde{m}$ like that simply because $\tilde{l}$ will depend on $l$ and $\widetilde{m}$ on $m$. In [6] $\tilde{l}$ is denoted as $2b$ and $\widetilde{m}$ as $2h$. In [17] $\widetilde{m}$ is denoted as $\mu$ and in [19] as $\lambda$. If we impose the following two conditions

$$0 \stackrel{!}{=} \widetilde{m}^2 - \widetilde{m}d - m^2 R_{\text{AdS}}^2 \qquad\qquad \widetilde{m}_\pm = \tfrac{d}{2} \pm \nu \qquad\qquad \nu := \sqrt{(d/2)^2 + m^2 R_{\text{AdS}}^2} \qquad (\text{B.2})$$

$$0 \stackrel{!}{=} \tilde{l}^2 + \tilde{l}(d-2) - l(l+d-2) \qquad\qquad \tilde{l}_+ = l \qquad\qquad \tilde{l}_- = -(l+d-2)\,, \qquad (\text{B.3})$$

then plugging the sine or cosine ansatz into the radial equation yields the hypergeometric differential equation DLMF[15.10.1] (wherein the primes denote derivation with respect to $\sin^2\rho$ respectively $\cos^2\rho$)

$$0 = \sin^2\rho \left(1-\sin^2\rho\right) \widetilde{S}''(\sin^2\rho) + \left[\gamma^{\text{s}} - \sin^2\rho\left(\alpha+\beta+1\right)\right] \widetilde{S}'(\sin^2\rho) - \alpha\beta\, \widetilde{S}(\sin^2\rho)\,, \qquad (\text{B.4})$$

$$0 = \cos^2\rho \left(1-\cos^2\rho\right) \widetilde{C}''(\cos^2\rho) + \left[\gamma^{\text{c}} - \cos^2\rho\left(\alpha+\beta+1\right)\right] \widetilde{C}'(\cos^2\rho) - \alpha\beta\, \widetilde{C}(\cos^2\rho)\,. \qquad (\text{B.5})$$

The parameters' dependences on $\tilde{l}, \widetilde{m}$ and $\omega$ are $\alpha = (\tilde{l}+\widetilde{m}-\omega)/2$, $\beta = (\tilde{l}+\widetilde{m}+\omega)/2$, $\gamma^{\text{s}} = \tilde{l} + d/2$ and $\gamma^{\text{c}} = \widetilde{m} - d/2 + 1$. The conditions imposed on $\widetilde{m}$ and $\tilde{l}$ are not motivated from physical principles here, and must be seen as part of the ansatz. They are justified by the fact that with these conditions holding, we get systems of solutions that are complete on the slice regions respectively on rod and tube regions. Moreover Breitenlohner and Freedman show after equation (11) of [17] that the $\widetilde{m}$-condition leads to a positive energy.

## 2. Sets of KG solutions for AdS

In this section we will give an overview of the Klein-Gordon solutions obtained via the two ansatzes. We use the following notation for the parameters:

$$\alpha_\pm = (l+\widetilde{m}_\pm - \omega)/2 \qquad \beta_\pm = (l+\widetilde{m}_\pm + \omega)/2 \qquad \gamma^{\text{s}} = l+d/2 \qquad \gamma^{\text{c}}_\pm = \pm\nu + 1\,. \qquad (\text{B.6})$$

We set $\tilde{l} = \tilde{l}_+ = l$ and $\widetilde{m} = \widetilde{m}_+$ without loss of generality because the other possible choices yield the same solutions. Although having fixed this choice, in some radial solutions we encounter linear combinations of the parameters $\alpha_+, \beta_+, \gamma^{\text{s}}, \gamma^{\text{c}}_+$ and $\widetilde{m}_+$ adding up to $\alpha_-, \beta_-, \gamma^{\text{s}}, \gamma^{\text{c}}_-$ and $\widetilde{m}_-$. These appearances do not depend on our choice, but are an intrinsic property of the respective solutions. In order to give a complete list of solutions we will rely heavily on DLMF §15.10.(i). Each of the hypergeometric equations (B.4) and (B.5) has two linear independent solutions which we shall denote by $^1\widetilde{S}_{\omega l}$ and $^2\widetilde{S}_{\omega l}$, respectively $^1\widetilde{C}_{\omega l}$ and $^2\widetilde{C}_{\omega l}$. Via the sine respectively cosine ansatz (B.1) this provides us the solutions of the radial part of the Klein-Gordon equation, $^1S_{\omega l}$ and $^2S_{\omega l}$, respectively $^1C_{\omega l}$ and $^2C_{\omega l}$. The form of the solutions of (B.4) depends on whether $\gamma^{\text{s}}$ is integer or not, and of (B.5) on whether $\gamma^{\text{c}}$ is integer or not. $\gamma^{\text{s}}$ is integer if the spatial dimension $d$ is even, and noninteger if not. $\gamma^{\text{c}}$ is integer if $\nu$ is integer, else noninteger. Where necessary we equip our solutions with labels distinguishing between these cases. The table below shows of which radial solutions we dispose in which case. The $S$-solutions and the $C$-solutions generically are not linear independent, their relations are examined in Section B 3. Since the parameters $\alpha, \beta, \gamma^{\text{s}}$ and $\gamma^{\text{c}}$ depend on $\omega$ and $l$, each radial solution is labeled by these parameters.

|  | d odd | | d even | |
|---|---|---|---|---|
| $\nu$ noninteger | $^1S_{\omega l}$    $^1C_{\omega l}$ | | $^1S_{\omega l}$    $^1C_{\omega l}$ | |
|  | $^2S^{\text{odd}}_{\omega l}$   $^2C^{\text{non}}_{\omega l}$ | | $^2S^{\text{eve}}_{\omega l}$   $^2C^{\text{non}}_{\omega l}$ | |
| $\nu$ integer | $^1S_{\omega l}$    $^1C_{\omega l}$ | | $^1S_{\omega l}$    $^1C_{\omega l}$ | |
|  | $^2S^{\text{odd}}_{\omega l}$   $^2C^{\text{int}}_{\omega l}$ | | $^2S^{\text{eve}}_{\omega l}$   $^2C^{\text{int}}_{\omega l}$ | |

Table B.7. Cases of radial solutions

Below we give explicit expressions for all radial solutions. The following parameter relations have been used:

$$\alpha_\pm + \beta_\pm - \gamma^{\text{S}} + 1 = \gamma^{\text{C}}_\pm \qquad 2 - \gamma^{\text{C}}_\pm = \gamma^{\text{C}}_\mp \qquad \alpha_\pm - \gamma^{\text{C}}_\pm + 1 = \alpha_\mp \qquad \beta_\pm - \gamma^{\text{C}}_\pm + 1 = \beta_\mp. \tag{B.8}$$

For certain frequencies the solutions show special behaviour. The magic frequencies [6] are defined by

$$\omega^\pm_{nl} := 2n + l + \tilde{m}_\pm = 2n + \gamma^{\text{S}} \pm \nu \qquad \begin{array}{l} \forall\, l \in \mathbb{N}_0 \\ \forall\, n \in \mathbb{N}_0 \end{array} \tag{B.9}$$

and what we shall call submagic frequencies by

$$\sigma^{\text{S}}_{nl} := 2n + l + \tilde{m}_\pm = 2n + \gamma^{\text{S}} + \gamma^{\text{C}}_+ - 1 \qquad \begin{array}{l} \forall\, l \in \mathbb{N}_0 \\ \forall\, n \in \{-(\gamma^{\text{S}}-1), \ldots, -1\} \end{array} \tag{B.10}$$

$$\sigma^{\text{C}}_{nl} := 2n + l + \tilde{m}_\pm = 2n + \gamma^{\text{S}} + \gamma^{\text{C}}_+ - 1 \qquad \begin{array}{l} \forall\, l \in \mathbb{N}_0 \\ \forall\, n \in \{-(\gamma^{\text{C}}-1), \ldots, -1\} \end{array}. \tag{B.11}$$

The magic frequencies $\omega^+_{nl}$ are always positive, while the sign of $\omega^-_{nl}$ depends on $n, l, \nu$ and $d$. However, we will see later that only for $\nu < 1$ we can make use of the $\omega^-_{nl}$, and for this case they are positive, too. The sign of $\sigma^{\text{S/C}}_{nl}$ depends on $n, l, \nu$ and $d$. The following solutions come from DLMF [15.10.2] for odd $d$ and noninteger $\nu$ (which is the only case we shall study in detail in this work):

$$^1S_{\omega l}(\rho) = \sin^l \rho \, \cos^{\tilde{m}_+}\rho \, F(\alpha_+, \beta_+; \gamma^{\text{S}}; \sin^2\rho) \tag{B.12}$$

$$^2S^{\text{odd}}_{\omega l}(\rho) = -(\sin\rho)^{2-l-d} \cos^{\tilde{m}_+}\rho \, F(\alpha_+ - \gamma^{\text{S}}+1, \beta_+ - \gamma^{\text{S}}+1; 2-\gamma^{\text{S}}; \sin^2\rho) \tag{B.13}$$

$$^1C_{\omega l}(\rho) = \sin^l \rho \, \cos^{\tilde{m}_+}\rho \, F(\alpha_+, \beta_+; \gamma^{\text{C}}_+; \cos^2\rho) \tag{B.14}$$

$$^2C^{\text{non}}_{\omega l}(\rho) = \sin^l \rho \, \cos^{\tilde{m}_-}\rho \, F(\alpha_-, \beta_-; \gamma^{\text{C}}_-; \cos^2\rho). \tag{B.15}$$

Moreover, DLMF 15.10.(a+b) tell us that the solutions $^1S_{\omega l}(\rho)$ and $^1C_{\omega l}(\rho)$ also hold for the other three cases where $d$ is even and/or $\nu$ is integer. The functions $^2S^{\text{eve}}$ and $^2C^{\text{int}}$ are different however, and need to be defined via case-by-case distinction. We indicate the corresponding case by bestowing yet another label (bottom left) to the solutions. Case 1 corresponds to DLMF [15.10.8], case 2 to 15.10(i)(a) with interpretation in the sense DLMF [15.2.5] to be DLMF [15.2.4], and case 3 to DLMF [15.10.9]. The case of DLMF [15.10.10], where both $\alpha_+$ and $\beta_+$ are nonpositive, cannot occur for our situation since independently of positive or negative $\omega$, either $\alpha_+$ or $\beta_+$ is positive. Below, ior denotes "inclusive or" and xor denotes "exclusive or".

$$^2S^{\text{eve}}_{\omega l} := \begin{cases} ^2_1S^{\text{eve}}_{\omega l} & \text{neither } \alpha_+ \text{ nor } \beta_+ \notin \mathbb{Z}^{\leq}(\gamma^{\text{S}}-1) \\ ^2_2S^{\text{eve}}_{\omega l} = {}^2S^{\text{odd}}_{\omega l} & \text{ior } \begin{array}{l}\alpha_+ = -n \in \{1, \ldots, \gamma^{\text{C}}-1\} \Leftrightarrow \omega = +\sigma^{\text{S}}_{nl} \\ \beta_+ = -n' \in \{1, \ldots, \gamma^{\text{C}}-1\} \Leftrightarrow \omega = -\sigma^{\text{S}}_{n'l}\end{array} \\ ^2_3S^{\text{eve}}_{\omega l} & \text{xor } \begin{array}{l}(\alpha_+ = -n \in \mathbb{Z}^{\leq}(0) \Rightarrow \beta_+ \notin \mathbb{Z}^{\leq}(\gamma^{\text{S}}-1)) \Leftrightarrow \omega = +\omega^+_{nl} \\ (\beta_+ = -n \in \mathbb{Z}^{\leq}(0) \Rightarrow \alpha_+ \notin \mathbb{Z}^{\leq}(\gamma^{\text{S}}-1)) \Leftrightarrow \omega = -\omega^+_{nl}\end{array} \end{cases} \tag{B.16}$$

$$^2C^{\text{int}}_{\omega l} := \begin{cases} ^2_1C^{\text{int}}_{\omega l} & \text{neither } \alpha_+ \text{ nor } \beta_+ \notin \mathbb{Z}^{\leq}(\gamma^{\text{C}}_+-1) \\ ^2_2C^{\text{int}}_{\omega l} = {}^2C^{\text{non}}_{\omega l} & \text{ior } \begin{array}{l}\alpha_+ = -n \in \{1, \ldots, \gamma^{\text{C}}-1\} \Leftrightarrow \omega = +\sigma^{\text{C}}_{nl} \\ \beta_+ = -n' \in \{1, \ldots, \gamma^{\text{C}}-1\} \Leftrightarrow \omega = -\sigma^{\text{S}}_{n'l}\end{array} \\ ^2_3C^{\text{int}}_{\omega l} & \text{xor } \begin{array}{l}(\alpha_+ = -n \in \mathbb{Z}^{\leq}(0) \Rightarrow \beta_+ \notin \mathbb{Z}^{\leq}(\gamma^{\text{C}}_+-1)) \Leftrightarrow \omega = +\omega^+_{nl} \\ (\beta_+ = -n \in \mathbb{Z}^{\leq}(0) \Rightarrow \alpha_+ \notin \mathbb{Z}^{\leq}(\gamma^{\text{C}}_+-1)) \Leftrightarrow \omega = -\omega^+_{nl}\end{array} \end{cases} \tag{B.17}$$

$$^2_1S^{\text{eve}}_{\omega l}(\rho) = \sin^l\rho \, \cos^{\tilde{m}_+}\rho \, \Big\{ \ln(\sin^2\rho) \, F(\alpha_+, \beta_+; \gamma^{\text{S}}; \sin^2\rho) \tag{B.18}$$

$$- \sum_{k=1}^{\gamma^{\text{S}}-1} \frac{(k-1)! \, (1-\gamma^{\text{S}})_k}{(1-\alpha_+)_k \, (1-\beta_+)_k} (\sin^2\rho)^{-k} + \sum_{k=0}^\infty \frac{(\alpha_+)_k \, (\beta_+)_k}{(\gamma^{\text{S}})_k \, k!} (\sin^2\rho)^k \, \widetilde{\psi}_k(\alpha_++k, \beta_++k, \gamma^{\text{S}}+k) \Big\}$$

$$^2_3S^{\text{eve}}_{\omega l}(\rho) = \sin^l\rho \, \cos^{\tilde{m}_+}\rho \, \Big\{ \ln(\sin^2\rho) \, F(\alpha_+, \beta_+; \gamma^{\text{S}}; \sin^2\rho) - \sum_{k=1}^{\gamma^{\text{S}}-1} \frac{(k-1)! \, (1-\gamma^{\text{S}})_k}{(1-\alpha_+)_k \, (1-\beta_+)_k} (\sin^2\rho)^{-k} \tag{B.19}$$

$$+ \sum_{k=0}^n \frac{(\alpha_+)_k \, (\beta_+)_k}{(\gamma^{\text{S}})_k \, k!} (\sin^2\rho)^k \, \widetilde{\psi}_k(1+n-k, n+l+\tilde{m}_++k, \gamma^{\text{S}}+k)$$

$$+ (-1)^n \, n! \sum_{k=n+1}^\infty \frac{(k-n-1)! \, (n+l+\widetilde{m}_+)_k}{(\gamma^{\text{S}})_k \, k!} (\sin^2\rho)^k \Big\}$$

$$\,_1^2C_{\omega l}^{\text{int}}(\rho) \;=\; \sin^l\rho\,\cos^{\tilde{m}_+}\!\rho\,\Big\{\ln(\cos^2\rho)\,F(\alpha_+,\beta_+;\gamma_+^{\text{C}};\cos^2\rho) \tag{B.20}$$

$$-\sum_{k=1}^{\gamma_+^{\text{C}}-1}\frac{(k-1)!\,(1-\gamma_+^{\text{C}})_k}{(1-\alpha_+)_k\,(1-\beta_+)_k}\,(\cos^2\rho)^{-k}+\sum_{k=0}^{\infty}\frac{(\alpha_+)_k\,(\beta_+)_k}{(\gamma_+^{\text{C}})_k\,k!}\,(\cos^2\rho)^k\,\widetilde{\psi}_k(\alpha_++k,\beta_++k,\gamma_+^{\text{C}}+k)\Big\}$$

$$\,_3^2C_{\omega l}^{\text{int}}(\rho) \;=\; \sin^l\rho\,\cos^{\tilde{m}_+}\!\rho\,\Big\{\ln(\cos^2\rho)\,F(\alpha_+,\beta_+;\gamma_+^{\text{C}};\cos^2\rho) - \sum_{k=1}^{\gamma_+^{\text{C}}-1}\frac{(k-1)!\,(1-\gamma_+^{\text{C}})_k}{(1-\alpha_+)_k\,(1-\beta_+)_k}\,(\cos^2\rho)^{-k} \tag{B.21}$$

$$+\sum_{k=0}^{n}\frac{(\alpha_+)_k\,(\beta_+)_k}{(\gamma_+^{\text{C}})_k\,k!}\,(\cos^2\rho)^k\,\widetilde{\psi}_k(1+n-k, n+l+\tilde{m}_++k, \gamma_+^{\text{C}}+k)$$

$$+(-1)^n\,n!\sum_{k=n+1}^{\infty}\frac{(k-n-1)!\,(n+l+\widetilde{m}_+)_k}{(\gamma_+^{\text{C}})_k\,k!}\,(\cos^2\rho)^k\Big\}$$

The exceptional radial solutions $\,_2^2 S^{\text{eve}}$, $\,_3^2 S^{\text{eve}}$ and $\,_2^2 C^{\text{int}}$, $\,_3^2 C^{\text{int}}$ can be presented here thanks to the extended DLMF section on the hypergeometric DEQ, we have not found them in the existing literature on AdS. The other solutions agree (up to small typos, hopefully we have none here) with those in e.g. [19] and [6].

Finally, a discrete set of special radial solutions arises for the magic frequencies $\omega_{nl}^\pm$ (see appendix B 3):

$$J_{nl}^{(+)}(\rho) \,:=\, \tfrac{n!}{(\gamma^{\text{S}})_n}\,\sin^l\rho\,\cos^{\tilde{m}_+}\!\rho\,P_n^{(\gamma^{\text{S}}-1,+\nu)}(\cos 2\rho) \qquad \text{for } \omega \,=\, \pm\omega_{nl}^+$$
$$J_{nl}^{(-)}(\rho) \,:=\, \tfrac{n!}{(\gamma^{\text{S}})_n}\,\sin^l\rho\,\cos^{\tilde{m}_-}\!\rho\,P_n^{(\gamma^{\text{S}}-1,-\nu)}(\cos 2\rho) \qquad \text{for } \omega \,=\, \pm\omega_{nl}^- \,. \tag{B.22}$$

$P_n^{(\cdot,\cdot)}$ are Jacobi polynomials. These Jacobi solutions are always of this form, independently of whether $d$ is even or odd and whether $\nu$ is integer or not. Since we use only real frequencies, and restrict to masses for which $m^2 \geq m_{\text{BF}}^2$, all parameters entering the hypergeometric functions (respectively Jacobi polynomials) are real, and therefore all radial solutions given here are real.

### 3. Linear (in)dependence of radial solutions

The solutions for the case of odd $d$ and noninteger $\nu$ are related by the following matrix equation, which can be obtained using AS[15.3.3+6]. The label "on" stands for odd-noninteger, and "no" for noninteger-odd.

$$\begin{pmatrix}\,^1 S_{\omega l}\\ \,^2 S_{\omega l}^{\text{odd}}\end{pmatrix} \;=\; M_{\omega l}^{\text{on}}\begin{pmatrix}\,^1 C_{\omega l}\\ \,^2 C_{\omega l}^{\text{non}}\end{pmatrix} \qquad\qquad \begin{pmatrix}\,^1 C_{\omega l}\\ \,^2 C_{\omega l}^{\text{non}}\end{pmatrix} \;=\; M_{\omega l}^{\text{no}}\begin{pmatrix}\,^1 S_{\omega l}\\ \,^2 S_{\omega l}^{\text{odd}}\end{pmatrix} \tag{B.23}$$

$$(M_{\omega l}^{\text{on}})_{11} = +\frac{\Gamma(\gamma^{\text{S}})\,\Gamma(1-\gamma_+^{\text{C}})}{\Gamma(\gamma^{\text{S}}-\alpha_+)\,\Gamma(\gamma^{\text{S}}-\beta_+)} \qquad\qquad (M_{\omega l}^{\text{no}})_{11} = +\frac{\Gamma(\gamma_+^{\text{C}})\,\Gamma(1-\gamma^{\text{S}})}{\Gamma(\gamma_+^{\text{C}}-\alpha_+)\,\Gamma(\gamma_+^{\text{C}}-\beta_+)}$$

$$(M_{\omega l}^{\text{on}})_{12} = +\frac{\Gamma(\gamma^{\text{S}})\,\Gamma(\gamma_+^{\text{C}}-1)}{\Gamma(\alpha_+)\,\Gamma(\beta_+)} \qquad\qquad (M_{\omega l}^{\text{no}})_{12} = -\frac{\Gamma(\gamma_+^{\text{C}})\,\Gamma(\gamma^{\text{S}}-1)}{\Gamma(\alpha_+)\,\Gamma(\beta_+)}$$

$$(M_{\omega l}^{\text{on}})_{21} = -\frac{\Gamma(2-\gamma^{\text{S}})\,\Gamma(1-\gamma_+^{\text{C}})}{\Gamma(1-\alpha_+)\,\Gamma(1-\beta_+)} \qquad\qquad (M_{\omega l}^{\text{no}})_{21} = +\frac{\Gamma(2-\gamma_+^{\text{C}})\,\Gamma(1-\gamma^{\text{S}})}{\Gamma(1-\alpha_+)\,\Gamma(1-\beta_+)}$$

$$(M_{\omega l}^{\text{on}})_{22} = -\frac{\Gamma(2-\gamma^{\text{S}})\,\Gamma(\gamma_+^{\text{C}}-1)}{\Gamma(\gamma_+^{\text{C}}-\alpha_+)\,\Gamma(\gamma_+^{\text{C}}-\beta_+)} \qquad\qquad (M_{\omega l}^{\text{no}})_{22} = -\frac{\Gamma(2-\gamma_+^{\text{C}})\,\Gamma(\gamma^{\text{S}}-1)}{\Gamma(\gamma^{\text{S}}-\alpha_+)\,\Gamma(\gamma^{\text{S}}-\beta_+)}$$

The two matrices $M_{\omega l}^{\text{no}}$ and $M_{\omega l}^{\text{on}}$ are the inverse of each other, and their determinants are

$$\det M_{\omega l}^{\text{on}} \;=\; (2l+d-2)/(2\nu) \qquad\qquad \det M_{\omega l}^{\text{no}} \;=\; (2\nu)/(2l+d-2)\,. \tag{B.24}$$

To show this, we write $\lambda := d/2+l-1$, giving

$$M_{\omega l}^{\text{on}}\,M_{\omega l}^{\text{no}} \;=\; \begin{pmatrix}1 & 0\\ 0 & 1\end{pmatrix} \;=\; M_{\omega l}^{\text{no}}\,M_{\omega l}^{\text{on}} \qquad\qquad \det M_{\omega l}^{\text{on}} \;=\; \frac{\lambda}{\nu} \;=\; (\det M_{\omega l}^{\text{no}})^{-1}\,. \tag{B.25}$$

Plugging the matrix elements into the matrix multiplication, the zeros in the unity matrix are obtained quickly using only the recursion relation AS[6.1.15]=DLMF[5.5.3] $z\,\Gamma(z) \;=\; \Gamma(1+z)$. Further, generously applying the reflection property of Euler's Gamma function AS[6.1.17]=DLMF[5.5.3]: $\Gamma(z)\,\Gamma(1-z) \;=\;$

$\pi/\sin(\pi z)$, we see that obtaining the ones in the unity matrix and the values of the determinant corresponds to showing $\sin(\pi\lambda)\sin(\pi\nu) = \cos([\lambda-\nu+\omega]\pi/2)\cos([\lambda-\nu-\omega]\pi/2) - \cos([\lambda+\nu+\omega]\pi/2)\cos([\lambda-\nu-\omega]\pi/2)$. This is done using first AS[4.3.42] (eliminating the $\omega$-terms) and then AS[4.3.41].

Next we have a look at exceptional cases of these dependencies. As stated in [6], $^1S$ is proportional to $^1C$ whenever $M_{12}^{\text{on}}$ and $M_{12}^{\text{no}}$ vanish, i.e., whenever $\alpha_+$ or $\beta_+$ are nonpositive integers. This happens precisely if the frequency is one of the "magic" frequencies $\pm\omega_{nl}^+$ from (B.9). $^1S$ is proportional to $^2C^{\text{non}}$ whenever $M_{11}^{\text{on}}$ and $M_{22}^{\text{no}}$ vanish, i.e., whenever $(\gamma^S-\alpha_+)$ or $(\gamma^S-\beta_+)$ are nonpositive integers. This happens iff the frequency is one of the "magic" frequencies $\pm\omega_{nl}^-$ from (B.9). In these cases the functions are related through:

$$\begin{aligned}
^1S_{\omega l} &= (-1)^n \frac{(\gamma_+^{\text{C}})_n}{(\gamma^{\text{S}})_n} {}^1C_{\omega l} = \frac{n!}{(\gamma^{\text{S}})_n}\sin^l\rho\,\cos^{\tilde{m}_+}\rho\, P_n^{(\gamma^{\text{S}}-1,+\nu)}(\cos 2\rho) &\Leftrightarrow\quad \omega &= \pm\omega_{nl}^+ \\
^1S_{\omega l} &= (-1)^n \frac{(\gamma_-^{\text{C}})_n}{(\gamma^{\text{S}})_n} {}^1C_{\omega l} = \frac{n!}{(\gamma^{\text{S}})_n}\sin^l\rho\,\cos^{\tilde{m}_-}\rho\, P_n^{(\gamma^{\text{S}}-1,-\nu)}(\cos 2\rho) &\Leftrightarrow\quad \omega &= \pm\omega_{nl}^- \quad \nu<1
\end{aligned}$$
(B.26)

The Jacobi polynomials $P_n^{(\cdot,\cdot)}$ arise from the hypergeometric function if the frequency is magic via DLMF [15.9.1]: $P_n^{(a,b)}(1-2x)\,n!/(a+1)_n = F(-n, a+b+n+1; a+1; x)$ for all $a,b>-1$. $^2S^{\text{odd}}$ is proportional to $^1C$ whenever $M_{11}^{\text{no}}$ and $M_{22}^{\text{on}}$ vanish, i.e., whenever $(\gamma_+^{\text{C}}-\alpha_+)$ or $(\gamma_+^{\text{C}}-\beta_+)$ are nonpositive integers. This happens exactly if the frequency is one of the frequencies $\pm\omega = -(l+d-2) + \tilde{m}_+ + 2n$ with $n \in \mathbb{N}_0$.

### 4. Wronskians

In this section we calculate the Wronskians $^1S_{\omega l}(\rho)\overleftrightarrow{\partial}_\rho {}^2S_{\omega l}^{\text{odd}}(\rho)$ and $^1C_{\omega l}(\rho)\overleftrightarrow{\partial}_\rho {}^2C_{\omega l}^{\text{non}}(\rho)$, which are needed for the symplectic structure. Starting with the former, for greater clarity we allow us to simplify our notation a bit for the moment: $\alpha_+ \to \alpha$, $\beta_+ \to \beta$, $\gamma^{\text{S}} \to \gamma$ and $\sin^2\rho \to x$. The calculation proceeds as follows:

$$\begin{aligned}
\tan^{d-1}\rho\left({}^1S_{\omega l}(\rho)\overleftrightarrow{\partial}_\rho {}^2S_{\omega l}^{\text{odd}}(\rho)\right) &= -2(1-x)^{\alpha+\beta-\gamma+1}\Big(F(\alpha,\beta;\gamma;x)\,F(\alpha-\gamma+1,\beta-\gamma+1;2-\gamma;x)(1-\gamma) \\
&\qquad + F(\alpha,\beta;\gamma;x)\,F(\alpha-\gamma+2,\beta-\gamma+2;3-\gamma;x)\,x\,\frac{(\alpha-\gamma+1)(\beta-\gamma+1)}{(2-\gamma)} \\
&\qquad - F(\alpha+1,\beta+1;\gamma+1;x)\,F(\alpha-\gamma+1,\beta-\gamma+1;2-\gamma;x)\,x\,\frac{\alpha\beta}{\gamma}\Big) \\
&= 2(1-x)^{\alpha+\beta-\gamma+1}\Big(aF(\alpha+1,\beta;\gamma;x)F(\alpha-\gamma+1,\beta-\gamma+1;2-\gamma;x) - F(\alpha,\beta;\gamma;x)F(\alpha-\gamma+2,\beta-\gamma+1;2-\gamma;x)(1+\alpha-\gamma)\Big) \\
&= F(\alpha+1,\beta;\gamma;x)\,F(1-\alpha,1-\beta;2-\gamma;x)(1-x)\,a - 2F(\alpha,\beta;\gamma;x)\,F(-\alpha,1-\beta;2-\gamma;x)(1+\alpha-\gamma) \\
&= -2F(\alpha,\beta;\gamma;x)\,F(-\alpha,-\beta;-\gamma;x)(1-x) + 2F(\alpha+1,1+\beta;2+\gamma;x)\,F(1-\alpha,1-\beta;2-\gamma;x)\,x^2\,\frac{\alpha\beta(\alpha-\gamma)(\beta-\gamma)}{\gamma^2(1+\gamma)} \\
&= 2(\gamma-1)\,.
\end{aligned}$$
(B.27)

Since here $\gamma = \gamma^S = l+d/2$, we obtain the result

$$^1S_{\omega l}(\rho)\overleftrightarrow{\partial}_\rho {}^2S_{\omega l}^{\text{odd}}(\rho) = 2(\gamma^S-1)\tan^{1-d}\rho = (2l+d-2)\tan^{1-d}\rho\,.$$
(B.28)

Thus $\tan^{d-1}\rho\,({}^1S_{\omega l}(\rho)\overleftrightarrow{\partial}_\rho {}^2S_{\omega l}^{\text{odd}}(\rho))$ is $\rho$-independent, which causes the $\rho$-independence of the symplectic structure (VI.4). In the calculation we make use of the following three hypergeometric contiguous relations:

$$F(\alpha+1,\beta;\gamma;x)(1-x)\gamma^2(1+\gamma) = F(\alpha,\beta;\gamma;x)\gamma(1+\gamma)[\gamma+x(\beta-\gamma)] - F(\alpha+1,\beta+1;\gamma+2;x)\,x^2\,\beta(\alpha-\gamma)(\beta-\gamma) \quad \text{(B.29)}$$

$$F(\alpha,\beta+1;\gamma+2;x)\gamma(1+\gamma-\alpha) = F(\alpha,\beta;\gamma;x)\gamma(1+\gamma) - F(\alpha+1,\beta+1;\gamma+2;x)\alpha[\gamma+x(\beta-\gamma)] \quad \text{(B.30)}$$

$$F(\alpha+1,\beta+1;\gamma+1;x)\,x\,\beta = -F(\alpha,\beta;\gamma;x)\gamma + F(\alpha+1,\beta;\gamma;x)\gamma\,. \quad \text{(B.31)}$$

The first equality in the Wronskian calculation (B.28) comes directly from plugging in the radial functions and cleaning up a bit. The second then is achieved by using the third contiguous relation for both $F(\alpha-\gamma+2,\beta-\gamma+2;3-\gamma;x)$ and $F(\alpha+1,\beta+1;\gamma+1;x)$. The third follows by using AS[15.3.3] = DLMF[15.8.1] $F(\alpha,\beta;\gamma;x) = (1-x)^{\gamma-\alpha-\beta}F(\gamma-\alpha,\gamma-\beta;\gamma;x)$ for both $F(\alpha-\gamma+2,\beta-\gamma+1;2-\gamma;x)$ and $F(\alpha-\gamma+1,\beta-\gamma+1;2-\gamma;x)$. The fourth results from using the second contiguous relation for $F(-\alpha,1-\beta;2-\gamma;x)$ and the first contiguous relation for $F(\alpha+1,\beta;\gamma;x)$. The last then can finally be read off from DLMF[15.16.4]:

$$1 = F(\alpha,\beta;\gamma;x)\,F(-\alpha,-\beta;-\gamma;x) + F(1+\alpha,1+\beta;2+\gamma;x)\,F(1-\alpha,1-\beta;2-\gamma;x)\,x^2\left(\alpha\beta(\alpha-\gamma)(\beta-\gamma)\right)/\left(c^2(1-c^2)\right)\,.$$

The three hypergeometric contiguous relations above were found using the Mathematica code file by Ibrahim and Rakha in [44], see also [45] by the same authors and Rathie. While greatly benefiting from their work, we remark that their code works well for *many sets* of parameter shifts, but *fails for others*, e.g., some of the shifts needed in Appendices B 6 and B 7. The results we obtain by hand for the latter shifts can be verified by Mathematica's FullSimplify command but disagree with the results given by the code.

The Wronskian for the radial $C$-solutions $\mathcal{W}^C_{\omega l}(\rho) := {}^1C_{\omega l}(\rho) \overleftrightarrow{\partial}_\rho {}^2C^{\text{non}}_{\omega l}(\rho)$ can be calculated in exactly the same way. To see this, we use the simplified notation $\alpha_+ \to \alpha$, $\beta_+ \to \beta$, $\gamma^C_+ \to \gamma$ and $\cos^2\rho \to y$. Then, using parameter relations (B.8), we can write the radial $C$-solutions in a similar form as the $S$-solutions: ${}^1C_{\omega l}(\rho) = \sin^l\rho \cos^{\tilde{m}_+}\rho\, F(\alpha, \beta; \gamma; y)$ and ${}^2C^{\text{non}}_{\omega l}(\rho) = \sin^l\rho \cos^{\tilde{m}_-}\rho\, F(\alpha-\gamma+1, \beta-\gamma+1; 2-\gamma; y)$. Plugging this form of the $C$-solutions into the Wronskian, we obtain the first line of the $S$-calculation, and thus can directly jump to its last line: $\tan^{d-1}\rho \left({}^1C_{\omega l}(\rho) \overleftrightarrow{\partial}_\rho {}^2C^{\text{non}}_{\omega l}(\rho)\right) = 2(\gamma-1)$. Since here $\gamma = \gamma^C_+ = 1+\nu$, we obtain

$$ {}^1C_{\omega l}(\rho) \overleftrightarrow{\partial}_\rho {}^2C^{\text{non}}_{\omega l}(\rho) = 2(\gamma^C_+ - 1)\tan^{1-d}\rho = 2\nu \tan^{1-d}\rho\,. \tag{B.32} $$

The two Wronskians ${}^1S_{\omega l}(\rho)\overleftrightarrow{\partial}_\rho {}^2S^{\text{odd}}_{\omega l}(\rho)$ and ${}^1C_{\omega l}(\rho)\overleftrightarrow{\partial}_\rho {}^2C^{\text{non}}_{\omega l}(\rho)$ are related as follows. In the definition of the $C$-Wronskian we can write the $C$-solutions as linear combinations of the $S$-solutions using the matrix $M^{\text{no}}_{\omega l}$ of (B.23). It is then straightforward to check that $\left({}^1C_{\omega l}(\rho)\overleftrightarrow{\partial}_\rho {}^2C^{\text{non}}_{\omega l}(\rho)\right) = \left({}^1S_{\omega l}(\rho)\overleftrightarrow{\partial}_\rho {}^2S^{\text{odd}}_{\omega l}(\rho)\right) \det M^{\text{no}}_{\omega l}$, which via inserting (B.28) and (B.32) implies that $\det M^{\text{no}}_{\omega l} = \left({}^1C_{\omega l}(\rho)\overleftrightarrow{\partial}_\rho {}^2C^{\text{non}}_{\omega l}(\rho)\right) / \left({}^1S_{\omega l}(\rho)\overleftrightarrow{\partial}_\rho {}^2S^{\text{odd}}_{\omega l}(\rho)\right) = \frac{2\nu}{2l+d-2}$. This agrees exactly with (B.24).

### 5. Flat limit of KG solutions

First we compute the flat limit of the radial functions $S^a_{\omega l}$ and $S^b_{\omega l}$. (The temporal part $e^{-i\omega t} = e^{-i\tilde\omega \tau}$ and the angular part $Y^{m_l}_l(\Omega)$ of the solutions remain unchanged in the flat limit.) Replacing $\rho$ by $r/R_{\text{AdS}}$, and then taking the flat limit $R_{\text{AdS}} \to \infty$, we can replace $\sin\frac{r}{R_{\text{AdS}}} \to \frac{r}{R_{\text{AdS}}}$ and $\cos^{\tilde{m}_+}\frac{r}{R_{\text{AdS}}} \to 1$ while $\tilde{m}_\pm, \pm\nu \to \pm m R_{\text{AdS}}$. Using that each $k, l$ is small compared to $R_{\text{AdS}}$ in the flat limit, we further get

$$ \frac{\Gamma(\alpha^a+k)\,\Gamma(\beta^a+k)}{\Gamma(\alpha^a)\,\Gamma(\beta^a)} \approx \frac{\Gamma(\alpha^b+k)\,\Gamma(\beta^b+k)}{\Gamma(\alpha^b)\,\Gamma(\beta^b)} \approx \frac{(-1)^k}{2^{2k}} R^{2k}_{\text{AdS}} \left(\tilde\omega^2 - m^2\right)^k = \frac{(-1)^k}{2^{2k}} R^{2k}_{\text{AdS}} \begin{cases} (\tilde{p}^{\mathbb{R}}_{\tilde\omega})^{2k} & \tilde\omega^2 \ge m^2 \\ (i\tilde{p}^{\mathbb{R}}_{\tilde\omega})^{2k} & \tilde\omega^2 < m^2 \end{cases}. $$

Now we can plug all this into the definitions of $S^a_{\omega l}$ and $S^b_{\omega l}$ and compare with the power series DLMF [10.53.1,2] of the spherical Bessel and Neumann functions

$$ j_l(z) = z^l \sum_{k=0}^\infty \frac{(-1)^k\, 2^{-k}\, z^{2k}}{k!\,(2l+2k+1)!!} $$

$$ n_l(z) = \frac{-1}{z^{l+1}} \sum_{k=0}^l \frac{(2l-2k-1)!!}{k!\,2^k} z^{2k} - \frac{(-1)^l}{z^{l+1}} \sum_{k=l+1}^\infty \frac{(-1)^k 2^{-k}/k!}{(2k-2l-1)!!} z^{2k} = -\frac{(2l-1)!!}{z^{l+1}} \sum_{k=0}^\infty \frac{(-1)^k z^{2k}}{2^k\, k!} \prod_{j=1}^k \frac{1}{(2j-2l-1)}. $$

With definitions (II.5), for $d=3$ we thus find the following flat limits of the radial functions

$$ \frac{(p^{\mathbb{R}}_\omega)^l}{(2l+d-2)!!} S^a_{\omega l}(\rho) \xrightarrow{\text{flat lim.}} \check{j}_{\tilde\omega l}(r) \qquad\qquad \frac{(2l+d-4)!!}{(p^{\mathbb{R}}_\omega)^{l+1}} S^b_{\omega l}(\rho) \xrightarrow{\text{flat lim.}} \check{n}_{\tilde\omega l}(r)\,. \tag{B.33} $$

Since $J^{(\pm)}_{nl}(\rho)$ is a special case of $S^a_{\omega l}$ we find the following flat limit for $d=3$

$$ \frac{(p^{\mathbb{R}}_\omega)^l}{(2l+d-2)!!} J^{(+)}_{nl}(\rho) \xrightarrow{\text{flat lim.}} j_l(\tilde{p}_{\tilde\omega} r)\,. \tag{B.34} $$

In order to compare now field expansions with the corresponding ones for Minkowski spacetime, we switch to the AdS coordinates $(\tau, r, \Omega)$ and parameter $\tilde\omega$, which correspond to the Minkowski coordinates $(t, r, \Omega)$ and energy $E$. The momentum representation for the new parameter is related to the original one by $\tilde\phi_{\tilde\omega l m_l} = R_{\text{AdS}}\phi_{\omega l m_l}$. First we consider our field expansion (IV.11) on the slice region. We can write the summation over $n$ as a sum over $\omega^+_{nl}$ (with $l$ fixed), which in the flat limit becomes an integral over $\tilde\omega$ (abbreviating $R = R_{\text{AdS}}$):

$$ \sum_{n=0}^\infty f(\omega^+_{nl}) = \sum_{\omega^+_{nl} = \gamma^S + \nu}^{\Delta\omega^+_{nl}=2} f(\omega^+_{nl}) = \sum_{\omega^+_{nl}/R = (\gamma^S+\nu)/R}^{\Delta\omega^+_{nl}/R=2/R} f(R\omega^+_{nl}/R) \approx \int_m^\infty d\tilde\omega\, f(R\tilde\omega) = \int_0^\infty d\tilde{p}\, \frac{\tilde p}{\tilde\omega_{\tilde p}} f(R\tilde\omega_{\tilde p})\,. $$

Rescaling the momentum representation $\phi^{\pm}_{n l m_l} = \tilde{\phi}^{\text{M},\pm}_{n l m_l} 2\omega^{+}_{nl}(p^{\mathbb{R}}_\omega)^l / [\sqrt{2\pi}\,(2l+d-2)!!]$ and using (B.34), for $d=3$ the flat limit of the field expansion becomes the one for a Minkowski slice region:

$$\phi(t,r,\Omega) \xrightarrow[\text{lim.}]{\text{flat}} \int_0^\infty d\tilde{p} \sum_{l,m_l} 2\tilde{p}\,(2\pi)^{-1/2} j_l(\tilde{p}r)\left\{\tilde{\phi}^{+}_{\tilde{p}l m_l} e^{-i\tilde{\omega}_{\tilde{p}}\tau} Y_l^{m_l}(\Omega) + \overline{\tilde{\phi}^{-}_{\tilde{p}l m_l}} e^{i\tilde{\omega}_{\tilde{p}}\tau}\,\overline{Y_l^{m_l}(\Omega)}\right\}. \tag{B.35}$$

For the tube region, with definitions (II.5) and rescaling the momentum representation $\phi^{a}_{\omega \underline{l} m_l} = \tilde{\phi}^{\text{M},a}_{\tilde{\omega} \underline{l} m_l}\tilde{p}^{\mathbb{R}}_{\tilde{\omega}}(p^{\mathbb{R}}_\omega)^l / [4\pi R_{\text{AdS}}(2l+d-2)!!]$ and $\phi^{b}_{\omega \underline{l} m_l} = \tilde{\phi}^{\text{M},b}_{\tilde{\omega} \underline{l} m_l}\tilde{p}^{\mathbb{R}}_{\tilde{\omega}}(p^{\mathbb{R}}_\omega)^{-(l+1)}(2l+d-4)!!/(4\pi R_{\text{AdS}})$, we get the flat limits for $d=3$:

$$\phi(t,r,\Omega) \xrightarrow[\text{lim.}]{\text{flat}} \int d\tilde{\omega} \sum_{l,m_l} \frac{\tilde{p}^{\mathbb{R}}_{\tilde{\omega}}}{4\pi}\left\{\tilde{\phi}^{\text{M},a}_{\tilde{\omega} l m_l} e^{-i\tilde{\omega}\tau} Y_l^{m_l}(\Omega)\,\check{\jmath}_{\tilde{\omega} l}(r) + \tilde{\phi}^{\text{M},b}_{\tilde{\omega} l m_l} e^{-i\tilde{\omega}\tau} Y_l^{m_l}(\Omega)\,\check{n}_{\tilde{\omega} l}(r)\right\}. \tag{B.36}$$

### 6. Jacobi recurrence relations for AdS

In this section we derive the ingredients necessary for calculating the action of $Z_d$ and $\overline{Z_d}$ on the AdS-KG solutions of Jacobi type. We recall the notation $\underline{l} = (l \equiv l_{d-1}, \tilde{\underline{l}})$ with $\tilde{\underline{l}} = (l_{d-2},...,l_2)$. In order to give an idea of the flow of our calculations, for the action of $\overline{Z_d}$ on the Jacobi modes we write down the essential details for obtaining the results. The first thing we note after applying $\overline{Z_d}$ to the KG mode $\mu^{(+)}_{n \underline{l} m_l}(t,\rho,\Omega)$ according to (IV.15) and (IV.26), is that there only appear terms with (magic) frequency $\omega^{+}_{nl}+1$. Since $\omega^{+}_{nl} = 2n + l + \widetilde{m}_+$, we have only two possibilities to realize an increase by 1 through adjusting $n$ and $l$: $\omega^{+}_{nl}+1 = \omega^{+}_{n+1,l-1} = \omega^{+}_{n,l+1}$. This hints to try and find out whether we can decompose $\overline{Z_d} \triangleright \mu^{(+)}_{n \underline{l} m_l}$ as some linear combination of $\mu^{(+)}_{n+1,l-1,\tilde{\underline{l}},m_l}$ and $\mu^{(+)}_{n,l+1,\tilde{\underline{l}},m_l}$. For the hyperspherical harmonics we already dispose of the necessary relations, see Section A. The remaining task is thus to decompose the Jacobi polynomials $P_n^{(\alpha,\beta)}(x)$. In our case we have $\alpha = \gamma^s - 1 = l + d/2 - 1$, while $\beta = \nu$ and $x = \cos 2\rho$. Thus $l \to l \pm 1$ induces $\alpha \to \alpha \pm 1$, and $\beta$ and $x$ remain unaffected. The Jacobi polynomials appear directly as $P_n^{(\alpha,\beta)}(x)$ and as derivative given by DLMF [18.9.15]: $\frac{d}{dx} P_n^{(\alpha,\beta)}(x) = \frac{1}{2}(n+\alpha+\beta+1)P_{n-1}^{(\alpha+1,\beta+1)}(x)$. (By definition, $P_n^{(\alpha,\beta)}(x) \equiv 0$ for all negative $n$.) We thus need to find the coefficients for the relations

$$P_n^{(\alpha,\beta)}(x) = a\,P_{n+1}^{(\alpha-1,\beta)}(x) + b\,P_n^{(\alpha+1,\beta)}(x) \qquad P_{n-1}^{(\alpha+1,\beta+1)}(x) = c\,P_{n+1}^{(\alpha-1,\beta)}(x) + d\,P_n^{(\alpha+1,\beta)}(x). \tag{B.37}$$

Instead of trying to puzzle together the various recurrence relations for Jacobi polynomials given in AS [22.7] and DLMF [18.9], we shall apply the following procedure: first write the Jacobi polynomials as hypergeometric functions using AS [22.5.42]: $P_n^{(\alpha,\beta)}(x) = \binom{n+\alpha}{n} F(-n,\, n+\alpha+\beta+1;\, \alpha+1;\, \frac{1-x}{2})$. We do this for all Jacobi polynomials in (B.37). Second, we use the program of Rakha et al. (see Appendix B 4) to determine the coefficients for the corresponding relations between hypergeometric functions. Third, we convert these relations back to Jacobi polynomials. We start writing the left side of (B.37) using hypergeometric functions:

$$P_n^{(\alpha,\beta)}(x) = \binom{n+\alpha}{n} F(-n,\, n+\alpha+\beta+1;\, \alpha+1;\, \tfrac{1-x}{2}) \quad = \binom{n+\alpha}{n} F(A+1,\, B;\, C+1;\, y) \tag{B.38}$$

$$P_{n+1}^{(\alpha-1,\beta)}(x) = \binom{n+\alpha}{n+1} F(-n-1,\, n+\alpha+\beta+1;\, \alpha;\, \tfrac{1-x}{2}) \quad = \binom{n+\alpha}{n+1} F(A,\, B;\, C;\, y) \tag{B.39}$$

$$P_n^{(\alpha+1,\beta)}(x) = \binom{n+\alpha+1}{n} F(-n,\, n+\alpha+\beta+2;\, \alpha+2;\, \tfrac{1-x}{2}) \quad = \binom{n+\alpha+1}{n} F(A+1,\, B+1;\, C+2;\, y) \tag{B.40}$$

(The placeholders $A, B, C, y$ will be used only to write the relations in a shorter way, their relation to $\alpha, \beta, n, x$ may be different in each calculation!) The program by Rakha et al. gives us the contiguous relation for the hypergeometric functions on the right hand side:

$$0 = C(1+C)\,F(A+1,\, B;\, C+1;\, y) + (-C)(C+1)\,F(A,\, B;\, C;\, y) + y\,B(A-C)\,F(A+1,\, B+1;\, C+2;\, y)\,. \tag{B.41}$$

It can be checked with Mathematica's FullSimplify command, like all such relations in this appendix. Using it, in equation (B.38) we can replace $F(A+1,\, B;\, C+1;\, y)$ by a linear combination of $F(A,\, B;\, C;\, y)$ and $F(A+1,\, B+1;\, C+2;\, y)$. Then converting the hypergeometric functions back to Jacobi polynomials we obtain the simple recurrence relation

$$P_n^{(\alpha,\beta)}(x) = \frac{n+1}{\alpha} P_{n+1}^{(\alpha-1,\beta)}(x) + (1-x)/2\,\frac{n+\alpha+\beta+1}{\alpha} P_n^{(\alpha+1,\beta)}(x)\,. \tag{B.42}$$

Next we write the right side of (B.37) using hypergeometric functions:

$$P_{n-1}^{(\alpha+1,\beta+1)}(x) = \binom{n+\alpha}{n-1} F(-n+1, n+\alpha+\beta+2; \alpha+2; \tfrac{1-x}{2}) \quad = \binom{n+\alpha}{n-1} F(A+2, B+1; C+2; y) \quad (B.43)$$

$$P_{n+1}^{(\alpha-1,\beta)}(x) = \binom{n+\alpha}{n+1} F(-n-1, n+\alpha+\beta+1; \alpha; \tfrac{1-x}{2}) \quad = \binom{n+\alpha}{n+1} F(A, B; C; y) \quad (B.44)$$

$$P_n^{(\alpha+1,\beta)}(x) = \binom{n+\alpha+1}{n} F(-n, n+\alpha+\beta+2; \alpha+2; \tfrac{1-x}{2}) \quad = \binom{n+\alpha+1}{n} F(A+1, B+1; C+2; y) \quad (B.45)$$

The program by Rakha et al. fails to return correct coefficients relating the three hypergeometric functions on the right hand side. However, using (B.30) with AS [15.2.10] $0 = (C-A) F(A-1, B; C; y) + (2A-C+y[B-A]) F(A, B; C; y) + A(y-1) F(A+1, B; C; y)$ with the shifts $A \to A+1$, while $B \to B+1$ and $C \to C+2$ we obtain the contiguous relation

$$0 = (y-1)C(A+1)F(A+2, B+1; C+2; y) + C(C+1)F(A, B; C; y) + \big(C[A-C] + yB[C-A]\big)F(A+1, B+1; C+2; y).$$

Using it, we can replace $F(A+2, B+1; C+2; y)$ by a linear combination of $F(A, B; C; y)$ and $F(A+1, B+1; C+2; y)$ in (B.43). Converting the hypergeometric functions back to Jacobi polynomials we get the recurrence relation

$$P_{n-1}^{(\alpha+1,\beta+1)}(x) = \tfrac{-2}{x+1}\tfrac{n+1}{\alpha} P_{n+1}^{(\alpha-1,\beta)}(x) + \tfrac{2/\alpha}{x+1}\big(\alpha + \tfrac{(x-1)}{2}(n+\alpha+\beta+1)\big) P_n^{(\alpha+1,\beta)}(x). \quad (B.46)$$

Now we can let $\overline{Z_d}$ act on the KG mode $\mu_{n\underline{l}m_l}^{(+)}(t,\rho,\Omega)$, and put to use relations (B.42), (B.46), (A.4) and (A.5). Cleaning up the large expressions resulting from this, we see that the terms containing $Y_{l+1}^{m_l} P_n^{(\alpha-1,\beta)}$ sum up to zero, and the same happens for those containing $Y_{l-1}^{m_l} P_n^{(\alpha+1,\beta)}$. Therefore, we only encounter terms containing either $Y_{l-1}^{m_l} P_{n+1}^{(\alpha-1,\beta)}$ (as needed for $\mu_{n+1,l-1,\tilde{l},m_l}^{(+)}$) or $Y_{l+1}^{m_l} P_n^{(\alpha+1,\beta)}$ (as needed for $\mu_{n,l+1,\tilde{l},m_l}^{(+)}$). This means that we can indeed decompose $\overline{Z_d} \triangleright \mu_{n\underline{l}m_l}^{(+)}$ as a linear combination of $\mu_{n+1,l-1,\tilde{l},m_l}^{(+)}$ and $\mu_{n,l+1,\tilde{l},m_l}^{(+)}$:

$$\overline{Z_d} \triangleright \mu_{n,l,\tilde{l},m_l}^{(+)} = +i\tilde{z}_{nl}^{(+)+-}\,\mu_{n+1,l-1,\tilde{l},m_l}^{(+)} + i\tilde{z}_{nl}^{(+)0+}\,\mu_{n,l+1,\tilde{l},m_l}^{(+)}$$
$$\tilde{z}_{nl}^{(+)+-} = +(2l+d-2)\,\chi_-^{(d-1)}(l,l_{d-2}) \quad (B.47)$$
$$\tilde{z}_{nl}^{(+)0+} = -2(n+l+\widetilde{m}_+)\,\frac{(n+l+\tfrac{d}{2})}{(l+\tfrac{d}{2})}\,\chi_+^{(d-1)}(l,l_{d-2}).$$

Since $\chi_-^{(d-1)}(l,l_{d-2})$ vanishes for $l=0$, so does $\tilde{z}_{nl}^{(+)+-}$, and we don't need to worry about defining modes with negative $l$. Negative $n$ cannot appear in this formula.

For the action of $Z_d$ on the Jacobi modes the calculations are completely analogous to the above steps, and thus we only give the results here. We obtain the recurrence relations

$$P_n^{(\alpha,\beta)}(x) = \tfrac{n+\alpha}{\alpha} P_n^{(\alpha-1,\beta)}(x) + \tfrac{1-x}{2}\tfrac{n+\beta}{\alpha} P_{n-1}^{(\alpha+1,\beta)}(x). \quad (B.48)$$

$$P_{n-1}^{(\alpha+1,\beta+1)}(x) = \tfrac{2/\alpha}{x+1}\tfrac{n(n+\alpha)}{(n+\alpha+\beta+1)} P_n^{(\alpha-1,\beta)}(x) + \tfrac{2/\alpha}{x+1}\tfrac{(n+\beta)+(1-x)n(n+\beta)/2}{\alpha(n+\alpha+\beta+1)} P_{n-1}^{(\alpha+1,\beta)}(x). \quad (B.49)$$

Again we can decompose $Z_d\,\mu_{n\underline{l}m_l}^{(+)}$ as a linear combination of $\mu_{n-1,l+1,\tilde{l},m_l}^{(+)}$ and $\mu_{n,l-1,\tilde{l},m_l}^{(+)}$:

$$Z_d \triangleright \mu_{n,l,\tilde{l},m_l}^{(+)} = \mu_{n,l,\tilde{l},m_l}^{(+)} + iz_{nl}^{(+)0-}\,\mu_{n,l-1,\tilde{l},m_l}^{(+)} + iz_{nl}^{(+)-+}\,\mu_{n-1,l+1,\tilde{l},m_l}^{(+)}$$
$$z_{nl}^{(+)0-} = -(2l+d-2)\,\chi_-^{(d-1)}(l,l_{d-2}) = -\tilde{z}_{nl}^{(+)+-} \quad (B.50)$$
$$z_{nl}^{(+)-+} = +2n\,\frac{(n+\nu)}{(l+\tfrac{d}{2})}\,\chi_+^{(d-1)}(l,l_{d-2}).$$

Since $\chi_-^{(d-1)}(l,l_{d-2})$ vanishes for $l=0$, so does $z_{nl}^{(+)0-}$, and we don't need to worry about defining modes with negative $l$. And $z_{nl}^{(+)-+}$ vanishes for $n=0$ so that we don't need to worry about defining modes with negative $n$ either. The action of $Z_d$ and $\overline{Z_d}$ on the complex-conjugated modes can be found by noting that $Z_d \triangleright \mu_{n,l,\tilde{l},m_l}^{(+)} = \overline{\overline{Z_d} \triangleright \mu_{n,l,\tilde{l},m_l}^{(+)}}$ and $\overline{Z_d} \triangleright \mu_{n,l,\tilde{l},m_l}^{(+)} = \overline{Z_d \triangleright \mu_{n,l,\tilde{l},m_l}^{(+)}}$, and thus results in $\overline{Z_d} \triangleright \overline{\mu_{n,l,\tilde{l},m_l}^{(+)}} = -iz_{nl}^{(+)0-}\,\overline{\mu_{n,l-1,\tilde{l},m_l}^{(+)}} - iz_{nl}^{(+)-+}\,\overline{\mu_{n-1,l+1,\tilde{l},m_l}^{(+)}}$ and $Z_d \triangleright \overline{\mu_{n,l,\tilde{l},m_l}^{(+)}} = -i\tilde{z}_{nl}^{(+)+-}\,\overline{\mu_{n+1,l-1,\tilde{l},m_l}^{(+)}} - i\tilde{z}_{nl}^{(+)0+}\,\overline{\mu_{n,l+1,\tilde{l},m_l}^{(+)}}$.

With equations (IV.26) it is now easy to write down the action of $K_{0,d}$ and $K_{d+1,d}$ on the Jacobi modes:

$$K_{0d} \triangleright \mu^{(+)}_{n,l,\tilde{l},m_l} = \tfrac{\mathrm{i}}{2}\tilde{z}^{(+)+-}_{nl}\mu^{(+)}_{n+1,l-1,\tilde{l},m_l} + \tfrac{\mathrm{i}}{2}\tilde{z}^{(+)0+}_{nl}\mu^{(+)}_{n,l+1,\tilde{l},m_l} + \tfrac{\mathrm{i}}{2}\tilde{z}^{(+)0-}_{nl}\mu^{(+)}_{n,l-1,\tilde{l},m_l} + \tfrac{\mathrm{i}}{2}\tilde{z}^{(+)-+}_{nl}\mu^{(+)}_{n-1,l+1,\tilde{l},m_l} \quad (B.51)$$

$$K_{d+1,d} \triangleright \mu^{(+)}_{n,l,\tilde{l},m_l} = -\tfrac{1}{2}\tilde{z}^{(+)+-}_{nl}\mu^{(+)}_{n+1,l-1,\tilde{l},m_l} - \tfrac{1}{2}\tilde{z}^{(+)0+}_{nl}\mu^{(+)}_{n,l+1,\tilde{l},m_l} + \tfrac{1}{2}\tilde{z}^{(+)0-}_{nl}\mu^{(+)}_{n,l-1,\tilde{l},m_l} + \tfrac{1}{2}\tilde{z}^{(+)-+}_{nl}\mu^{(+)}_{n-1,l+1,\tilde{l},m_l} \quad (B.52)$$

$$K_{0d} \triangleright \overline{\mu^{(+)}_{n,l,\tilde{l},m_l}} = -\tfrac{\mathrm{i}}{2}\tilde{z}^{(+)+-}_{nl}\overline{\mu^{(+)}_{n+1,l-1,\tilde{l},m_l}} - \tfrac{\mathrm{i}}{2}\tilde{z}^{(+)0+}_{nl}\overline{\mu^{(+)}_{n,l+1,\tilde{l},m_l}} - \tfrac{\mathrm{i}}{2}\tilde{z}^{(+)0-}_{nl}\overline{\mu^{(+)}_{n,l-1,\tilde{l},m_l}} - \tfrac{\mathrm{i}}{2}\tilde{z}^{(+)-+}_{nl}\overline{\mu^{(+)}_{n-1,l+1,\tilde{l},m_l}} \quad (B.53)$$

$$K_{d+1,d} \triangleright \overline{\mu^{(+)}_{n,l,\tilde{l},m_l}} = -\tfrac{1}{2}\tilde{z}^{(+)+-}_{nl}\overline{\mu^{(+)}_{n+1,l-1,\tilde{l},m_l}} - \tfrac{1}{2}\tilde{z}^{(+)0+}_{nl}\overline{\mu^{(+)}_{n,l+1,\tilde{l},m_l}} + \tfrac{1}{2}\tilde{z}^{(+)0-}_{nl}\overline{\mu^{(+)}_{n,l-1,\tilde{l},m_l}} + \tfrac{1}{2}\tilde{z}^{(+)-+}_{nl}\overline{\mu^{(+)}_{n-1,l+1,\tilde{l},m_l}}. \quad (B.54)$$

### 7. Hypergeometric recurrence relations for AdS

In this section we derive the $d$-boost actions on the AdS-KG solutions (IV.1) of hypergeometric type. Since the calculation technique is similar to the one for the Jacobi solutions, we give only the results. Using $\frac{\mathrm{d}}{\mathrm{d}x}F(A,B;C;x) = \frac{AB}{C}F(A+1,B+1;C+1;x)$, we obtain the contiguous relations

$$F(A,B;C;x) = F(A-1,B;C-1;x) + x\frac{B}{C}\frac{C-A}{C-1}F(A,B+1;C+1;x) \quad (B.55)$$

$$F(A+1,B+1;C+1;x) = \frac{C/A}{1-x}F(A-1,B;C-1;x) + \frac{(C-1)(A-C)+xB(C-A)}{(1-x)A(C-1)}F(A,B+1;C+1;x). \quad (B.56)$$

Letting $\overline{Z_d}$ act on the KG modes $\mu^{(a,b)}_{\omega l m_l}(t,\rho,\Omega)$, and using (B.55), (B.56), (A.4) and (A.5), results in that we can decompose $\overline{Z_d}\triangleright\mu^{(a,b)}_{\omega l m_l}$ as a linear combination of $\mu^{(a,b)}_{\omega+1,l-1,\tilde{l},m_l}$ and $\mu^{(a,b)}_{\omega+1,l+1,\tilde{l},m_l}$:

$$\overline{Z_d}\triangleright\mu^{(a,b)}_{\omega,l,\tilde{l},m_l} = +\mathrm{i}\tilde{z}^{(a,b)+-}_{\omega l}\mu^{(a,b)}_{\omega+1,l-1,\tilde{l},m_l} + \mathrm{i}\tilde{z}^{(a,b)++}_{\omega l}\mu^{(a,b)}_{\omega+1,l+1,\tilde{l},m_l}$$

$$\tilde{z}^{(a)+-}_{\omega l} = +(2l+d-2)\chi^{(d-1)}_{-}(l,l_{d-2}) \qquad \tilde{z}^{(b)+-}_{\omega l} = +2\beta^b\frac{\alpha^b-\gamma^b}{\gamma^b}\chi^{(d-1)}_{-}(l,l_{d-2}) \quad (B.57)$$

$$\tilde{z}^{(a)++}_{\omega l} = +\frac{2\beta^a(\alpha^a-\gamma^a)}{\gamma^a}\chi^{(d-1)}_{+}(l,l_{d-2}) \qquad \tilde{z}^{(b)++}_{\omega l} = -(2l+d-2)\chi^{(d-1)}_{+}(l,l_{d-2}).$$

Since $\chi^{(d-1)}_{-}(l,l_{d-2})$ vanishes for $l=0$, so do $\tilde{z}^{(a)+-}_{\omega l}$ and $\tilde{z}^{(b)+-}_{\omega l}$, and we don't need to worry about defining modes with negative $l$. Thus for $l=0$ there appears no $(l-1)$-term. While $\tilde{z}^{(a)+-}_{\omega l}$ is always finite, $\tilde{z}^{(a)++}_{\omega l}$ vanishes if either $\beta^a=0$ or $(\gamma^a-\alpha^a)=0$, which happen for the magic frequencies $-\omega^+_{0l}=-(l+\widetilde{m}_+)$ and $-\omega^-_{0l}=-(l+\widetilde{m}_-)$. Further, while $\tilde{z}^{(b)++}_{\omega l}$ is always finite, $\tilde{z}^{(b)+-}_{\omega l}$ vanishes only if either $\beta^b=0$ or $(\alpha^b-\gamma^b)=0$, which happen for $\omega=(l-2)+\widetilde{m}_+$ and $\omega=(l-2)+\widetilde{m}_-$.

Letting $Z_d$ act on the KG modes $\mu^{(a,b)}_{\omega l m_l}(t,\rho,\Omega)$, and using relations (B.55), (B.56), (A.4) and (A.5), again we can decompose $Z_d\triangleright\mu^{(a,b)}_{\omega l m_l}$ as a linear combination of $\mu^{(a,b)}_{\omega-1,l-1,\tilde{l},m_l}$ and $\mu^{(a,b)}_{\omega-1,l+1,\tilde{l},m_l}$:

$$Z_d\triangleright\mu^{(a,b)}_{\omega,l,\tilde{l},m_l} = +\mathrm{i}z^{(a,b)--}_{\omega l}\mu^{(a,b)}_{\omega-1,l-1,\tilde{l},m_l} + \mathrm{i}z^{(a,b)-+}_{\omega l}\mu^{(a,b)}_{\omega-1,l+1,\tilde{l},m_l}$$

$$z^{(a)--}_{\omega l} = -(2l+d-2)\chi^{(d-1)}_{-}(l,l_{d-2}) \qquad z^{(b)--}_{\omega l} = +2(\gamma^b-\beta^b)\frac{\alpha^b}{\gamma^b}\chi^{(d-1)}_{-}(l,l_{d-2}) \quad (B.58)$$

$$z^{(a)-+}_{\omega l} = +\frac{2\alpha^a(\gamma^a-\beta^a)}{\gamma^a}\chi^{(d-1)}_{+}(l,l_{d-2}) \qquad z^{(b)-+}_{\omega l} = +(2l+d-2)\chi^{(d-1)}_{+}(l,l_{d-2}).$$

Since $\chi^{(d-1)}_{-}(l,l_{d-2})$ vanishes for $l=0$, so does $\tilde{z}^{+-}_{nl}$, and we don't need to worry about defining modes with negative $l$. While $z^{(a)--}_{\omega l}$ is always finite except for $l=0$, $z^{(a)-+}_{\omega l}$ vanishes if either $\alpha^a=0$ or $(\gamma^a-\beta^a)=0$, which happen for the magic frequencies $\omega^+_{0l}=l+\widetilde{m}_+$ and $\omega^-_{0l}=l+\widetilde{m}_-$. Further, while $z^{(b)-+}_{\omega l}$ is always finite, $z^{(b)--}_{\omega l}$ vanishes if either $\alpha^b=0$ or $(\gamma^b-\beta^b)=0$, which happen for the frequencies $\omega=-(l-2)-\widetilde{m}_+$ and $\omega=-(l-2)-\widetilde{m}_-$.

With equations (IV.26) it is now easy to write down the action of $K_{0,d}$ and $K_{d+1,d}$:

$$K_{0d}\triangleright\mu^{(a)}_{\omega,l,\tilde{l},m_l} = +\tfrac{\mathrm{i}}{2}\tilde{z}^{(a)+-}_{\omega l}\mu^{(a)}_{\omega+1,l-1,\tilde{l},m_l} + \tfrac{\mathrm{i}}{2}\tilde{z}^{(a)++}_{\omega l}\mu^{(a)}_{\omega+1,l+1,\tilde{l},m_l} + \tfrac{\mathrm{i}}{2}z^{(a)--}_{\omega l}\mu^{(a)}_{\omega-1,l-1,\tilde{l},m_l} + \tfrac{\mathrm{i}}{2}z^{(a)-+}_{\omega l}\mu^{(a)}_{\omega-1,l+1,\tilde{l},m_l} \quad (B.59)$$

$$K_{d+1,d}\triangleright\mu^{(a)}_{\omega,l,\tilde{l},m_l} = -\tfrac{1}{2}\tilde{z}^{(a)+-}_{\omega l}\mu^{(a)}_{\omega+1,l-1,\tilde{l},m_l} - \tfrac{1}{2}\tilde{z}^{(a)++}_{\omega l}\mu^{(a)}_{\omega+1,l+1,\tilde{l},m_l} + \tfrac{1}{2}z^{(a)--}_{\omega l}\mu^{(a)}_{\omega-1,l-1,\tilde{l},m_l} + \tfrac{1}{2}z^{(a)-+}_{\omega l}\mu^{(a)}_{\omega-1,l+1,\tilde{l},m_l} \quad (B.60)$$

$$K_{0d}\triangleright\mu^{(b)}_{\omega,l,\tilde{l},m_l} = -\tfrac{\mathrm{i}}{2}\tilde{z}^{(b)+-}_{\omega l}\mu^{(b)}_{\omega+1,l-1,\tilde{l},m_l} - \tfrac{\mathrm{i}}{2}\tilde{z}^{(b)++}_{\omega l}\mu^{(b)}_{\omega+1,l+1,\tilde{l},m_l} - \tfrac{\mathrm{i}}{2}z^{(b)--}_{\omega l}\mu^{(b)}_{\omega-1,l-1,\tilde{l},m_l} - \tfrac{\mathrm{i}}{2}z^{(b)-+}_{\omega l}\mu^{(b)}_{\omega-1,l+1,\tilde{l},m_l} \quad (B.61)$$

$$K_{d+1,d}\triangleright\mu^{(b)}_{\omega,l,\tilde{l},m_l} = -\tfrac{1}{2}\tilde{z}^{(b)+-}_{\omega l}\mu^{(b)}_{\omega+1,l-1,\tilde{l},m_l} - \tfrac{1}{2}\tilde{z}^{(b)++}_{\omega l}\mu^{(b)}_{\omega+1,l+1,\tilde{l},m_l} + \tfrac{1}{2}z^{(b)--}_{\omega l}\mu^{(b)}_{\omega-1,l-1,\tilde{l},m_l} + \tfrac{1}{2}z^{(b)-+}_{\omega l}\mu^{(b)}_{\omega-1,l+1,\tilde{l},m_l}. \quad (B.62)$$

Since $\mu^{(a,b)}_{-\omega,\underline{l},-m_l} = \overline{\mu^{(a,b)}_{\omega,\underline{l},m_l}}$ and both $K_{d+1,d}$ and $K_{0d}$ are real, we can write $K_{d+1,d} \triangleright \mu^{(a,b)}_{-\omega,\underline{l},-m_l} = K_{d+1,d} \triangleright \overline{\mu^{(a,b)}_{\omega,\underline{l},m_l}} = \overline{K_{d+1,d} \triangleright \mu^{(a,b)}_{\omega,\underline{l},m_l}}$ and $K_{0d} \triangleright \mu^{(a,b)}_{-\omega,\underline{l},-m_l} = K_{0d} \triangleright \overline{\mu^{(a,b)}_{\omega,\underline{l},m_l}} = \overline{K_{0d} \triangleright \mu^{(a,b)}_{\omega,\underline{l},m_l}}$.

---

[1] G. Holzegel and J. Smulevici, [arxiv:gr-qc/1103.3672v1] (2011).
[2] K. Yagdjian and A. Galstian, Rend. Sem. Mat. Univ. Pol. Torino, **Vol. 67, No.2**, 271 (2009).
[3] J. Maldacena, Adv. Theor. Math. Phys., **2**, 231 (1998), [hep-th/9711200].
[4] E. Witten, Adv. Theor. Math. Phys., **2**, 253 (1998), [arXiv:hep-th/9802150v2].
[5] S. Avis, C. Isham, and D. Storey, Phys. Rev. D, **Vol. 18, Number 10**, 3565 (1978).
[6] V. Balasubramanian, P. Kraus, and A. Lawrence, [arxiv:hep-th/9805171] (1998).
[7] P. Dirac, Ann. of Math. (2nd series), **Vol.36 No. 3**, 657 (1935), [jstor.org/stable/1968649].
[8] C. Fronsdal, Rev. of Mod. Phys, **Vol. 37 Number 1**, 221 (1965).
[9] R. Raczka, N. Limic, and J. Niederle, J. Math. Phys., **7**, 1861 (1966).
[10] V. Balasubramanian, S. Giddings, and A. Lawrence, JHEP, **9903**, 001 (1999).
[11] N. Limic, J. Niederle, and R. Raczka, J. Math. Phys., **7**, 2026 (1966).
[12] C. Fronsdal, Phys. Rev. D, **Vol. 10 Number 2**, 589 (1974).
[13] J. Podolsky and J. Griffiths, Phys. Rev D, **56 No. 8**, 4756 (1997).
[14] I. Bengtsson, "Anti-de Sitter space," [physto.se/∼ingemar/] (1998), (Lecture Notes).
[15] O. Aharony, S. Gubser, J. Maldacena, H. Ooguri, and Y. Oz, Phys.Rept., **323:183-386** (2000).
[16] M. Dohse, (Master Thesis) (2007), [arxiv:0706.1887].
[17] P. Breitenlohner and D. Freedman, Physics Letters, **115B, Number 3**, 197 (1982).
[18] P. Breitenlohner and D. Freedman, Annals of Physics, **144**, 249 (1982).
[19] L. Mezincescu and P. Townsend, University of Texas Preprint, **UTTG-8-84** (1984).
[20] C. Burgess and C. Lutken, University of Texas Preprint, **UTTG-29-84** (1984).
[21] C. Dullemond and E. van Beveren, J. Math. Phys., **Vol. 26 Nr. 8**, 2050 (1985).
[22] S. Giddings, Phys.Rev.Lett., **83:2707-2010** (1999), [arxiv:hep-th/9903048].
[23] M. Gary and S. Giddings, arXiv:0904.3544v3 (2009), [arxiv:0904.3544].
[24] H. Dorn, G. Jorjadze, C. Kalousios, and J. Plefka, J.Phys.A (IOP), **44:095402** (2011), [arxiv:1011.3416].
[25] M. Abramowitz and I. Stegun, *Handbook of mathematical functions* (Dover Publ., 1998).
[26] NIST, "Digital library of mathematical functions," [dlmf.nist.gov/] (2010), release date 2010-05-07.
[27] N. Woodhouse, *Geometric quantization (2nd edition)* (Oxford University Press, 1991).
[28] R. Oeckl, SIGMA, **8**, 050 (2012), [arxiv:1009.5615].
[29] N. Vilenkin and A. Klimyk, *Representations of Lie groups and special functions (Vol. 2)* (Kluwer, 1993).
[30] R. Oeckl, J. Geom. Phys., **62**, 1373 (2012), [arxiv:1104.5527].
[31] P. Di Francesco, P. Mathieu, and D. Sénéchal, *Conformal Field Theory* (Springer, 1997).
[32] C. Warnick, ArXiV (2012), [arXiv:1202.3445v2].
[33] R. Oeckl, J. Math. Phys., **53**, 072301 (2012), [arxiv:1109.5215].
[34] R. Oeckl, Adv.Theor.Math.Phys., **12**, 319 (2008), [arxiv:hep-th/0509122].
[35] R. Oeckl, Conference Proceedings: Quantum Theory: Reconsideration of Foundations - 6, Växjö, 2012 (2012), [arxiv:1210.0944].
[36] D. Colosi and R. Oeckl, Phys. Rev., **D 78:025020** (2008), [arxiv:0802.2274].
[37] D. Colosi and R. Oeckl, Phys. Lett., **B665**, 310 (2007), [arxiv:0710.5203].
[38] D. Colosi, [arxiv:1010.1209] (2010).
[39] D. Colosi and M. Dohse, ArXiV (2010), [arxiv:1011.2243].
[40] D. Colosi, M. Dohse, and R. Oeckl, JPCS, **360** (2012), [iopscience.iop.org/1742-6596/360/1/012012].
[41] J. Avery, *Hyperspherical harmonics and generalized Sturmians* (Kluwer Acad. Publ., 2000).
[42] V. Aquilanti, S. Cavalli, and C. Coletti, Chem. Phys., **214**, 1 (1997).
[43] E. Wigner, *Group theory and its applications to the quantum mechanics of atomic spectra* (Acad. Press, 1959).
[44] A. Ibrahim and M. Rakha, Computers and Mathematics with Applications, **56**, 1918 (2008).
[45] M. Rakha, A. Ibrahim, and A. Rathie, Commun. Korean Math. Soc., **24 No. 2**, 291 (2009).